\documentclass[letterpaper,twocolumn,10pt]{article}
\usepackage{usenix,epsfig,endnotes}
\usepackage{float}
\usepackage{multirow}
\usepackage{makecell}
\usepackage{adjustbox}
\usepackage{array}
\usepackage{float}
\usepackage{url}
\usepackage{amsmath}
\usepackage{mathtools}

\usepackage{amssymb}
\usepackage{xspace}
\usepackage{graphicx}
\usepackage{grffile}
\usepackage{longtable}
\usepackage{amsmath}
\usepackage{xcolor}

\usepackage{enumitem}
\usepackage{cite}
\usepackage{microtype}
\usepackage{xspace}
\usepackage{pifont}
\usepackage{hyperref}
\usepackage{caption}
\usepackage{subcaption}
\usepackage{algorithm}
\usepackage{algpseudocode}
\usepackage{float}
\usepackage{comment}
\usepackage{multicol}

\usepackage{breakurl}
\usepackage{titlesec}
\usepackage{booktabs}
\titlespacing*{\section}{0pt}{0.5\baselineskip}{0.5\baselineskip}
\titlespacing*{\subsection}{0pt}{0.2\baselineskip}{0.2\baselineskip}
\titlespacing*{\subsubsection}{0pt}{0.1\baselineskip}{0.1\baselineskip}

\PassOptionsToPackage{hyphens}{url}
\let\oldding\ding% Store old \ding in \oldding
\renewcommand{\ding}[2][1]{\scalebox{#1}{\oldding{#2}}}% Scale \oldding via optional argument
\usepackage{array}
\newcolumntype{L}[1]{>{\raggedright\let\newline\\\arraybackslash\hspace{0pt}}m{#1}}
\newcolumntype{C}[1]{>{\centering\let\newline\\\arraybackslash\hspace{0pt}}m{#1}}
\newcolumntype{R}[1]{>{\raggedleft\let\newline\\\arraybackslash\hspace{0pt}}m{#1}}

\newcolumntype{P}[1]{>{\centering\arraybackslash}p{#1}}

\newcommand{\imtiaz}[1]{\textcolor{blue}{Imtiaz: #1}}
\newcommand{\kazi}[1]{\textbf{\color{teal}Kazi: #1}}

\newcommand{\M}[1]{\footnotesize\textit{#1}\normalsize\xspace}

\newcommand{\MTable}[1]{\scriptsize\textit{#1}\scriptsize\xspace}

\newcommand{\PacketState}[1]{\footnotesize\texttt{}{#1}\normalsize\xspace}

\newcommand{\system}{\ensuremath{\mathsf{FBSDetector}}\xspace}

\newcommand{\fbsdataset}{\ensuremath{\mathsf{FBSAD}}\xspace}
\newcommand{\msadataset}{\ensuremath{\mathsf{MSAD}}\xspace}

\newcommand{\TrackingAreaUpdateRequest}{\footnotesize\textit{TrackingAreaUpdateRequest}\normalsize\xspace}

\newcommand{\totalLinesOfCodeAddedInsrsRANforFBS}{\ensuremath{\mathsf{316}}\xspace}
\newcommand{\totalLinesOfCodeAddedInsrsRANforMultiStep}{\ensuremath{\mathsf{542}}\xspace}
\newcommand{\totalLinesOfCodeinDataProcessing}{\ensuremath{\mathsf{72}}\xspace}
\newcommand{\totalLinesOfCodeinTraingMLModels}{\ensuremath{\mathsf{40}}\xspace}
\newcommand{\totalLinesOfCodeinSequentialLSTM}{\ensuremath{\mathsf{42}}\xspace}
\newcommand{\totalLinesOfCodeinGraphLearning}{\ensuremath{\mathsf{70}}\xspace}
\newcommand{\totalLinesOfCodeinAppDevelopment}{\ensuremath{\mathsf{290}}\xspace}

\newcommand{\totaluserstwentyfour}{\ensuremath{\mathsf{4.88}}\xspace}
\newcommand{\totaluserstwentyfive}{\ensuremath{\mathsf{5.28}}\xspace}
\newcommand{\totaldevicestwentyfour}{\ensuremath{\mathsf{17.72}}\xspace}
\newcommand{\totaldevicestwentyfive}{\ensuremath{\mathsf{18.22}}\xspace}
\newcommand{\rateofincinusers}{\ensuremath{\mathsf{4.9\%}}\xspace}

\newcommand{\totalnumofattacks}{\ensuremath{\mathsf{21}}\xspace}

\newcommand{\totalsizeofdata}{\ensuremath{\mathsf{9.2}}\xspace}

\newcommand{\totalfbsbenignnas}{\ensuremath{\mathsf{907}}\xspace}
\newcommand{\totalfbsmaliciousnas}{\ensuremath{\mathsf{755}}\xspace}
\newcommand{\totalfbsnas}{\ensuremath{\mathsf{1662}}\xspace}
\newcommand{\totalfbsbenignrrc}{\ensuremath{\mathsf{338195}}\xspace}
\newcommand{\totalfbsmaliciousrrc}{\ensuremath{\mathsf{1618}}\xspace}
\newcommand{\totalfbsrrc}{\ensuremath{\mathsf{339813}}\xspace}

\newcommand{\totalmultistepbenignnas}{\ensuremath{\mathsf{502}}\xspace}
\newcommand{\totalmultistepmaliciousnas}{\ensuremath{\mathsf{15962}}\xspace}
\newcommand{\totalmultistepnas}{\ensuremath{\mathsf{16464}}\xspace}
\newcommand{\totalmultistepbenignrrc}{\ensuremath{\mathsf{391810}}\xspace}
\newcommand{\totalmultistepmaliciousrrc}{\ensuremath{\mathsf{1874}}\xspace}
\newcommand{\totalmultisteprrc}{\ensuremath{\mathsf{393684}}\xspace}

\newcommand{\fbsdetectionaccuracy}{\ensuremath{\mathsf{96\%}}\xspace}
\newcommand{\msadetectionaccuracy}{\ensuremath{\mathsf{86\%}}\xspace}

\newcommand{\fbsdetectionfpr}{\ensuremath{\mathsf{2.96\%}}\xspace}
\newcommand{\msadetectionfpr}{\ensuremath{\mathsf{3.28\%}}\xspace}

\newcommand{\enseblelearningimprovement}{\ensuremath{\mathsf{1\sim2\%}}\xspace}

\newcommand{\powerusage}{\ensuremath{\mathsf{2}}\xspace}
\newcommand{\memoryusage}{\ensuremath{\mathsf{835}}\xspace}

\usepackage{fancyhdr}

\fancypagestyle{firstpage}{%
  \fancyhf{} % Clear header and footer
  \fancyhead[C]{\small This paper has been accepted to appear at USENIX Security ‘25.} % Centered header text
}

\usepackage[available, functional]{usenixbadges}
\begin{document}
\thispagestyle{firstpage} % Apply the custom header style to the first page

%\title{\system: Machine Learning Based Fake Base Station and Multi Step Attack Detection in Cellular Networks}

\title{Gotta Detect 'Em All: Fake Base Station and Multi-Step Attack Detection in Cellular Networks}
% \author{\IEEEauthorblockN{Kazi Samin Mubasshir}
% \IEEEauthorblockA{Purdue University\\
% kmubassh@purdue.edu}
% \and
% \IEEEauthorblockN{Imtiaz Karim}
% \IEEEauthorblockA{Purdue University\\
% karim7@purdue.edu}
% \and
% \IEEEauthorblockN{Elisa Bertino}
% \IEEEauthorblockA{Purdue University\\
% bertino@purdue.edu}}

\author{
% Anonymous Submission
{\rm Kazi Samin Mubasshir*, Imtiaz Karim*, Elisa Bertino}\\
Purdue University
% \and
% {\rm Imtiaz Karim*}\\
% Purdue University
% \and
% {\rm Elisa Bertino}\\
% Purdue University
}
\maketitle

\begin{abstract}
Fake base stations (FBSes) %, also known as rogue base stations, stingrays, and IMSI catchers,
pose a significant security threat by impersonating legitimate base stations (BSes). %and compromising cellular communication's integrity, privacy, and availability. 
Though efforts have been made to defeat this threat, up to this day, the presence of FBSes and the multi-step attacks (MSAs) stemming from them can lead to unauthorized surveillance, interception of sensitive information, and disruption of network services. % for legitimate users. %Different multi-step attacks on the cellular network use FBSes in their threat model.
% motivation
Therefore, detecting these malicious entities is crucial to ensure the security and reliability of cellular networks. 
% What do we do currently
% However, 
%Traditional detection methods often rely on additional hardware, predefined rules, signal scanning, changes to protocol specifications, or cryptographic mechanisms that have limitations and incur huge infrastructure costs in accurately identifying FBSes. 
% What we did in this paper
In this paper, we develop \system--an effective and efficient detection solution that can reliably detect FBSes and MSAs from layer-3 network traces using machine learning (ML) at the user equipment (UE) side. %enhancing the overall security of cellular networks. Our approach is computationally efficient, uses low resources in terms of memory and power, does not require modifications to mobile phones, and takes advantage of the unique fingerprints exhibited by FBSes. 
% results
% \elisa{I have made some minor changes to the sentence below.}
% To develop \system, we create \fbsdataset and \msadataset, the \emph{first-ever} high-quality and large-scale datasets \ %using the POWDER testbed by 
% for training machine learning models able to detect FBSes and MSAs. 
To develop \system, we create \fbsdataset and \msadataset, the \emph{first-ever} high-quality and large-scale datasets incorporating instances of FBSes and \totalnumofattacks MSAs. %for training machine learning models capable of detecting FBSes and MSAs.
These datasets capture the network traces in different real-world cellular network scenarios (including mobility and different attacker capabilities) incorporating legitimate BSes and FBSes. 
%
% The combined network trace has a volume of \totalsizeofdata GB containing \totalnumberofpackets packets. 
%
%
Our novel ML framework, 
specifically designed to detect FBSes in a multi-level approach for packet classification using stateful LSTM with attention and trace level classification and MSAs using graph learning, 
% \imtiaz{Add here.}
can effectively detect FBSes with an accuracy of \fbsdetectionaccuracy and a false positive rate of \fbsdetectionfpr, and recognize MSAs with an accuracy of \msadetectionaccuracy and a false positive rate of \msadetectionfpr. 
We deploy \system as a real-world solution to protect end-users through a mobile app and extensively validate it in real-world environments. Compared to the existing heuristic-based solutions that fail to detect FBSes, \system can detect FBSes in the wild in real time.
\end{abstract}

\renewcommand{\thefootnote}{}
\footnotetext{* Equal contribution. The student author’s name is given first.}
\renewcommand{\thefootnote}{\arabic{footnote}}


% \imtiaz{Include the number of MSA in the abstract.}

% \fi
%%\IEEEpeerreviewmaketitle

\section{Introduction}
\looseness = -1
The widespread adoption of cellular networks has brought about unprecedented improvements in data rates, latency, and device connectivity, resulting in a surge in the number of connected devices worldwide. 
In 2024, the number of mobile devices is assessed at \totaldevicestwentyfour billion, having an estimated \totaluserstwentyfour billion global users, 
% \imtiaz{Use 2024 data.}
marking a \rateofincinusers annual increase~\cite{oberlo-total-smartphones-in-2024}, and the number of mobile devices is expected to reach \totaldevicestwentyfive billion by 2025 having \totaluserstwentyfive billion users~\cite{statista-number-of-mobile-devices-worldwide-from-2020-to-2025}.
% \imtiaz{Revise the numbers, and add citation!}
% Cellular Network components
%The cellular network infrastructure consists of three main components: core network, base stations, and mobile devices, each playing a crucial role in enabling wireless communication. %The core network is the central hub, managing call and data traffic and authenticating users. Mobile devices, such as smartphones, tablets, and IoT devices, are the end-users of the network, connecting to base stations for communication. They are equipped with radios to send and receive signals, allowing users to make calls, send messages, and access data services.
% Base Stations
%Among them, Base stations (BSes) constitute the basic infrastructure of today’s cellular networks, providing essential services for wireless communication. They ensure radio coverage, facilitate signal transmission and reception for connected devices, manage call and data traffic, and optimize resource allocation for reliable communication. Additionally, they play a crucial role in mobility management, enabling seamless connectivity as devices move within the network. At present, millions of base stations exist all over the world, serving billions of mobile devices. 
% Therefore, the security of the base stations is critical to ensure the seamless operation of this vast worldwide infrastructure. 
Given the extensive usage and dependence on cellular networks, they become desirable targets for malicious entities. %seeking to exploit vulnerabilities for ransom. 
These entities try to disrupt cellular network services by exploiting different types of vulnerabilities in cellular network protocols. % with the help of sophisticated hardware. 
% \imtiaz{Removed IMSI catchers according to R4.}
Of all the attacks and threat models targeting cellular networks, Fake Base Stations (FBSes), a.k.a. false base stations and rouge base stations, are among the most widespread and represent a significant threat.
% Vulnerabilities in the cellular network protocol enable the creation of Fake Base Stations (FBSes), which are unauthorized devices impersonating legitimate base stations.
Due to the lack of authentication of initial broadcast messages as well as the unprotected connection setup in the bootstrapping phase~\cite{PHOENIX, Insecure-Connection-Root-of-All-Evil, Yomna2020PreAuthentication},  adversaries can install FBSes %and compromise existing BSs 
that can lure unsuspecting devices to connect to them and then launch sophisticated Multi-Step Attacks (MSAs)~\cite{FBSSpecifications, LTEFuzz, Hussain2018LTEInspectorAS, 5GReasoner, Hussain2019PrivacyAT, altaf2016, shaik2019new, altaf_wisec18, 5G-Spector:NDSS24, kotuliak2022ltrack}. %~\cite{Hussain2018LTEInspectorAS, 5GReasoner, DIKEUE, prochecker, TouchingtheUntouchables, ChlostaLTESecurityDisabled, Borgaonkar2019NewPT, CallMeMaybe, rupprecht2019breaking}.
%FBSes, also known as rogue base stations, stingrays, and IMSI catchers, impersonate legitimate base stations to compromise cellular communication’s integrity~\cite{DoLTEst, LTEFuzz}, privacy~\cite{Hussain2018LTEInspectorAS, altaf2016, 5GReasoner, Hussain2019PrivacyAT}, and availability~\cite{Hussain2018LTEInspectorAS, erni2022adaptover, altaf2016, 5GReasoner}. %Typical cellular devices, in the presence of multiple available networks, tend to choose to connect to the one with the highest signal strength~\cite{Chen2014}. Malicious third parties exploit this phenomenon by setting up their high signal-strength FBSes and making the nearby clients want to attach to them. With FBSes, attackers can gather private information of the user~\cite{Hussain2018LTEInspectorAS, altaf2016, 5GReasoner, Hussain2019PrivacyAT}, track the physical location of mobile devices~\cite{Hussain2018LTEInspectorAS, kotuliak2022ltrack}, and launch denial-of-service attacks~\cite{Hussain2018LTEInspectorAS, erni2022adaptover, altaf2016, 5GReasoner}. 
% A Persistent Problem

\noindent \textbf{Motivation.} The threat posed by FBSes is not new and has been around for a while, but they are still extensively used by attackers worldwide~\cite{FBS-News, Stingray-THAI-NEWSROOM}. The key motivations 
% \elisa{I would replace \system below with "our work", as so far \system has not been introduced.}
of our work are: 
(1) \underline{\textit{A persistent problem.}} 
% \elisa{I'm not sure saying "be created" is correct here. Can we say "be deployed and used for attacks"?}
% Despite numerous efforts, FBSes are still one of the most persistent problems in cellular network security and can be deployed and used for attacks in 5G cellular networks, despite the use of the encrypted permanent ID SUCI compared to the unencrypted IMSI in 4G, which the UE sends to the base station for authentication
Despite considerable efforts, FBSes remain a persistent challenge in cellular network security. They can still be deployed for attacks in 5G cellular networks, even with the use of the encrypted permanent ID (SUCI), in contrast to the unencrypted IMSI used in 4G, which the UE transmits to the base station for authentication~\cite{chlosta20215g}.
Recent efforts have introduced certificate-based solutions and digital signatures~\cite{BARON, Look-Before-You-Leap, FBSSpecifications} to address such problems. Nonetheless, these proposals are still in their infancy and would impose substantial overhead and require changes to the specifications to be applicable in a real-world setting. Moreover, the proposed certificate-based solutions make the design of roaming difficult for cellular network providers. When a user travels from one place to another place or from one country to another country, the different network providers will have to share their secret keys for authentication, which is very challenging considering the security aspects of the sharing process~\cite{MobileAtlas}.  %Also if the keys are updated at one point the providers will have to communicate it with all the providers they shared it with. 
%
% As a result of 5G deployments, the number of BSs will substantially increase; in addition, many BSs will cover small cells and support specialized application domains, such as manufacturing, healthcare, critical infrastructures, etc. Note that while recent efforts toward open radio access network (O-RAN) enhance flexibility and resource optimization, they also open the way to attacks leveraging weaknesses in the implementation of the BS software functions and the platforms hosting these functions. Fake BSs and legitimate BSs, compromised because of vulnerabilities in their code and in other protocol implementations, such as DNS, IPSec, and MQTT~\cite{Lin2021pdf}, could be used to launch disruptive attacks in those domains. It is thus critical to design and deploy techniques to detect FBSs and compromised BSs.
%
%\imtiaz{Revise citations here.}
(2) \underline{\textit{Billions of unprotected devices. }} The 3GPP is planning to incorporate several defense mechanisms in the protocol to defend against FBSes~\cite{FBSSpecifications} in future generations. However, these defense mechanisms will still take several years to roll out in the specification and then to the implementation. Currently, billions of devices worldwide are vulnerable to FBS-based attacks, and billions of new devices released in the coming years will be vulnerable to attacks until the new protocol is implemented. These devices need to be secured through in-device solutions because replacing them is impractical and would incur huge costs. 
%
% \imtiaz{Please incorporate the discussion about roaming in the first section when you talk about these authentication-based solutions. This will basically show that though 3GPP and researchers are talking about signature or authentication-based defenses they are far from being practical.}
%
%Recent machine learning (ML) techniques have been shown to be effective in detecting attacks and anomalies from traces \cite{Trustin5G, Onetraceisallittakes, NakarmiML2022}. Different researchers show that information related to the connection between a UE and a BS can be collected from the UE’s traces and used to reason about the authenticity of the BS\cite{Arslan2019, Hamad2019ICCAIS, Dabrowski2016TheMS, CatchMeIfYouCan, Huang2018IdentifyingTF, Li2017FBSRadarUF, Murat}. \imtiaz{Add one sentence to show the issues of these two groups of works.}
%
% Therefore, our goal is to use ML to detect fake/compromised BSs. 
%We discuss the overview of our plan in Section \ref{sec:overview} and the detailed design of our framework in Section \ref{sec:detailed_design}.
%
(3) \underline{\textit{Impracticality and high cost of existing detection mechanisms.}}
Although efforts~\cite{ Trustin5G, Karaçay2021, Murat, CatchMeIfYouCan, Cooper2021, Arslan2019, Hamad2019ICCAIS, Huang2018IdentifyingTF, NakarmiML2022, Li2017FBSRadarUF, Darshak, AIMSICD, PHOENIX, SnoopSnitch,5G-Spector:NDSS24, OVERWATCH, SeaGlass, CellDAM} have been undertaken to detect %and take down 
FBSes, 
% based on lower-layer traces/fingerprinting, 
% several challenges have hampered these efforts.
they suffer from at least one of the following limitations:
% The current solutions are complicated and costly, which is challenging to incorporate for the companies that run the networks. 
% Although prior solutions aim to %defend, 
% detect and take down FBSes and MSAs, 
%\imtiaz{Citation merging is not working properly. You might need to include some package as far as I remember. Let me know if it does not work. I will have a look.}
%(A) The solutions based on digital signatures for base station authentication~\cite{Look-Before-You-Leap, Singla2020ProtectingT4, FBSSpecifications, Trustin5G, BARON, Insecure-Connection-Root-of-All-Evil} incur substantial overhead and require major changes to the specifications to be applicable in a real-world setting and are not practical for roaming.
(A) Heuristic~\cite{Darshak, AIMSICD, SnoopSnitch} and signature/rule~\cite{PHOENIX,5G-Spector:NDSS24, Murat} based solutions fail to adapt to the detection of ever-evolving attacks.
(B) Some solutions depend on crowd-sourced data~\cite{Li2017FBSRadarUF}, which is impractical to scale up to a system that has to protect billions of devices.
(C) Some detection solutions~\cite{CatchMeIfYouCan, Cooper2021, OVERWATCH, SeaGlass} require installing expensive hardware; in most cases, they are proprietary. For example, CellDAM~\cite{CellDAM} 
% detects attacks via verifying data signaling correctness with state-dependent model checking. However, it 
% works in the data plane and 
requires a separate companion node near 
% \elisa{by device do you refer to the UE?}
the UEs to capture the signaling messages.
% from/to the protected device. 
This is not a practical solution for protecting billions of devices.
(D) Lower layer based solutions~\cite{ Arslan2019, Hamad2019ICCAIS, Huang2018IdentifyingTF, Karaçay2021, NakarmiML2022, Trustin5G} cannot detect sophisticated FBSes, and are not inherently able to detect MSAs. 

Recently, both Google and Apple have adopted new approaches to defeat 
%the prevalence of 
FBSes~\cite{New-Ways-to-Defeat-Cell-Site-Simulators}. 
% In 2021, Google released an optional feature for Android to turn off the ability to connect to 2G cell sites.
% Apple announced that in iOS 17, out September 18, iPhones will not connect to insecure 2G mobile towers if placed in Lockdown Mode.
% These approaches are very promising within their scope, however,
% these measures have little impact on 
These approaches are very promising within their scope; however, a knowledgeable and well-equipped adversary beyond their scope can still continue to operate. 
% and protecting against 2G FBSes is insufficient. 
% \elisa{do we discuss later in the paper the limitations of those approaches? As these approaches are by companies that are big players in the mobile phone area, readers may be interested to know what these limitations are. }
% What we do
Therefore, it is essential to design effective and efficient solutions to detect FBSes and MSAs.  
In this paper, we aim to address 
this
% such a need
%solve the problem of FBS detection 
by developing a low-overhead, no-cost, and in-device solution that can effectively detect FBSes and MSAs that use FBSes in their threat model, from the network traces in the UEs.

\noindent\textbf{A practical solution.}
Machine learning (ML) can be a practical in-device solution to address all the existing problems in detecting and defending against the 
% \elisa{what do you mean by "evolving FBSes"? before we talked about evolving attacks? Or will we have FBS technology rapidly changing?}
FBSes and MSAs. On a high level, 
an ML-based solution has the following benefits: %To detect FBSes and multi-step attacks from network traces in the UEs, more specifically using upper (layer 3) protocol traces, machine learning adds the following benefits: 
\ding[1.2]{182} It can protect all the existing end-user devices vulnerable to the attacks.
\ding[1.2]{183} No change is required in the protocol. Using the network traces, especially the higher layer (layer-3) traces, the ML algorithms can determine whether there are any FBSes in the network and recognize MSAs.
\ding[1.2]{184} No additional hardware is required.
% to detect FBSes or MSAs. 
% \elisa{I'm not sure that we need the sentence below. We have already said before that we focus on a in-device solution. So it is clear that the algorithm will run on the UE.}
% The ML algorithms can run on the UEs.
\ding[1.2]{185} ML algorithms add little overhead regarding memory and power consumption.
\ding[1.2]{186} As the attack patterns are similar worldwide, ML algorithms can detect FBSes and MSAs anywhere in the world and thus can support roaming of the device they are deployed in. %A critical issue when applying ML techniques to FBS and MSA detection is which features to use. Most previous solutions rely on features extracted from the physical layer traces. \imtiaz{Add citation}. However, as mentioned earlier, such approaches have several limitations. Our insight about this issue is that features extracted from traces at the layer-3 protocols are much more informative--which is critical for being able to detect not only FBSes but also detect and classify MSAs, especially when dealing with sophisticated attackers.

% ===============================================
% Critical challenge to an ML-based solution
% ===============================================
\noindent\textbf{Challenges.} 
The design of an ML-based system for the detection of FBSes, however, requires addressing several major challenges: 
(1) Acquiring a comprehensive and high-quality dataset encompassing a wide range of real-world cellular network scenarios.
(2) Incorporating the surrounding context into consideration. %Unlike traditional problems where packets can be inherently malicious, in this situation, the true nature of a packet hinges on \emph{when} and \emph{in which context} it was sent.
(3) Capturing, representing, and learning the unique characteristics that define the MSAs. %The actual challenge lies in learning these complex attack signatures. This task requires the deployment of sophisticated ML algorithms that not only recognize known attack patterns but also adapt to evolving attacks. 
% \elisa{ (4) is a nice contribution, but it appears as if someone else has already proposed, and here we worry about how to combine the predictions of the layer-3 protocols; so it would be good if in the previous paragraph (that is, a practical solution), we say something like  "A critical issue when applying ML techniques to FSB and MSA detection is which features to use. Most previous solutions rely on features extracted from the physical layer traces. However, as mentioned earlier, such approaches have several limitations. Our insight about this issue is that features extracted from traces at the layer-3 protocols are much more informative - which is critical for being able to detect not only FBSes, but also detect and classify MSAs, and when dealing with sophisticated attackers". In case other approaches use traces at layer 3, we can discuss them in the related work section.}
(4) Combining the predictions of layer-3 protocols. Layer-3, also known as the network layer, has two key protocols: (i) NAS (Non-Access Stratum)~\cite{NAS} and (ii) RRC (Radio Resource Control)~\cite{RRC}. These protocols operate within the control plane, each serving distinct purposes in facilitating communication between the User Equipment (UE) and the network infrastructure. To have a unified detection and improve robustness, the predictions made separately for the two protocols need to be combined.
(5) Enabling real-time detection of the attacks. The framework needs to be an in-device solution that captures, processes and analyzes incoming packets promptly to detect 
% and 
% \elisa{I'm not sure we should keep "respond" in this sentence as this means that our approach will take some actions to deal with the presence of FBS and the attack. }
% respond to 
the presence of the attacks effectively.

% ==============================
% Our approach
% ==============================
% \elisa{Reviewers may say that there could be other solutions non-UE-centric; should we discuss this somewhere before mentioning the requirements below?}

\noindent\textbf{Our approach.} 
To detect FBSes effectively, in this paper, we present the design, implementation, and deployment of \system, an ML-based framework for FBS detection and MSA recognition.
%===============================
% Addressing the challenges
%===============================
% (1) Creating a dataset 
As it is illegal to create FBSes in public areas and there are no publicly available datasets of FBS and MSA traces, we create \fbsdataset and \msadataset, the \emph{first comprehensive} and \emph{high-quality} FBS and MSA datasets, respectively. To achieve this, we utilize different facilities at POWDER~\cite{powder}. POWDER is a city-scale and end-to-end software-defined platform to support mobile and wireless research.
% Motivate here 
% \imtiaz{Motivate here: 2 lines. Question is: Why POWDER created dataset is a analogous to a real-world dataset and why it is not just simulation. Ans: 1. Over-the-air packets, 2. Spectrums, 3. POWDER paper+Other papers}
The dataset created using POWDER is analogous to real-world datasets.
% distinguishes itself by POWDER's commitment to authenticity, moving beyond traditional simulations. 
This is ensured by including over-the-air actual packets transmitted within its spectrum between actual devices 
% that closely mirror the intricacies of genuine network behavior 
instead of simulated wire transmissions~\cite{powder-usecase1, powder-usecase2-nexran, powder-usecase3-rf-propagation, powder-usecase4-WiMatch, powder-usecase5-ZCNET}. %\fbsdataset and \msadataset incorporate different networking scenarios (static and mobile endpoints) and different attacker capabilities (from less sophisticated to more sophisticated).
To tackle the second challenge, incorporating the surrounding context into detection, we design a two-step detection framework: a packet-level classification followed by a trace-level classification, ensuring both granular and contextual analysis. 
% \imtiaz{I have removed some stuff that is repeated in the challenge section. Please recheck.}
%
%For the packet-level classification, we retrofit a stateful LSTM model with attention specialized for network traffic data~\cite{wang-etal-2016-attention-based-lstm, Understanding-LSTM}. Following our intuition that the length of the sequence of FBS-generated messages follows a specific distribution (see Section~\ref{subsec:fbs-det-eval}), the statefulness captures
%long-term dependencies that span across sequences, and the attention mechanism focuses on the parts of each sequence that affect the classification outcome the most.
%
%
% Sequential-LSTM model that takes sequence length as a hyperparameter and trains on chunks of packets. 
% This ensues from our intuition that the length of the sequence of FBS-generated messages follows a specific distribution (see Section~\ref{subsec:RQ1}). %Another important objective of this design is incorporating the surrounding context into consideration. 
%This approach for FBS detection improves the detection accuracy and lowers the false positive rates than other state-of-the-art models.
% Unlike traditional methods, in this problem, the true nature of a packet hinges on the context in which it was sent.
% (3) Multi-step attack recognition
For MSA detection, we use graph learning--derived from our intuition that all the MSAs follow a specific pattern %(see Section~\ref{subsubsec:Multi-Step-Attack-Recognition-using-Graph})
, and that to recognize MSAs successfully, it is necessary to capture these patterns. 
% To this end, 
%We 
% use graph data structures to 
%construct a directed graph from the packets to capture the patterns and train a graph learning model on these graphs to learn the patterns. %Graph learning models perform better than any other state-of-the-art model in recognizing multi-step attacks. 
%Moreover, by using maximum overlapping sub-graphs, we can detect unseen attacks and recognize the MSAs, even if they evolve or the attackers try to evade detection by re-shaping traffic.
%, using maximum overlapping sub-graphs %This approach also enables us to detect unseen attacks in our dataset 
%(see Section~\ref{subsubsec:Multi-Step-Attack-Recognition-using-Graph}).
% (4) Ensemble learning
To obtain a unified prediction %of the layer-3 protocols, 
we combine predictions made on NAS and RRC packets using a weighted confidence-based fusion method.
% of the best-performing ML models for each layer. 
%This method of combining predictions enhances detection accuracy beyond individual predictions.
% (5) Deployment
For the deployment of \system, we create a mobile app that 
% can scan the network and 
analyzes the packet traces by running the pre-trained models in the device to detect FBSes and MSAs effectively in real time.

%===========
% Results
%===========
% (1) Dataset stats
%\elisa{his paragraph seems to be disconned from the previous paragrah. I was wondering whether to introduce a paragraph title like "Key Experimental Results".}
\noindent \textbf{Experimental results.} 
% \elisa{I would remove "To the end" from here.}
% To the end, 
The unprocessed \fbsdataset and \msadataset datasets have a combined size of \totalsizeofdata GB.
% After the initial filtering, we have \totalnumberofpackets packets in total. %There are total \totalfbsnas NAS layer packets in the FBS dataset, \totalfbsbenignnas benign and \totalfbsmaliciousnas malicious, and there are \totalfbsrrc RRC layer packets in total, \totalfbsbenignrrc benign and \totalfbsmaliciousrrc malicious. For the multi-step-attack dataset, there are \totalmultistepnas NAS layer packets, \totalmultistepbenignnas benign and \totalmultistepmaliciousnas malicious, and there are \totalmultisteprrc RRC layer packets, \totalmultistepbenignrrc benign and \totalmultistepmaliciousrrc malicious. 
% (2) FBS detection performance
Trained on this combined dataset, our experiments show that our FBS detection framework can detect FBS with \fbsdetectionaccuracy accuracy and a false positive rate (FPR) of \fbsdetectionfpr.
% (3) Multi-step attack recognition performance
Similarly, 
% \elisa{should we replace "we can detect" with our graph-based model can"?}
our graph learning model can detect \totalnumofattacks MSAs with \msadetectionaccuracy accuracy and an FPR of \msadetectionfpr.
% using graph learning.
% \imtiaz{Include how many multi-step attacks we can detect.}
% (4) Ensemble learning performance
Our experiments also show that combining NAS and RRC  predictions 
% using weighted confidence-based fusion method 
improves the performance 
% beyond the individual models trained on NAS and RRC layers separately 
by \enseblelearningimprovement.
% in FBS detection and MSA recognition.
% (5) Deployment performance
% To evaluate \system's performance in real-world scenarios, we instantiate an Android application for 4G UEs. 
% POWDER's approach is a testament to its dedication to offering a dataset that authentically represents the dynamic landscape of real networks, empowering researchers to tackle real-world challenges more effectively. 
% We will discuss the details of POWDER in Section~\ref{subsec:powder}.
Furthermore, \system detects unseen Overshadow attacks with 86\% accuracy.
To validate \system's fidelity and evaluate its performance, memory and power consumption in real-world scenarios, we instantiate 
% an Android application 
a mobile app
for 4G UEs and set up FBSes and MSAs in a controlled lab environment. %to prove the fidelity of the solution. 
Using our lab setup, we spawn FBSes and run experiments with different FBS detection and MSA recognition scenarios. Lastly, we run longer evaluations with the \system app in multiple countries and areas with varying population densities with diverse use cases.
% Furthermore, we experiment \system in an open environment with a commercial SIM card and create scenarios with high mobility and handover to induce false positive behavior. 
%The results are analogous to the similar experiments we conducted at POWDER.
% , supporting our claim of POWDER's ability to generate real-world datasets. 
%Deployed in the real-world environment, 
The experimental results show that compared to previous signature/heuristic-based approaches
\system can detect FBSes 
and MSAs effectively 
% in all the tested scenarios without false positives 
using an average \memoryusage KB of memory and less than \powerusage mW of power. Furthermore, we discuss how \system can be deployed and combined with network side defenses to create a robust ecosystem to prevent attacks in cellular networks in Section~\ref{section:discussion}.

% Because of the extensive infrastructure already in place for 4G networks and billions of 4G devices all over the world that are vulnerable to FBSes, we focus on detecting FBSes and multi-step attacks in the context of 4G which can be ported to 5G easily by the carefully designed pipeline.

% Using the dataset, we trained ML models to detect the presence of FBSes in cellular networks.
% The machine learning pipeline performs two major tasks. (1) FBS detection, and (2) Multi-step attack recognition. 
% \elisa{the previous two paragraphs provide a lot of details (such as the use of PCA) which are quite standard when one uses ML. However, we do not say much about multi-step attack detection, which is a new result of our work (also compared with the recent paper by Ziqiang. Why multi-step detection is important? Is there any technical challenge for this? We emphasize a lot using POWDER but do not emphasize multi-step.}

% ==============================
% Contribution
% ==============================
\noindent \textbf{Contributions.}
This paper makes the following contributions:
% \imtiaz{Follow this contribution outline:}
\begin{itemize}[leftmargin=*,noitemsep,nolistsep]
    \item We develop a new framework--\system to detect FBSes and MSAs from network traces using ML. For this, we create \fbsdataset and \msadataset, the \emph{first-ever} large-scale, high-quality, real-world datasets containing FBS and MSA traces in different scenarios. %We introduce many real-world phenomena in the dataset such as mobility with the help of mobile endpoints available at POWDER, and sophisticated attacker capabilities like cloning legitimate base stations and changing signatures. The unprocessed dataset has a size of \totalsizeofdata Gigabytes containing \totalnumberofpackets packets. 
    %To the best of our knowledge, this is the \emph{first-ever} large-scale public dataset prepared for this task.

    \item We design a two-step detection framework: a packet-level classification followed by a trace-level classification, ensuring granular and contextual analysis. For the packet level classification, we design a stateful LSTM with attention utilizing stateful training and attention in parallel layers, 
    % a Sequential-LSTM model for the FBS detection task, 
    which improves the detection accuracy and reduces the false positive rates.
    % compared to other state-of-the-art models. 
    % It can detect FBS with \fbsdetectionaccuracy accuracy and a false positive rate of \fbsdetectionfpr.
    For MSAs, we innovate by converting the attack signatures to graphs and using a graph-based learning approach to detect the attacks. Graph learning models perform better than any other state-of-the-art model in recognizing MSAs. Moreover, even when MSAs evolve,
    %the MSAs, even if they evolve, and 
    unseen and reshaped MSAs can still be detected by this approach by using maximum overlapping sub-graphs. 
    % It can detect multi-step attacks with \msadetectionaccuracy accuracy and a false positive rate of \msadetectionfpr.

    %\item We combine RRC and NAS layer predictions using ensemble learning of the best-performing ML models from each layer. This ensemble model performs better than any individual model trained on NAS and RRC layers separately.
    
    \item 
    % \imtiaz{add stuff for validation.}
    We deploy the solution in 
    % an Android 
    a mobile 
    app and validate its performance in real-world setups.
    %by creating a testbed in our controlled lab setup.
    %Compared to the available end-user FBS detection solutions, which fail to detect FBSes and MSAs, our approach effectively detects FBSes and MSAs in all the tested real-world scenarios.
    Compared to the available end-user FBS detection solutions, including signature-based solutions, our approach significantly improves the performance of FBS and MSA detection. 
    %\item We will open-source the dataset, our experiments, and the Android app for further research and help end-users protect from FBS-related attacks.
\end{itemize}

%\noindent \textbf{Open-source.} We will open-source the datasets, our models, and the mobile app for further research and help end-users protect themselves from FBS-related attacks. 

% We provide part of the dataset in the anonymous link for the reviewers~\footnote{The partial dataset for \system is available at \url{https://drive.google.com/drive/folders/182oGoGmarf6sslm0fF4RW3K081B_ldf6?usp=sharing}}.

% Criminal gangs are now using FBSes to directly attack users. Using an FBS, an attacker can send SMS messages to users from spoofed phone numbers, including privileged numbers associated with mobile carriers, government agencies, public services, banks, etc. These messages can contain spam advertisements, phishing links, and solicitations for high-fee premium services. In China alone, users received over 2.9B (B = billion), 4.2B, and 5.7B spam/fraud messages from FBSes in 2013, 2014, and 2015, respectively, causing estimated losses of billions of dollars. Surprisingly, an attacker with a \$700 FBS that is small enough to mount inside a car can earn up to \$1400 a day.

\section{Background}
In this section, we introduce relevant background about 4G, FBSes, MSAs and POWDER--the platform we use for dataset generation. 
% \imtiaz{One general comment, define Fake Base Stations (FBS) once in the intro, then use FBS later on throughout. Be consistent.}
\subsection{4G Cellular Networks}
%The emergence of 4G, the Fourth Generation of wireless mobile communication technology, brought faster data transfer rates, reduced latency, enhanced capacity for simultaneous connections, and improved multimedia services, including high-quality voice calls and HD video streaming. 
In a 4G network, cellular devices are called User Equipments (\textbf{UE}). The core network is called the Evolved Packet Core (\textbf{EPC}). Geographic locations are partitioned into hexagonal cell areas, each of which is serviced by a designated BS (\textbf{eNodeB}), which enables connectivity of UEs in that cell to the EPC.  The Mobility Management Entity (\textbf{MME}) manages the connectivity and mobility of UEs in a particular tracking area (a set of cell areas). %Figure~\ref{fig:lte-architecture} illustrates the architecture of a cellular network. 
% \imtiaz{Figure 1 is missing MME and EPC.}
% \imtiaz{Include a discussion about RRC and NAS.}

\noindent \textbf{Non-Access Stratum (NAS).}
% \elisa{Before you used 4G; here you use LTE. It would be better to use the same term.}
In 4G, the Non-Access Stratum (NAS)~\cite{NAS} protocol is a layer-3 (Network Layer) protocol specified by 3GPP that serves as a functional layer between the core network and the UEs.
Its primary role is to manage the
%revolves around the management of
communication sessions and seamlessly maintain
%and the seamless maintenance of 
the connections with the UE, even when the UE roams. 
% NAS operates through a protocol that facilitates the exchange of messages between the UE and Core Nodes.
% \elisa{do we need the text below? If not, I would remove it.}
% with these messages traversing the radio network transparently. 
% These messages encompass a range of functions, including updates and attachments, authentication, service requests, and more. 
% Once the UE establishes a radio connection, it harnesses this link to engage in communications with the core nodes, orchestrate the provisioning of services and ensure uninterrupted connectivity.
%Examples of NAS messages include \textit{Update} and \textit{Attach} messages, \textit{Authentication} messages, \textit{Service Requests}, and so forth. 

\noindent \textbf{Radio Resource Control (RRC).} The Radio Resource Control (RRC)~\cite{RRC} protocol is another layer-3 protocol used between the UEs and the BS. The major functions of the RRC protocol include connection establishment and release functions, broadcast of system information, radio bearer establishment, reconfiguration and release, and
% RRC connection mobility procedures, 
paging notification and release.
\subsection{Fake Base Station (FBS)}
An FBS %, a.k.a. rogue base station, cell-site simulator, IMSI catcher, and stingray, 
is an unauthorized device an attacker uses to impersonate a legitimate BS within a cellular network. % (shown in Figure~\ref{fig:fbs-arch}.b). 
% It operates by exploiting vulnerabilities in the cellular network infrastructure to intercept and manipulate communications. 
FBSes typically consist of a radio transceiver capable of broadcasting signals at legitimate BSes' frequencies. By emitting these signals, FBS creates a cell or coverage area, attracting nearby mobile devices to connect to it. 
With the deployed FBS, attackers carry out Multi Step Attacks (MSAs), resulting in DoS, location tracking, bidding down attacks, and traffic monitoring. Detecting FBSes can essentially stop these MSAs, because FBSes are the key stepping stones for these attacks. However, MSA detection provides fine-grained information about the attack and attacker, which is essential for forensics and defense design. %Here, we discuss some of the main categories of MSAs based on their impact.
In the following, we discuss an MSA done with an FBS--Tracking Area Update Reject (TAU) Reject attack~\cite{altaf2016}.

\begin{figure}
    \centering
    \includegraphics[width=\linewidth]{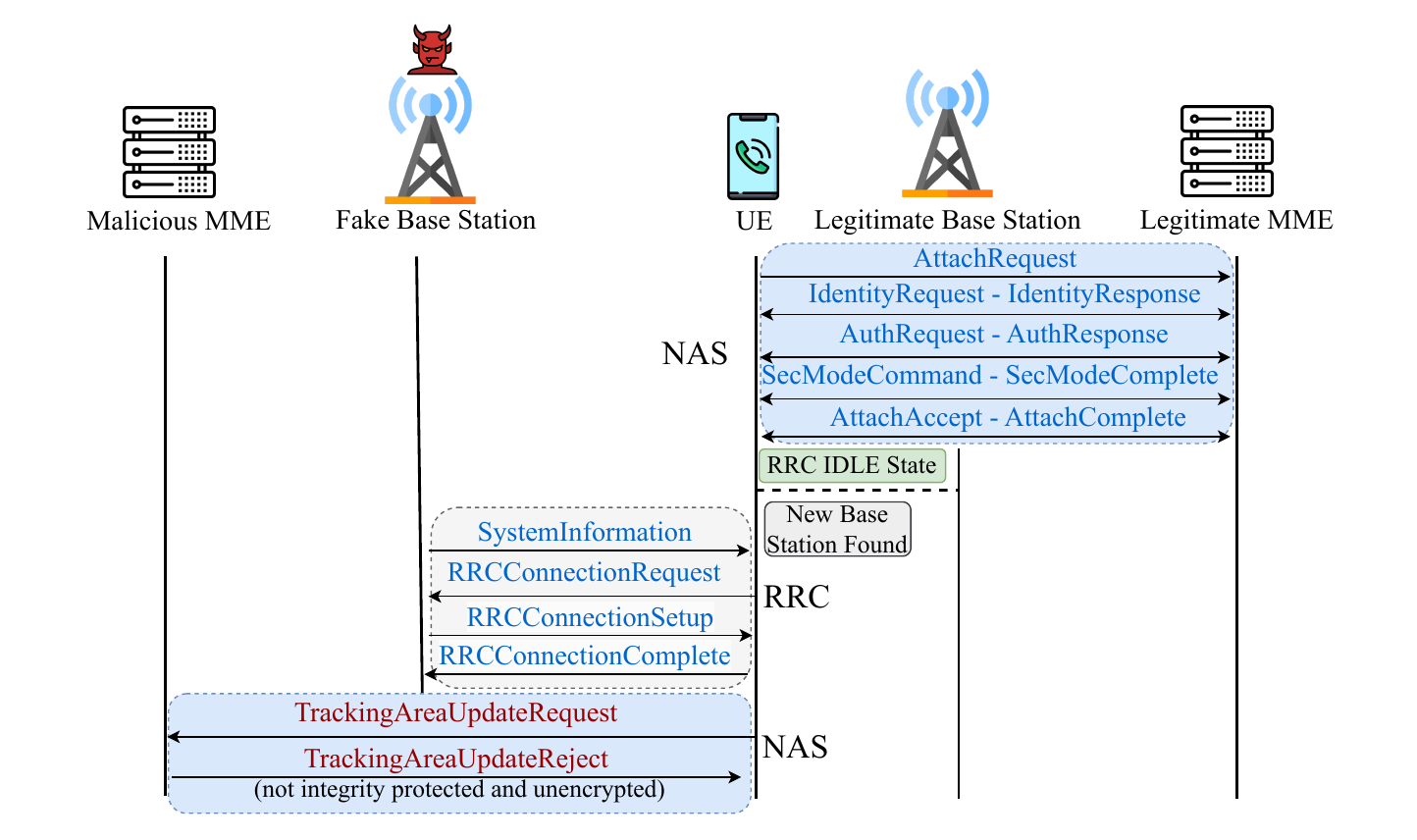}
    \vspace{-0.5cm}
    \caption{Communication process of an FBS with an MSA}
    \vspace{-0.5cm}
    \label{fig:fbs-comm-setup}
\end{figure}

\noindent \textbf{TAU Reject attack.} 
\label{subsec:taureject}
% \imtiaz{Revised. Recheck.}
To deploy an FBS and to interrupt the existing connections between nearby user devices and legitimate BSes, the attacker would adjust the signal strength of the FBS to guarantee that it offers a much higher signal strength than the legitimate BS. Furthermore, the FBS broadcasts MCC and MNC numbers identical to the network operator of targeted subscribers to impersonate the real network operator. Once the attacker has properly configured the FBS, the attacker usually does the following steps for a FBS and TAU Reject attack:  
   \ding[1.2]{172} The FBS broadcasts its \M{SystemInformation} using the configured radio frequency. To overcome different UE functionalities, the FBS exploits a feature named \emph{absolute priority-based cell reselection}. The principle of priority-based reselection is that UEs in the IDLE state should periodically monitor and try to connect to BSes operated with high-priority frequencies. Hence, even if the UE is close to a real eNodeB, operating the
   FBS on a frequency with the highest reselection priority would force UEs to attach to it. These priorities are
   defined in SIB Type number 4, 5, 6, and 7 messages broadcast by the real BSes. Using a passive attack setup, the attacker can sniff
   these priorities and configure the FBS accordingly.
    \ding[1.2]{173} When a UE receives the system information of the FBS, it detects it as a new BS 
    % in a new coverage area 
    with a higher signal strength, and generally, when UE detects a new TA, it initiates a \M{TrackingAreaUpdateRequest} to the FBS. In order to trigger such a request, the FBS operates on a TAC that is different from the real BS.
    %In other cases, the FBS can send a \M{PagingRequest} to trigger the service procedure to establish the communication.
     \ding[1.2]{174} For the TAU Reject attack Upon receiving the \M{TrackingAreaUpdateRequest} the FBS sends a \M{TrackingAreaUpdateReject} message. This attacker can utilize different EMM causes to either deny the LTE network (downgrade) or deny all network services (shown in Figure~\ref{fig:fbs-comm-setup}).
     %or , the FBS first usually sends an \M{IdentityRequest} to the UE to acquire its IMSI information. %and then sends another \M{IdentityRequest} to acquire its IMEI information. %When both types of information are obtained, the FBS returns a Location Updating Accept to it.
     %\ding[1.2]{175}  Subsequently, the FBS sends different messages to the UE. This action can be repeated multiple times. For instance, the FBS can send a plain-text \M{TAUReject} or \M{ServiceReject} or \M{AttachReject} with different to the UE to cause a Bidding-Down/Downgrade/DoS~\cite{altaf2016} attack.
     \ding[1.2]{176}  When the FBS completes sending messages, it cuts off the cellular connections with its UEs by lowering its signal strength or shutting down the signal. After that, the affected UEs may or may not automatically re-connect to a legitimate BS (some UE's require manual restarting to connect back).
%A simplified diagram of this communication process of FBS with a bidding down attack with \M{TAUReject} is shown in Figure~\ref{fig:fbs-comm-setup}.

% \imtiaz{Distinguish NAS and RRC message. Include SIB and MIB.}

\subsection{Multi-Step Attacks (MSAs)}
% \kazi{Clarify Bidding down}
% \imtiaz{My first pass done please revise.}
\noindent \textbf{Location tracking.}
% Apart from IMSI and SUCI catching there are even more ways to track users using MSAs. For instance 
%An attacker can obtain the IMSIs of nearby devices by exploiting vulnerabilities in the paging protocol~\cite{Hussain2019PrivacyAT}. Similarly, temporary identifiers such as GUTI can track users as they are not refreshed that often~\cite{hong2018guti}. As mentioned earlier in the 5G protocol, this issue has been addressed by requiring UE to encrypt the SUPI (alternative to IMSI). Apart from being vulnerable to the SUPI catching attacks, these 4G attacks are still possible because an attacker first can do a bidding-down attack to downgrade the connection to 4G and then carry out the attacks. 
IMSI catchers, StingRays, or Cell-site simulators have been widely used to collect user IMSIs and track them~\cite{CatchMeIfYouCan}. 
An attacker can also use Measurement Reports to track user location~\cite{altaf2016}. For this attack, an attacker forces a subscriber who is initially attached to a legitimate BS to connect to an FBS with a similar approach to the TAU Reject attack discussed in the previous sub-section. The subscriber UE completes the RRC connection and initiates a TAU procedure with the FBS. Next, UE enters into CONNECTED state. 
At this moment, the attacker turns off the first FBS and starts a second FBS. Meanwhile the UE detects it has lost synch and starts Radio Link Failure (RLF) timer. When the RLF timer expires, UE creates an RLF report
and goes into IDLE mode. In this mode, UE starts cell selection
procedure to attach to second FBS. The attacker sends an unprotected \M{UEInformationRequest} message through the second FBS and gets the \M{UEInformationResponse} message with the RLF report with the failure events and signal strength of neighboring BSes.

%The attacker creates a \M{RRCReconfiguaration} message with different cell IDs and necessary frequencies and sends it to the UE without any protection. After receiving this message the UE computes the signal power from neighboring cells and frequencies and sends an unprotected \M{MeasurementReport} to the FBS. 

\noindent \textbf{Activity monitoring.} 
% Using FBS 
% as a MiTM relay 
%Attackers can learn the user's activity patterns using FBS. %This includes the number of calls and SMSs sent at a given time~\cite{shaik2019new}. 
%\imtiaz{Does activity monitoring require FBS? It does not seems to. Please reverify.}
To monitor users' activity patterns, an attacker can create an FBS, connect to the devices and gather the UE capabilities. This is possible because the UE transfers its capabilities without performing authentication. An attacker can acquire the UE's core network capabilities but not the radio capabilities because they are exchanges after the RRC security setup~\cite{shaik2019new}. Therefore, an FBS setup is needed to monitor all the capabilities.
Other attacks such as authentication relay attack~\cite{Hussain2018LTEInspectorAS} allows an attacker's malicious UE to impersonate a legitimate UE and poison location history or profile network usage.

\noindent \textbf{Bidding down attacks.} These attacks allow an FBS to force a UE to use an older version of cellular protocols~\cite{BiddingDownAttacks}. Such attacks can be carried out in different ways. During the TAU procedure the FBS can send a \M{TAUReject} %or a \M{ServiceReject} or an \M{AttachReject} 
message to force the UE to start searching for 2G and 3G networks in the area~\cite{altaf2016}. Furthermore, a recent study has found that most new devices are vulnerable to bidding down attacks, which can be divided into inter-generation and intra-generation bidding down attacks manipulating different messages and message parameters~\cite{BiddingDownAttacks}. The most common way is to utilize NAS Reject messages. These messages include a specific cause that informs the UE about how to behave when rejected by the network. Since the UE is allowed to accept unprotected reject messages if it receives them before the establishment of the security context an attacker with an FBS can use the reject messages to completely disable support for the current network generation and do a downgrade attack. Another option is to use BS redirections. The BS uses \M{RRCRelease} to release the radio connection with a UE, e.g., if the UE switches into IDLE mode. As the release procedure can be initiated before the radio connection is secured, an FBS can cause RRC redirection from 4G to 3G. %Since release 15.3.0, the core network can explicitly forbid
%the UE to accept an unauthenticated \M{RRCRelease} with a redirection field in 4G by using an optional NAS
%flag during the attach procedure. If the flag is not used, an insecure
%redirection from 4G to 2G is possible.
Lastly, 3G to 2G redirection is possible since the specification does not prevent a pre-authenticated RRC redirection.
The recent measures by Google and Apple to disable 2G connectivity and prevent using a "null-cipher" still do not resolve the downgrade attacks to 3G and are limited to newer devices and operating systems (Android-14, iOS-17). 

\noindent \textbf{DoS attacks.} There are numerous MSAs that an attacker can utilize to cause both short and long-term DoS. The easiest option is to use reject messages such as \M{AuthReject}. \M{RRCReject}, and \M{NASReject}~\cite{FBSSpecifications,Hussain2018LTEInspectorAS,altaf_wisec18}. 
This forces the UE to disconnect from the network. Another set of DoS attacks can be caused by paging channel hijacking~\cite{Hussain2018LTEInspectorAS}. For hijacking the paging channel, the FBS operates the same frequency band as the legitimate BS and broadcasts fake empty paging messages in the shared paging channel. One pre-requisite for the attack is to know the victims paging cycle. The FBS broadcasts paging messages with higher signal power, ensuring attack success. The victim is unable to receive legitimate paging messages from the core network.

\noindent \textbf{Energy depletion and power drain.} 
% \imtiaz{Add citations.}
These attacks aims to make the victim UE perform expensive cryptographic operations~\cite{Hussain2018LTEInspectorAS,5GReasoner}. One way to achieve this is to force the UE to keep carrying out the expensive attach procedure repeatedly by sending a paging message with IMSI between two successive attach procedures. Other ways are to force victim device to release existing connection and spend energy on further cryptographic operations~\cite{5GReasoner}. %In case the adversary knows the GUTI of the victim~\cite{kune2012location}, it can send a paging message with GUTI, to which the UE responds with a cryptographically involved \M{ServiceRequest} message.

%We assume the attacker knows about the deployed defense and can change the discussed FBS and MSA patterns or reshape attack traffic for both FBS and MSAs.

\iffalse
\subsection{Multi-Step Attack}
Multi-step attacks (MSAs) are cyber threats that unfold in stages, typically involving coordinated steps to achieve a malicious objective. These attacks can be intricate and may exploit vulnerabilities at various layers of the cellular network architecture. An MSA might begin with an initial compromise, such as gaining unauthorized access to a network node or exploiting a vulnerability in a specific protocol. Subsequently, the attacker can escalate their privileges, move laterally within the network, and strategically compromise additional elements. MSAs can target not only the core infrastructure but also mobile devices and their communication protocols. Examples include exploiting vulnerabilities in signaling protocols, compromising BSes, or manipulating mobile device settings.
\fi

% POWDER
\subsection{POWDER}\label{subsec:powder}
% \imtiaz{Before jumping into what POWDER is add one line to remind reviewers why POWDER is useful for us. For instance: One of the critical challenges of FBS detection is the lack of a real-world dataset containing FBS traces. Due to regulatory reasons, it is not possible to deploy FBSes in real-world scenarios including mobility. POWDER resolves all these challenges and provides us with a way to generate high-quality, real-world, over-the-air datasets of FBSes with all the real-world scenarios involved especially mobility.}
% \elisa{I would remove this paragraph, as this has already been said in section 1 and section 2.}
% One of the critical challenges for detecting FBS using anomaly detection is the lack of a real-world dataset containing FBS traces. Due to regulatory reasons, it is not possible for researchers to deploy FBSes in real-world scenarios including mobility ones. In most cases, they are deployed in 
% Faraday cages to understand their behavior in a small laboratory setup. This is a fundamental limitation for collecting real-world datasets as there is a certain upper bound on the number of devices and BSes in a laboratory setup. Moreover, it is not possible to create handovers or scenarios involving mobility in a lab setup.
% \elisa{I would also remove the paragraph below. In this part of the paper, we are supposed to introduce the technical background required to understand the rest of the paper. Commenting on how great POWDER is can be moved in the conclusions or when talking about the implementation.}
POWDER (\textbf{P}latform for \textbf{O}pen \textbf{W}ireless \textbf{D}ata-driven \textbf{E}xperimental \textbf{R}esearch)~\cite{powder} is a cityscale, remotely accessible, end-to-end software-defined platform funded by NSF to support mobile and wireless research and provides advances in scale, realism, diversity, flexibility, and access.

%provides us a platform to generate high-quality, real-world, over-the-air datasets incorporating FBSes in a controlled environment with real-world scenarios (especially mobility and sophisticated attacker capabilities) without violating the law against creating FBSes in open spaces. 
%POWDER encompasses three key components: (1) a core component, (2) an edge component, and (3) a radio access component. All these components are deployed in real-world devices, making POWDER a unique platform to generate real-world data, unlike the other platforms that generate simulated data.
%An architectural overview of POWDER is given in Figure \ref{fig:powder-overview}, and a detailed overview is discussed in appendix section \ref{apndx-sec:powder}. 

\noindent \textbf{Fidelity of POWDER to the real-world.}
% \imtiaz{Add discussion here}
POWDER's fidelity extends beyond conventional simulation due to the following reasons:
% stands out as a robust platform for networking research. 
(1) POWDER provides a dedicated frequency band and real devices~\cite{powder}. This access to real devices while maintaining scalability and mobility makes POWDER equivalent to real-world testbeds. 
(2) POWDER's fidelity is demonstrated through rigorous testing and evaluation, establishing that solutions developed using POWDER, which mirrors the complexity and challenges found in real network environments, perform effectively well in real-world scenarios~\cite{powder-usecase1, powder-usecase2-nexran, powder-usecase3-rf-propagation, powder-usecase4-WiMatch, powder-usecase5-ZCNET}. 
% This ensures that the knowledge gained and solutions crafted within POWDER have practical applicability and reliability when deployed against real-world cyber threats.
Because of incorporating over-the-air actual packets in a dedicated spectrum instead of simulated 
% \elisa{is this "wire" or should it be "wireless"?}
wire transmissions, datasets created using POWDER are analogous to real-world datasets.
%The datasets created using POWDER are analogous to real-world datasets because of incorporating over-the-air actual packets, POWDER captures real-world network behavior.
% , providing researchers with data that closely mirrors the complexities of genuine network conditions. 
%The dedicated spectrum further ensures that the datasets are relevant to real-world scenarios. 
%POWDER facilitates the creation of profiles (network topologies) with different parameters encompassing hardware and network configurations, signal propagation characteristics, mobility patterns, interference scenarios, and more. Profiles make it possible to create real-world conditions for experimentation, allowing one to create diverse scenarios, including those involving FBSes for our experiments. 
A detailed overview of POWDER is given in the appendix Section~\ref{apndx-sec:powder}.
\section{Overview of \system}\label{sec:overview}
In this section, we discuss the threat model, deployment scope, challenges, and requirements of \system.
% Our approach, although based on 4G, is equally applicable to 5G. As an example, similar to 4G, 5G has a multi-layer design with most of the procedures unchanged from 4G. Thus the insights will largely remain the same when adopting \system to 5G.
%the high similarity of the high-level protocols of 4G and 5G.
%that are working towards supporting all 5G functionalities, and our work will be extended to 5G as soon as support for all 5G functionalities arrives in POWDER. The extension to 5G will take minimal effort as our approach is easily portable to 5G by the carefully designed pipeline.
% However, it's essential to note that the transition to 5G is ongoing. As the technology matures and becomes more widespread, we can expect to see a shift towards a more balanced landscape in the future. 
% One of the core modules of this work is the handover of the UEs between the BSes. The open source implementations, srsRAN\cite{srsRAN}, Open5GS\cite{Open5GS} and OpenAirInterface\cite{OpenAirInterface} support the handover in 4G but not in 5G. 
\subsection{Threat Model}
For our FBS attacker threat model, we consider the adversary 
% to have the capability to establish a man-in-the-middle relay~\cite{Hussain2018LTEInspectorAS, ALTER}, which in turn may allow the adversary to drop, modify, eavesdrop and forward messages transmitted between benign protocol entities (legitimate UE and BSes) in the public channel. The active attacker, which can run and install its own FBS, 
can impersonate the legitimate BS and thus force a victim UE to initiate a reselection with a higher signal strength than the legitimate BS. We assume the adversary can learn and mimic the legitimate values of the original BS by eavesdropping the public channels.
We also assume that the adversary cannot break the cryptographic assumptions and cannot tamper with SIM cards, BSes or core network components. For instance, an attacker can only create plain-text packets but is unable to create integrity-protected or encrypted packets other than just replaying them. Furthermore, the attacker may employ various techniques to evade detection. This includes rapidly changing the parameters of the FBSes, adjusting transmission power, or adopting sophisticated obfuscation methods to mask its activities. %Apart from deploying FBS, we also consider the MSAs that can be enabled by FBS. 

\subsection{Deployment Scope}\label{subsec:dep-scope}
The current deployment scope of \system is detecting FBSes and MSAs in the context of 4G cellular networks. There are two significant reasons for this. \emph{First,} because of the extensive infrastructure already in place,  4G networks are widely accessible to a larger portion of the population. As of 2024, 4G adoption stands at $59\%$ among $8.6$ billion SIM connections~\cite{The-Mobile-Economy-2024, 4g-5g-subscribers-march-2022-quarterly-update, ericsson_news_Mobile_subscriptions_outlook, telegeography}. 4G adoption is predicted to stay above 50\% until 2027, and in 2029 5G is expected to overtake 4G~\cite{The-Mobile-Economy-2024, statista}. Therefore, until 2027, many end-user devices using 4G are at risk of FBS attacks. \emph{Second,} the platform we use for real-world dataset generation for FBS and MSAs (i.e., POWDER) supports 4G functionalities and real-world experiments can be run in POWDER for different scenarios only in 4G. POWDER currently does not support all the 5G functionalities (for instance, handover). 
%
% \imtiaz{Discuss overshadow attacks and say it is not within the scope because of the limitations of powder and overshadow attacks being done mostly in lab environments.}
In a related discussion, recently, the research community has uncovered a new kind of attack called \emph{signal overshadowing attacks}~\cite{yang2019hiding, erni2022adaptover, sigunder} that can be an alternative for attacks without requiring an FBS. However, conducting physical signal overshadowing attacks at a large scale in the POWDER testbed presents significant challenges due to the need for sophisticated and fine-grained control over network devices and precise physical placement. %POWDER's shared, large-scale nature, while excellent for many types of wireless research, does not easily accommodate the specific and demanding requirements of overshadowing attack experiments. 
Researchers aiming to explore such attacks still need specialized, controlled lab environments with complete control over the physical and radio environment to achieve the necessary precision to experiment with these attacks. Thus, it is out of scope to include overshadowing attacks in POWDER. 
% \imtiaz{Include discussion that we defend zero shot against overshadow attacks. Also introduce Alter MiTM relay attacks.}

\looseness = -1
Despite not being trained on an overshadowing attack dataset, since being trained on layer-3 data, the \system model performs relatively well against our controlled lab environment Overshadowing~\cite{yang2019hiding} attack dataset. We conduct a zero-shot evaluation with this dataset and discuss the results in Section~\ref{subsec:overshadow}.

%Our discussion with the POWDER team revealed that support for all 5G functionalities and supporting overshadowing attacks are in their plans. 
%Nonetheless, extending \system to support 5G and newer threat models would require minimal effort once these supports are available in POWDER because the high-level protocols of 4G and 5G are very similar, and the only requirement would be to generate more data. Therefore, the design of \system models will remain the same.

\subsection{Challenges}
% The detection of FBSes require addressing several challenges. These challenges arise because of the dynamic and evolving nature of cellular networks, the complexity of network traffic, and the tactics employed by adversaries. 
The challenges in designing our ML-based framework for detecting FBSes and MSAs are:
% \imtiaz{Include the challenge of multiple L3 data source.}

\noindent \textbf{C1. Dataset availability and quality.}
Acquiring a comprehensive and high-quality dataset that encompasses a wide range of real-world cellular network scenarios is a significant challenge. The dataset must include instances of legitimate BSes, FBSes, and execution traces of different MSAs.
% ensuring a balanced representation for effective training and evaluation of the machine learning models. 
Currently, there is no such dataset publicly available. There are several reasons behind this: (1) The law prohibits the deployment of FBSes in public areas. If someone wants to deploy a FBS for research and experimentation, they must do so in a controlled RF environment. (2) Incorporating different real-world cellular network scenarios, such as handovers and mobility, is a difficult task and would require a lot of specialized hardware and other equipment and facilities.

\looseness = -1
\noindent \textbf{C2. Detecting FBSes from packet traces.}
% Detecting FBSes from packet traces presents a unique challenge.
Unlike traditional methods where packets themselves are inherently malicious, in the case of detecting FBSes, the true nature of a packet hinges on \emph{when} and \emph{in which context} the packet was sent. For instance, a legitimate BS or an FBS can send the same packet with the same contents. 
One solution is to perform only a single trace-level classification, where the ML model takes the entire trace as input and outputs whether an FBS is present. 
This approach for detecting FBSes would be at a very high granularity level and thus miss a lot of context at the packet level. A well-equipped adversary can bypass the detection by keeping the trace the same as benign but changing the packet configurations and thus executing different attacks. Also, detecting unseen and reshaped attacks would not be possible at trace-level classification. In order to accurately detect FBSes and MSAs, we need to design an approach with two different granularity levels: at the single packet level and at the packet sequence or trace level.
%Elisa: I have changed the text below (see the previous sentence). I have commented out the previous text, so if you prefer the previous version, you can re-instate it.
%to go more granular at the sequence and packet levels for more accurate detection of FBSes and MSAs.
% Simply labeling packets as purely benign or malicious would not suffice; we must consider the surrounding context. 
% So, the presence of such packets in a trace does not guarantee the presence or absence of an FBS.
%
%======  New Addition for NDSS'25  ===========
% \imtiaz{Add the challenge of why just trace level identification won't work.}
We need to 
% extract a comprehensive set of features from the packets and 
classify each packet individually as suspicious or benign, serving as a preliminary filter to identify potential FBS activity. Subsequently, sequences of packets, or traces, containing packets flagged as suspicious need to undergo another contextual analysis at the trace level, which examines the order of packets and sequence patterns to discern characteristics indicative of FBS transmissions.
% , thereby considering the broader context in which packets are sent.
%=================
%
% Moreover, a critical constraint is maintaining the original sequence of packets as they were sent and received throughout the process to preserve the context information. 
% This distinctive approach acknowledges that the ``when'' and ``how'' of packet transmission hold the key to distinguishing between benign and malicious activity, making it essential to factor in this context information for accurate detection.

\noindent \textbf{C3. Detecting MSAs.}
Recognizing MSAs from packet traces is even more challenging than FBS detection. These attacks have unique characteristics that define them. Moreover, MSAs often exhibit complex, evolving patterns that require careful observation to distinguish them from legitimate traffic. %The real complexity lies in the subsequent step – learning these complex attack signatures. This task requires the deployment of sophisticated ML algorithms that not only recognize known attack patterns but also adapt to emerging threats and unseen attacks. 
An adversary can improve the attacks adaptability by constantly changing to evade detection. Consequently, our approach must evolve alongside these threats. We must capture these attack characteristics and represent them effectively in a structured data format.
% , making it a demanding yet essential endeavor.

\noindent \textbf{C4. Combining NAS and RRC predictions.}
Training our models separately on NAS and RRC layer packets is a necessary step due to their distinct features and characteristics. However, a challenge arises when we must consolidate these separate model predictions into a unified model. 
%Elisa: is this combination done only for the packet sequences or also for the single packet?
% NAS and RRC layers have inherently different sets of features, making this ensemble a complex task. 
% The challenge lies in harmonizing these diverse sets of predictions, ensuring that our final model can effectively capture the information of both layers while providing robust, accurate predictions. %It is a crucial step, demanding careful consideration of how to blend these distinct sources of information into a single, coherent predictive framework.

\noindent \textbf{C5. Real-time detection.} 
Enabling real-time detection of FBSes and MSAs is a challenge due to the large volume and velocity of network traffic. The framework needs to be an in-device solution that captures, processes, and analyzes incoming packet traces promptly to detect 
% and respond 
the presence of FBSes and MSAs effectively.

\subsection{Proposed Solution}
% \imtiaz{Add a line.}
In this section, we present and analyze our proposed solutions to the discussed challenges.
%To address those challenges, we carefully design and develop \system, a ML-based FBS and and multi-step attack detector for cellular networks. 
% \system addresses these challenges in the following manner.

\begin{table}[]
    \centering
    \renewcommand{\arraystretch}{1}
    \fontsize{6}{6}\selectfont
    \begin{tabular}{|c|L{3cm}|L{1cm}|L{2.5cm}|}
    \hline
       Sl & Attack & Attack Category & Impact \\
        \hline
        1 & Authentication relay attack~\cite{Hussain2018LTEInspectorAS} & Activity monitoring; DoS & Complete or selective DoS; Location history poisoning; Network profiling\\
        \hline
        2 & Bidding down with \MTable{AttachReject}~\cite{altaf2016} & DoS & Selective DoS \\ 
        \hline
        3 & Paging channel hijacking attack~\cite{Hussain2018LTEInspectorAS} & DoS & Complete DoS\\
        \hline
        4 & Location tracking via measurement reports~\cite{altaf2016} & Location tracking & Leak fine-grained location\\
        \hline
        5 & Capability Hijacking~\cite{shaik2019new} & DoS, Downgrading & Selective DoS and downgrading \\
        \hline
        6 & Incarceration with \MTable{rrcReestablishReject}~\cite{5GReasoner} & DoS & Complete DoS \\ 
        \hline
        7  & Lullaby attack using \MTable{rrcReestablishRequest}~\cite{5GReasoner} & Battery drain & Force state change, battery draining \\ 
        \hline
        8 & Bidding down with \MTable{ServiceReject}~\cite{altaf2016} & DoS & Selective DoS \\
        \hline
        9 & Mobile Network Mapping (MNmap)~\cite{shaik2019new} & Device Identification & Identify devices on a mobile network \\
        \hline
        10 & Energy Depletion attack~\cite{Hussain2018LTEInspectorAS} & Battery drain & Battery draining \\
        \hline
        11 & Lullaby attack with \MTable{rrcResume}~\cite{5GReasoner} & Battery drain & Force state change, battery draining \\
        \hline
        12 & Stealthy Kickoff Attack~\cite{Hussain2018LTEInspectorAS} & DoS & Complete DoS \\
        \hline
        13 & Incarceration with \MTable{rrcReject} and \MTable{rrcRelease}~\cite{5GReasoner} & DoS & Complete DoS\\
        \hline
        14 & IMSI catching~\cite{CatchMeIfYouCan} & Information Leak & Leaking sensitive information\\
        \hline
        15 & NAS counter Desynch attack~\cite{5GReasoner}  & DoS & Complete, prolonged DoS \\
        \hline
        16 & X2 signalling flood~\cite{altaf_wisec18} & Resource waste & Waste resources for the network \\
        \hline
        17 & Handover hijacking~\cite{altaf_wisec18} & DoS, Energy Depletion & Complete DoS and battery draining \\
        \hline
        18 & RRC replay attack~\cite{FBSSpecifications} & DoS & Complete DoS \\
        \hline
        19 & Lullaby attack with \MTable{rrcReconfiguration}~\cite{5GReasoner} & Battery drain & Force state change, battery draining \\
        \hline
        20 & Bidding down with \MTable{TAUReject}~\cite{altaf2016} & DoS & Selective DoS\\
        \hline
        21 & Panic Attack~\cite{Hussain2018LTEInspectorAS} & Misinformation & Artificial emergency \\
        \hline
    \end{tabular}
    \caption{MSAs detected by \system}
    \vspace{-0.7cm}
    \label{tab:multi-step-attack}
\end{table}

\noindent \textbf{S1. }% Dataset Generation.
% To address the challenge of dataset unavailability, 
We use different facilities available at POWDER~\cite{powder} %(see Section~\ref{subsec:powder}) 
to create \fbsdataset and \msadataset, real-world datasets to detect FBSes and MSAs.
%We have discussed the overview of POWDER in detail in Section~ \ref{subsec:powder}. 
We design different networking scenarios, incorporating legitimate BSes and FBSes, and collect data from these scenarios.
%We prepare a diverse dataset consisting of different networking scenarios incorporating both legitimate and fake base stations. 
We also collect network traces of different MSAs that use FBSes in their threat model. Different adversaries are incorporated for both FBSes and MSAs, from less sophisticated to more sophisticated, with the ability to clone legitimate BSes and change signatures.
The dataset is then processed to make it appropriate for training different ML models. This includes protocol filtering, feature extraction and feature alignment.
% , and replacing missing values. 
% \elisa{I would also mention that we collect traces for different types of adversaries, and we say which ones they are.}

\noindent \textbf{S2. }%FBS Detection.
To address the challenge of granular classification of packets and incorporating the context for detection, we propose a two-level ML-based detection framework.  The approach integrates packet-level classification with trace-level classification, leveraging the strengths of machine learning at both granular and sequential analysis levels. The high-level overview of both the detection models is discussed below: (i) \underline{\textit{Packet-Level Classification.}}  We perform a packet-level classification by classifying each packet individually as suspicious or benign.
% The lengths of the sequence of packets generated by FBSes follow a specific distribution.
% (shown in Figure~\ref{fig:fbs-seq-len-distr-nas} and \ref{fig:fbs-seq-len-distr-rrc}). 
% Based on this observation,
% we devise a solution incorporating the sequence length as a hyperparameter of the ML model for FBS detection. 
% we design a 
% Sequential-LSTM model that takes sequence length as a hyperparameter and trains on chunks of packets. 
% stateful LSTM with attention to classify the packets as benign and malicious.
We leverage a stateful LSTM model with attention for this packet-level classification to model long-term dependencies that span the fixed-size sequences of the packets. We utilize the attention mechanism to learn each training sequence optimally by focusing on the parts of each sequence that affect the classification outcome the most. An essential objective this design helps us achieve is incorporating the surrounding context into consideration while giving attention to only relevant information while classifying packets.
(ii) \underline{\textit{Trace-Level Classification.}} Subsequently, traces containing packets flagged as malicious or benign undergo another contextual analysis. 
One possible solution is to flag the trace as malicious in case one of the packets in the packet-level classification is inferred to be malicious. However, such simple heuristics lack contextual sensitivity, fail to adapt to evolving attack strategies and are not flexible.
% , leading to high rates of false positives and negatives. 
Therefore, in this stage, we employ a simple classification model to examine and classify the trace. The model examines the order of packets and sequence patterns to discern characteristics indicative of FBS transmissions.

\noindent \textbf{S3. }% MSA Recognition.
% To recognize MSAs within packet traces, 
Each MSA shows a unique pattern if MSAs are represented as a directed graph.
% (see Figure~\ref{fig:paging_channel_hijacking_attack} in Section~\ref{subsubsec:Multi-Step-Attack-Recognition-using-Graph}). 
Based on this observation, we devise an innovative solution centered around graph data structures and learning techniques. We transform the packet traces into directed graphs. This graph representation captures the relationships inherent in the trace, which is ideal for detecting MSAs with complex and evolving patterns. We train a graph learning model specialized in learning complex patterns within the graph and learn the patterns of the MSAs. 
% \elisa{From the previous sentence, it seems that we use different graph learning algorithms, thus having different learned graph models. On the other hand, the next sentence seems to suggest that we use one graph model.}
With the trained model, by using a maximum overlapping sub-graphs approach, we can recognize MSAs even when they are unseen or reshaped from known attacks.
%as unseen or reshaped attacks %in our dataset 
%by using a maximum overlapping sub-graphs approach. 
%This way we solve the problem of recognizing complex and evolving multi-step attacks using graph data structures and graph learning. 

% \imtiaz{Add one more line on how graph-based machine learning helps with adaptability. On the whole S2 and S3 are buzzword-filled but with little content.}
\noindent \textbf{S4. }% Ensemble Learning on NAS and RRC Predictions.
To combine the 
% model 
predictions for RRC and NAS traces,
% in FBS and MSA detection
we design a weighted confidence-based fusion method, a widely used technique in the multi-sensor information fusion field~\cite{ZHANG2022202, Tang2023}. 
The weights are assigned to the best trace-level classification model of each layer.
% based on the model performance.
% we turn to ensemble learning as a viable solution. 
% \elisa{just to be sure; also for the graph model, we have one for the RRC layer and one for the NAS; is this correct?}
%The aim of \system is to detect FBS and multi-step attacks from upper layer (layer 3) packet traces (i.e., NAS and RRC layer traces). To achieve this we train models on both NAS and RRC layer packets. For consolidating predictions from separately trained NAS and RRC layer models, we turn to ensemble learning as a viable solution. 
% Ensemble methods offer a strategic approach to harmonize the distinct features and characteristics of these two layers. 
% Stacking employs a meta-learner to intelligently combine predictions from both models, while voting methods leverage the consensus among individual predictions to arrive at a final decision~\cite{Ensemble-Deep-Learning:A-Review}. 
% Weighted averaging is a practical approach for ensembling, assigning weights based on model performance to ensure each layer contributes appropriately to the final prediction~\cite{Ensemble-Deep-Learning:A-Review, Woniak2014}. 
This method offers a robust means to blend the predictions for NAS and RRC layer.
% , facilitating the creation of a unified and accurate predictive framework that captures the strengths of each layer.

\noindent \textbf{S5. }% Mobile App for Real-Time Detection.
The development of a mobile app emerges as a pragmatic solution for deploying \system for real-time detection. Such a dedicated app would serve as an in-device solution, capable of swiftly capturing, processing, and analyzing incoming packet traces with
%Within the app, various packet parsers optimized for mobile devices would efficiently parse baseband packets. 
on-device ML models, specifically tailored for detecting FBSes and MSAs. Regular model updates would ensure adaptability to 
% threats and 
new and unseen attacks.
\section{Detailed Design}\label{sec:detailed_design}
\begin{figure*}
    \centering
    \includegraphics[scale=0.7]{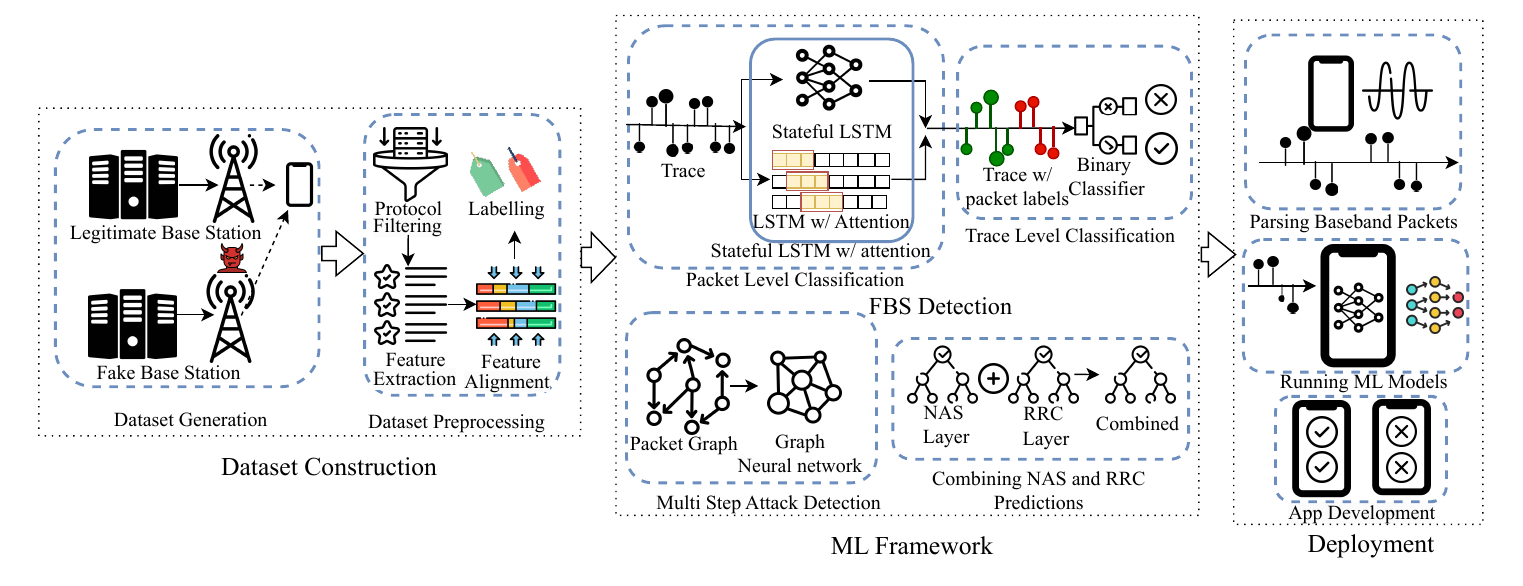}
    \vspace{-0.9cm}
    \caption{Overview of FBS-Detector}
    \label{fig:fbsdetector}
    \vspace{-0.5cm}
\end{figure*}

In this section, we discuss the detailed design of \system (see Figure~\ref{fig:fbsdetector} for an overview). On a high-level, the design of \system is divided into three components: (1) Dataset Construction, (2) ML Framework, and (3) Deployment.

% \imtiaz{Add formalism on multi-step attack detection, labeling, etc.} \\
% \imtiaz{Discuss in much more detail on multi-step attack detection, add figures and examples of flow diagrams and how it is detecting multi-step attacks, why this is not needed for fake base stations.}

\subsection{Dataset Construction}
%To train ML models for the detection of FBSes, we need high-quality datasets that cover all the possible scenarios in typical cellular networks. 
Our dataset construction process is described below.
\subsubsection{Dataset Generation}
% We used POWDER to create a 4G/5G network 
% To generate the dataset, we set up different cellular networks with different combinations.
% for example, static and mobile UEs, legitimate base stations, and FBSes. 
We create cellular networks in POWDER incorporating legitimate BSes, FBSes, and MSAs
% , run experiments, 
and capture the packets from all the cellular network components to generate the dataset. 

\noindent \textbf{Mobility. }
One real-world phenomenon in cellular networks is the mobility of the UEs. Signal strengths of BSes received at UEs moving from one place to another vary, causing the UEs to be handed over from one BS to another. Incorporating this scenario in the dataset is important; otherwise, a benign handover due to mobility might be interpreted as malicious. With the help of mobile endpoints available at POWDER, we incorporate mobility scenarios into our dataset.
% \imtiaz{Add a few more lines about these endpoints. Like they follow some routes, they are couriers or something like that.}
%to make the dataset more sophisticated and close to real-world scenarios.
% Diversity in mobility is provided by using mobile couriers that have relatively predictable movement patterns (e.g., buses), less predictable but bounded mobility (e.g., maintenance vehicles), and couriers that are controllable (e.g., backpacks/portable endpoints that can be moved by researchers that come on-site). 
% We use different couriers, such as buses, providing relatively predictable movement patterns. 
% Additionally, we incorporate less predictable but bounded mobility through endpoints like maintenance vehicles. 
% For further versatility, we include couriers that are controllable, such as backpacks or portable endpoints that researchers can manually move on-site. 
% This approach ensures a comprehensive representation of mobility scenarios in our dataset, capturing the complexity of real-world cellular network dynamics.
% We use diverse couriers, including buses with predictable routes and maintenance vehicles for less predictable but bounded mobility. This strategy ensures a comprehensive representation of mobility scenarios.

\noindent \textbf{Attacker ability. }
\label{subsec:attacker_ability}
The attackers we consider %in our dataset
have a diverse set of abilities.
% , from less sophisticated to more sophisticated. 
Based on their abilities, we rank them in five levels, level 0 being the least sophisticated and level 4 being the most sophisticated. 
% We uniformly incorporate all five levels of attackers while creating the datasets.
\begin{itemize}[leftmargin=*,noitemsep,nolistsep]
    \item \textbf{(Level 0)} Attackers only set up FBSes naively with a high signal strength.
    \item \textbf{(Level 1)} Attackers set up FBSes with an optimal signal strength sufficient to trigger a handover in the UEs.
    \item \textbf{(Level 2)} Attackers can clone all the parameters of a legitimate BS and impersonate the legitimate BS.
    % \imtiaz{Which parameters?} 
    Parameters such as the Cell ID, Mobile Network Code (MNC), Mobile Country Code (MCC), Tracking Area Code (TAC), and Physical Cell ID (PCI) can be cloned to impersonate a legitimate base station. They can also replicate radio frequency parameters like carrier frequency, bandwidth, and transmission power. Protocol-specific information such as the System Frame Number (SFN), Timing Advance (TA), Synchronization Signal Block (SSB), and Random Access Configuration may also be copied. Additionally, network-specific details like the Public Land Mobile Network (PLMN) ID and neighbor cell information can be cloned. %as well as authentication request and response mechanisms in weak systems, can be exploited for successful impersonation.
    \item \textbf{(Level 3)} Attackers use level 2 FBS and can carry out MSAs with the usual signatures.
    \label{subsec:adaptive}
    \item \textbf{(Level 4)} At the most sophisticated level, the attacker is aware of typical defenses and actively reshapes attacks to evade them. 
    % To bypass these defenses, the attacker can modify traditional attack signatures associated with Fake Base Stations (FBSes) and Multi-Step Attacks (MSAs) by reshaping attack traffic. 
    Adaptive adversaries employ two primary strategies: \ding[1.2]{172} changing fields of malicious messages. In the first strategy, attackers manipulate non-critical fields—those that do not impact the success of the attack—within malicious messages to evade signature-based detection~\cite{3GPP, Hussain2018LTEInspectorAS, 5GReasoner, LTEFuzz, DIKEUE, altaf2016}. 
    For instance, an attacker can modify the cause field in reject messages or adjust optional fields in attach responses without affecting the attack’s functionality~\cite{altaf2016}. %\imtiaz{Add more}.  
    Similarly, they can use different reserved values for security headers in messages to cause the same impact~\cite{NAS, RRC}.%For detecting such fields and values we manually go through the specifications and use 
   \ding[1.2]{173} The second strategy involves altering the temporal sequence of malicious messages to evade detection systems that rely on identifying standard message patterns. For instance, an attacker might inject multiple \M{IdentityRequest}, \M{AuthenticationRequest} messages or other extraneous protocol messages before executing an attack, creating many variant sequences that avoid detection.
   To find the fields and the messages that alter the temporal attack sequence but do not affect the attack success, we manually go through the specifications~\cite{3GPP} and follow the prior works~\cite{Hussain2018LTEInspectorAS, 5GReasoner, LTEFuzz, DIKEUE, altaf2016}. Furthermore, before deploying these adaptive reshaped attacks to POWDER, we run some of the attacks in our lab setup and manually validate the attack's effectiveness.
    %It is known that there are typical defenses deployed and the attackers actively aim to evade them.
    % \imtiaz{What do we mean by reshaping?} 
\end{itemize}
% Less sophisticated attackers only set up an FBS with high signal strength. 
% In contrast, more sophisticated attackers can clone legitimate base stations, change signatures, and turn on and off to evade detection.
% \elisa{And then, how do we represent these attackers? Is this easy to do? See my comment before.}
% \imtiaz{Define in detail what is meant by less sophisticated and what is meant by more sophisticated.}

% \noindent \textbf{Implementation of Tracking Area Update Request Message. } When a UE detects a better signal from a neighboring BS, it sends a Tracking Area Update Request Message to its currently serving BS and initiates the handover. But currently, srsRAN\cite{srsRAN}, the open-source implementation of cellular network protocols,  doesn't have the Tracking Area Update Request Message implemented. We had to implement it ourselves and use the modified version in POWDER to initiate handovers during experiments.

\noindent \textbf{MSA data generation. }
To create \msadataset, we chose several attacks that are executed in multiple steps and use FBS in their threat model. As shown in Table~\ref{tab:multi-step-attack}, we have selected 
%evaluated 
\totalnumofattacks attacks, 
% which are representative of the most popular MSAs. These attacks 
covering a wide range of practical threats, including DoS, privacy leakage, and downgrade.
% , which can be quickly launched to compromise practical networks and devices 
% with COTS SDRs 
% severely. 
% All attacks were successfully reproduced end-to-end.
% \elisa{The reviewers may ask whether these attacks are sufficient to have a dataset with good coverage of all multi-step attacks.}
% \elisa{I would remove the following sentence.}
% Detection of these attacks is crucial because they are more dangerous than standalone FBSes. 
% \elisa{The sentence above seems redundant with respect to the sentence below. I would remove the sentence above.}
We implement and execute these attacks in the cellular networks in POWDER and incorporate instances of each attack in \msadataset.

\subsubsection{Dataset Preprocessing}
In order to make \fbsdataset and \msadataset suitable for training ML models, we pre-process them in several steps.
%We have to preprocess the dataset captured from POWDER to make it suitable for model training. 
%We need to process the dataset that we captured during our experiments in POWDER to train the machine learning models. 
% Data preprocessing includes several steps such as data cleaning, normalization, and feature extraction. Data cleaning involves removing any noise or irrelevant information from the packet traces, ensuring the dataset is of high quality. Normalization ensures that the data is transformed into a standardized format, allowing for fair comparisons between different features. Feature extraction involves extracting meaningful characteristics or patterns from the packet traces that can be used as inputs for the ML models. 
%We process the NAS and RRC data separately due to their different structures and prepare two separate datasets.
%to train machine learning models.

% \elisa{It would be good to include some introduction to these paragraphs.}
\noindent \textbf{Protocol filtering and field extraction.}
% and alignment.}
Each packet in the network trace contains multiple protocol information. %Different protocol information is required for different applications. To extract meaningful information for FBS and multi-step attack detection, we focus on only the upper (layer 3), NAS, and RRC layers. 
We focus only on NAS and RRC data and use protocol filtering to isolate the relevant packets precisely. We further extract the values of fields associated with these packets. 

\noindent \textbf{Dataset features for training the ML Models. }
After the protocol filtering,
% and missing values handling, 
we find $119$ fields in NAS layer packets and $183$ fields in RRC layer packets. 
% Each field has four values: \textit{size}, \textit{value}, \textit{show} and \textit{unmaskedvalue}.
% \elisa{what does unmaskedvalue mean?}
We use these fields as features to train our ML models. A subset of these fields are listed in Table~\ref{tab:list-of-features}.

\subsubsection{Dataset Labelling} 
% \imtiaz{There are issues. Revise!}
The datasets are labeled according to their reason for generation. 
% Let $\mathcal{D}_{FBS}$ represent \fbsdataset and $\mathcal{D}_{MSA}$ represent \msadataset.
% $\mathcal{P}$ represent the set of packets, and $\mathcal{L}_{\mathbb{FBS}}$ and $\mathcal{L}_{\mathbb{MSA}}$ are the set of labels for FBS detection and MSA respectively. 
% For FBS detection, if the packet is generated from an FBS, we label it as $\mathbb{FBS}$; otherwise, we label it benign. 
Formally, 
% for 
% \elisa{Just to be sure: is P a packet or a set of packets?}
% a packet $\mathcal{P}$ 
we define %the following formulation: 
$\fbsdataset \coloneqq <X_{\fbsdataset}, Y_{\fbsdataset}>$,
%
%Let,
%\begin{equation*}
%    \fbsdataset \coloneqq <X_{\fbsdataset}, Y_{\fbsdataset}> 
%\end{equation*}
where $X_{\fbsdataset}$ are the packet features, $Y_{\fbsdataset}$ are the packet labels. %and,
% \elisa{I'm not sure about the notation below. Are you saying that i takes a value between 1 and a number equal to the number of features of the packet? If so, the notation would need to be slightly changed.}
%\begin{equation*}
% \begin{split}
%    i \in \{1, n_{\fbsdataset}\},  where \;   n_{\fbsdataset}\ = |X_{\fbsdataset}| \\
%    {\overrightarrow{x}_{\fbsdataset}}^{(i)} \in X_{\fbsdataset}, {\overrightarrow{y}_{\fbsdataset}}^{(i)} \in Y_{\fbsdataset}
%    \end{split}
%\end{equation*} 
Then for each packet $\mathcal{P}$ we label $\overrightarrow{y}_{\fbsdataset}^{(i)}$ as:
\[ {\overrightarrow{y}_{\fbsdataset}}^{(i)}= \begin{cases} 
      0 & \text{if ${\overrightarrow{x}_{\fbsdataset}}^{(i)}$ is a benign packet} \\
      1 & \text{if ${\overrightarrow{x}_{\fbsdataset}}^{(i)}$ is generated  from the FBS}
   \end{cases}
\]

For \msadataset, we define the set of MSAs detected by \system as $\mathcal{A} \coloneqq \{attack_1, attack_2, \cdots, attack_k\}$ %as the set of MSAs detected by \system 
where $k$ is the number of MSAs that can be detected and $ \mathcal{L}_{\mathcal{A}} \coloneqq \{0, 1, 2, \cdots, k\}$ as the set of labels for the MSAs. $\msadataset \coloneqq <X_{\msadataset}, Y_{\msadataset}>$,
%Let,
%\begin{equation*}
%    \msadataset \coloneqq <X_{\msadataset}, Y_{\msadataset}> 
%\end{equation*}
where $X_{\msadataset}$ are the packet features, $Y_{\msadataset}$ are the packet labels.
%\begin{equation*}
% \begin{split}
%    i \in \{1,n_{\msadataset}\}, where \; n_{\msadataset}\ = |X_{\msadataset}| \\
%    {\overrightarrow{x}_{\msadataset}}^{(i)} \in X_{\msadataset}, {\overrightarrow{y}_{\msadataset}}^{(i)} \in Y_{\msadataset}
%\end{split}
%\end{equation*} 
Then for each packet $\mathcal{P}$ we label $\overrightarrow{y}_{\msadataset}^{(i)}$ as:

% TODO: change att to attacks
% change the definition of i

\[ {\overrightarrow{y}_{\msadataset}}^{(i)}= \begin{cases} 
      0 & \text{if ${\overrightarrow{x}_{\msadataset}}^{(i)}$ is a benign packet} \\
      \mathcal{L}_{\mathcal{A}}[attack_j] & \text{if ${\overrightarrow{x}_{\msadataset}}^{(i)}$ is generated} \\ 
      &  \text{by $attack_j \in \mathcal{A}$}
   \end{cases}
\]

\subsection{Machine Learning Framework}
%Our ML pipeline is designed to detect 
%After we have the dataset, we need to design the Machine Learning Pipeline that can 
%detect the presence of FBSes and recognize multi-step attacks from captured packet traces. The pipeline has multiple components that we detail in what follows.
In what follows, we detail the ML framework used in \system.

\subsubsection{FBS Detection}\label{subsubsec:fbsdetection}
%The task of FBS detection can be described as follows. Let $\mathcal{P}$ represent a packet, $\mathcal{X}$ represent the feature vector extracted from the packet $\mathcal{P}$, and $\mathcal{Y}$ represent the class label. $\mathcal{Y}$ can be defined as
% where $\mathcal{Y} = 1$ indicates a packet generated due to a FBS, and $ \mathcal{Y} = 0 $ indicates a packet generated due to a legitimate base station. 
%\[
%    \mathcal{Y} = \begin{cases} 
%                      $1$ & \text{if packet generated  from FBS} \\
%                      $0$ & \text{otherwise}
%                   \end{cases}
%\]
%$\Theta$ represents the model parameters. The classification task can be formally represented as:
%$$\mathcal{Y} = Classifier(\mathcal{X}, \Theta)$$
% We train several 
% Classification models 
% for the FBS detection task. 
% They all 
% show good accuracy in detecting malicious packets; however, as those models have no memory, their false positive rates are much higher than attention-based models. 
%(more details are given in the evaluation section~\ref{subsec:fbs-det-eval}). 
% The reason is that 
% FBSes 
% typically 
% generate sequences of messages, and classifying a single packet standalone often leads to classification errors due to lack of context. 
% Furthermore, we observe that 
% \imtiaz{The next line is said before.}
% The lengths of the sequence 
% of packets 
% generated by FBSes follow specific distributions, albeit different for NAS and RRC. 
% Following these observations, 
We design a two-step framework for FBS detection.

\noindent \textbf{Packet-level classification.}
In the first step, we perform a packet-level classification using a stateful LSTM model with attention. 
% that takes the length of the sequence 
% of packets 
% into account 
% (shown in Algorithm~\ref{alg:fbs-msa-det-algo}). 
% The algorithm essentially divides the dataset into chunks of sequences using sequence length $(len_{seq})$ as a hyperparameter and returns a Seq-LSTM model trained on these sequences. 
% This custom model enhances the ability of LSTM networks to both focus on importance within a sequence and remember long term patterns, thereby not only increasing accuracy but also reducing or eliminating the need for extensive data preparation. 
This model utilizes stateful training and attention in parallel layers, merged and fed forward for the final combined output for the packet class prediction. 
% \imtiaz{Why are you doing this?}
The statefulness models long-term dependencies that span across sequences and the attention mechanism focuses on the parts of each sequence that affect the classification outcome the most.
The algorithm is shown in  Algorithm~\ref{alg:fbs-det-algo}.
The algorithm begins by taking the dataset and $len_{seq}$ hyperparameter as input. It then defines procedures for the Stateful LSTM and LSTM with Attention. The Stateful LSTM procedure initializes LSTM parameters, sets the stateful property to true to maintain state across batches, and then iterates over each timestep, feeding input $x_t$ and previous hidden state $h_{t-1}$ and cell state $c_{t-1}$ into the LSTM. It then returns the final hidden state $h_t$,
% \imtiaz{What is the need for this final hidden state?}
which enables state continuity across batches, ensuring that temporal dependencies are maintained even when sequences span multiple batches. 
The LSTM with Attention procedure initializes LSTM parameters, sets the return sequences property to true to output sequences instead of just the last timestep, processes the input sequence $x_t$ through the LSTM, computes a context vector from an attention mechanism over the LSTM outputs, and calculates the attended output $h'_t$. After defining the procedures, in lines $19-22$, the main algorithm sets the input sequence $x_t$ using $len_{seq}$ as the sequence length, computes the output of the Stateful LSTM and LSTM with Attention modules, concatenates their outputs, and applies a dense layer to produce the final output $y_t$. 
The modules' outputs are concatenated to provide a richer input to the dense layer, facilitating more informed and potentially more accurate predictions. The Stateful LSTM preserves temporal continuity, while the Attention module highlights relevant sequence parts.
The model is then trained using the calculated loss between the predicted output $y$ and the ground truth $\hat{y}$, and the gradients are propagated back through the network for parameter updates.

\noindent \textbf{Trace-level classification.}
% \imtiaz{Need a lot more details. Discuss the packet and trace level classifications in separate paragraphs with textbf.}
% In the trace-level classification step, traces containing packets flagged as benign and malicious undergo another contextual analysis. A simple binary classification model examines the order of packets and sequence patterns to discern characteristics indicative of FBS transmissions.
In the trace-level classification phase, traces comprising packets identified as benign or malicious are subjected to additional contextual analysis. This step applies a simple binary classification model, which analyzes the temporal sequence of packets to discern distinctive characteristics of FBS transmissions. By analyzing the packet traces, this model differentiates between FBS-related activity and benign network behavior, and gives the final prediction about the presence of an FBS in the traces.
%Elisa: do we have traces in which no packet is labeled as malicious or benign? Or can we have traces which contains some packets flagged as malicious and some as benigns?
% This stage employs sequence modeling techniques to capture temporal dependencies and sequence patterns. 
% thereby considering the broader context in which packets are sent.

\subsubsection{MSA Recognition}\label{subsubsec:Multi-Step-Attack-Recognition-using-Graph}
% For MSA recognition, we train all the classification models, the attention models, and the Sequential-LSTM discussed in Section~\ref{subsubsec:fbsdetection}, but all of them perform poorly. 
% The reason is that 
% As different MSAs follow different patterns, recognizing them successfully requires 
%it is necessary 
% capturing these patterns. Following this intuition, we use the graph learning approach to detect MSAs. 
% \imtiaz{This can not be a topic sentence.}
MSAs can be uniquely represented as directed graphs and detected using graph learning.
We describe the steps to create a directed graph from the traffic dataset and graph learning in Algorithm~\ref{alg:msa-det-algo}. 
% In this approach, 
The algorithm constructs a directed graph from the packets - each node denoting a packet, each packet having a directed edge towards the next packet. 
Edges are labeled with their reason for generation, the same as the packet label, generated as part of a benign flow or in the path of an MSA. 
% MSAs follow a specific path in this graph. 
Then, a graph learning model learns and generalizes this knowledge from the graphs, and the model is finally returned.
Recognizing MSAs using graph learning approaches can be formally represented as follows: 
let $G = (V, E)$ be the graph constructed from the packets, where each node $V$ denotes a packet. $E$ represents the directed edges between nodes in the flow graph. 
% $\mathcal{R(E)}$ represent the reason for the generation of each edge $\mathcal{E}$, where $R(\mathcal{E})$ = $\mathbb{BENIGN}$ if the edge is generated as part of a benign process and $R(\mathcal{E})$ = $\mathcal{A}$ if the edge is in the path of an MSA $\mathcal{A}$. 
%The process of recognizing multi-step attacks through graph learning can be expressed formally:
An MSA is recognized by identifying a specific path $P(G)$ corresponding to $\mathcal{A}$ within the graph $G$. These paths represent the packets' sequences that follow an MSA pattern. If there is an evolving and unseen attack, the attack would deviate from path $P(G)$ and follow a different path $P'(G)$.
Due to the nature of the vulnerabilities exploited by the attackers, $P'(G)$ and $P(G)$ will not be completely edge-disjoint and will have overlaps. From these overlaps, we can detect evolving, unseen and reshaped attacks.

\begin{figure}[]
\centering
\includegraphics[width=\linewidth]{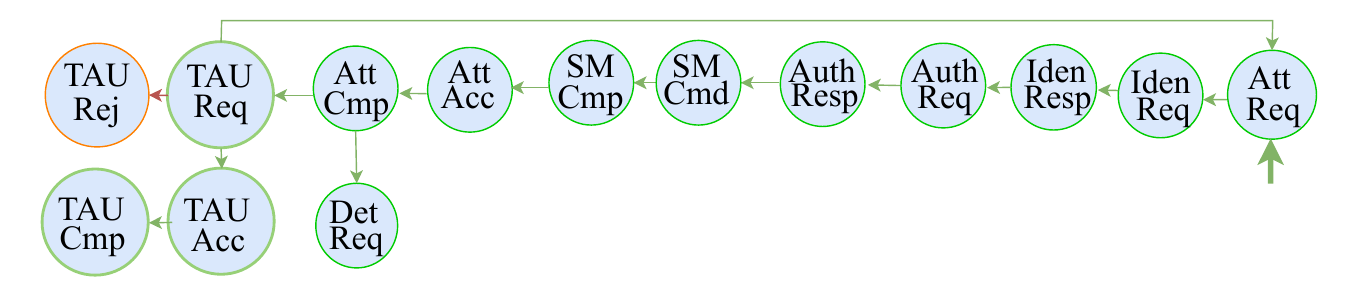}
\caption{Graph of TAU Reject Attack}
\label{fig:tau-reject-attack-graph}
\vspace{-0.7cm}
\end{figure}

\noindent \textbf{MSA detection graph example. }
To illustrate the process of MSA detection, we circle back to the \M{TAUReject} attack discussed in Section~\ref{subsec:taureject}. For this attack the FBS connects the legitimate BS and injects a Reject message. The graph generated from the NAS attack traces is shown in Figure~\ref{fig:tau-reject-attack-graph}. The communication starts when a UE sends \M{AttachRequest} to the legitimate BS, which is the starting node of the graph. An incoming edge is added to the node that represents the subsequent messages. The graph enters into an attack sequence when a \M{TAUReject} is sent after a \M{TAURequest}.

\subsubsection{Combining NAS and RRC Predictions}
% \imtiaz{This is not right.}
% In our approach, we leverage Dempster–Shafer theory (DST) to facilitate the fusion of predictions. We employ a Weighted Confidence-based Fusion method, where weights are assigned to the best models of each layer based on their performance. Additionally, when there's a disagreement between the models, the Dempster–Shafer theory is applied to prioritize predictions based on their confidence scores. This combination of DST and weighted confidence ensures a robust and effective approach to prediction fusion.
To combine the predictions from the NAS and RRC layer trace-level classification models, we leverage the Dempster–Shafer theory (DST)~\cite{DempsterShafertheory} to facilitate the fusion of predictions. We employ a weighted confidence-based fusion method, where weights are assigned to the best model of each layer based on the model performance.
% to ensure each layer's contribution is appropriate.
% The task of combining predictions from separately trained NAS and RRC layer models using ensemble techniques can be stated as follows: 
% Weighted averaging
% Weighted averaging is expressed as:
% \begin{equation}\label{eq:ensemble-formula}
%     \mathcal{P}_\mathbb{W} = \mathcal{W}_\mathbb{NAS} \cdot \mathcal{P}_\mathbb{NAS} + \mathcal{W}_\mathbb{RRC} \cdot \mathcal{P}_\mathbb{RRC}
% \end{equation}
% \begin{equation*}
%    P_W = \sum W \cdot P
% \end{equation*}
% where $W = \{\frac{W_{NAS}}{W_{NAS}+W_{RRC}}, \frac{W_{RRC}}{W_{NAS}+W_{RRC}}\}$. 
\looseness=-1
Let $W_{NAS}$ and $W_{RRC}$ represent the weights assigned to the best trace-level classification models of NAS and RRC layer, $P = \{P_{NAS}, P_{RRC}\}$ represent the predictions made by those models, and $P_W$ is the final prediction.
% , which combines the predictions in the NAS and RRC layer and gives a unified prediction.
% \imtiaz{Add some more lines on how these weights are selected. You can say these edges are empirically selected based on experiments or something like that.}
% $\mathcal{E}$ represents the ensemble method used to combine predictions from both models.
% \elisa{we do not say anything about the values of those weights. Are they the same?}
% We take the best-performing model from each layer, and each selected model independently predicts the outcomes for the incoming packets in its layer.
We assign weights $W_{NAS}$ and $W_{RRC}$ proportional to their support scores on the inference. The more confident model among the two models receives a higher weight, indicating a more significant influence on the combined prediction. Mathematically, this is expressed as:
\looseness=-1
\begin{align*}
 W_{NAS} \propto \text{Support\_Score($P_{NAS}$)} \\
 W_{RRC} \propto \text{Support\_Score($P_{RRC}$)}
\end{align*}
When there's a disagreement between the models, the Dempster–Shafer theory is applied, and the prediction with higher confidence gets priority.
% and is set as the combined prediction. 
This process is formalized as:

% \[ \text{$P_W$} = \begin{cases} 
% \text{$P_{NAS}$} & \text{if } \text{$W_{NAS}$} > \text{$W_{RRC}$} \\
% \text{$P_{RRC}$} & \text{otherwise}
% \end{cases}
% \]

%\[ \text{$P_W$} = \begin{cases} 
%\text{$P_{NAS}$} & \text{if } \text{$P_{NAS}$} = \text{$P_{RRC}$} %\text{ or } \text{$W_{NAS}$} > \text{$W_{RRC}$} \\
%\text{$P_{RRC}$} &  \text{otherwise}
%\end{cases}
%\]

\[ \text{$P_W$} = \begin{cases} 
\text{$P_{NAS}$} & \text{if } \text{$P_{NAS}$} = \text{$P_{RRC}$}\\
\text{$P_{NAS}$} & \text{if } \text{$P_{NAS}$} \neq \text{$P_{RRC}$} \text{ and }  \text{$W_{NAS}$} > \text{$W_{RRC}$} \\
\text{$P_{RRC}$} & \text{if } \text{$P_{NAS}$} \neq \text{$P_{RRC}$} \text{ and } \text{$W_{NAS}$} < \text{$W_{RRC}$} 
\end{cases}
\]

Combining the predictions in this way yields a better detection accuracy than the individual predictions.
\section{Implementation}\label{sec:implementation}
Now, we discuss the implementation details of each component of \system.
% We develop two models, one for FBS detection and another for MSA recognition.
We implement two distinct models: one to detect FBSes and another specifically tailored to recognize MSAs. Implementation efforts of \system are summarized in Table~\ref{tab:imp-efforts}.

% \imtiaz{Add a table showing all the implementation efforts.}
% \elisa{It seems to me that there is some overlap and repetitions between Sections 5 and 6. I wonder whether we can merge the two sections into one titled "Detailed Design and Implementation".}
\subsection{Dataset Construction}

%\subsubsection{Dataset Generation}

%\noindent\textbf{Appropriate Network topology. }
% \begin{itemize}
%     \item gini code
%     \item powder profiles
%     \item fake base station incorporation
% \end{itemize}
%POWDER has different profiles already set up 
%in their platform 
%that 
%users 
%one can use for experiments. Among the profiles, we choose the \textit{srsran-handover}~\cite{srsran-handover} profile because it is the closest one to our requirements. To support handover, it uses the Open5GS~\cite{Open5GS} implementation for the core network which serves two base stations. One UE can be handed over between the base stations by tuning the profile parameters. Details can be found in the documentation of the profile. Both the base stations and UE use the srsRAN~\cite{srsRAN} implementation.

\noindent \textbf{Incorporation of FBSes. }
% \imtiaz{This whole paragraph seems filled with trivial stuff. Instead, focus on this reviewer's comments:  The delivery of methodology is somehow broad. For instance, in Section 5.2, FBSDetector introduces LSTM and GNN to trace the packages during attack performing, while most details on how to utilize and bridge the two algorithms are missed, in which inputs and outputs are not clear. Besides, the ensemble learning process in 5.2.3 is not elaborated, just an equation but nothing is explainable other than experimental results in Table 3 and Table 17, which can lead to confusion.}
% \imtiaz{The implementation should be modified in a way that should clarify all these questions.}
% POWDER has different profiles already set up that one can use for experiments.
% Among the profiles, we choose the \textit{srsran-handover}~\cite{srsran-handover} profile because it is the closest one to our requirements.
% It uses Open5GS~\cite{Open5GS} for the core network, which supports handover.
% The core network serves two base stations, both the base stations and UE use srsRAN~\cite{srsRAN} implementation.
To incorporate FBSes, 
% we modify the \textit{srsran-handover} as follows: 
% we create a new core network and base station pair, 
% cloning the legitimate core network and base station from \textit{srsran-handover} profile.
% The modified profile has
we create (i) a legitimate core network and BS pair, and (ii) a fake core network and BS pair in POWDER.
The legitimate BS serves UEs that the FBS can attack by spawning near it with higher signal strength.
%\imtiaz{Mention srsran and OAI and mention this is due to reduce bias on srsran.}
The core networks use Open5GS~\cite{Open5GS} to support handover, and the UE uses srsRAN~\cite{srsRAN} and OpenAirInterface(OAI)~\cite{OpenAirInterface}--the two available open-source implementations. We equally use srsRAN and OAI for the dataset generation to reduce bias on a specific implementation. %Following our threat model, the FBS core network does not know the cryptographic keys.
% We experimented with diverse scenarios of varying attacker capability, mobility, and combinations of benign and fake handovers.
% We create and experiment with attackers with diverse abilities by changing different parameters in the base station and core network configurations across our experiments in POWDER. 
%To incorporate less sophisticated attackers, we only set up a FBS with a fixed high signal strength value in the configuration. 
%In contrast, to incorporate more sophisticated attackers, we clone legitimate BS IDs and different parameters, change signatures and signal strength, and turn them on and off to evade detection. 
We introduce mobility in the dataset by using UEs with the mobile endpoints in POWDER that are mounted on the campus shuttles. Benign handovers are initiated when a shuttle with a mobile endpoint moves from the vicinity of one BS to another, and the signal quality received at the UE changes. %To initiate a fake handover through FBSes, we spawn a FBS near a UE with high signal strength and make the UE initiate a fake handover. 
% We combine all these scenarios in our experiments, listed in Table~\ref{tab:fbs-exp-desc}.

%We save this profile as \textit{srsran-fbs-handover}. 
% We change the capabilities of the cloned FBS to ensure the setup follows the assumptions in the threat model. All the traces from different components, such as the legitimate core network and base station, fake core network and FBS, and the UE, are captured for both the NAS and RRC layers.
% \imtiaz{Add more if I am missing anything.}
%\cite{srsran-fbs-handover}.
\noindent \textbf{Implementation of handover capability in the srsUE. }
The UEs usually initiate handovers by sending a \TrackingAreaUpdateRequest message. This message is not implemented in the current release of srsRAN and OAI. We implement the \TrackingAreaUpdateRequest message in UEs and make it able to initiate a handover request by sending a \TrackingAreaUpdateRequest message.

\noindent \textbf{Implementation and execution of MSAs. }
We implement all the attacks listed in Table~\ref{tab:multi-step-attack}, including all the attacker levels in srsRAN and OAI and execute them in the experimental setup we created in POWDER. 
% Similar to the FBS experiments, we experiment with varying attacker capabilities. 
%For a less sophisticated attacker, we implement the MSAs as their definition, but for a more sophisticated attacker, we reshape the packets and change the attack patterns by changing the packet sending orders.
% Same as FBS data collection, 
All the traces from different components
% , such as the legitimate core network and base station, fake core network and FBS, and the UE,
are captured for both NAS and RRC layer.
% In total \totalLinesOfCodeAddedInsrsRANforMultiStep lines of code were added to the srsRAN project for the implementation of all the attacks.

\noindent \textbf{Data processing.}
%As our approach is to identify the fake base stations and multi-step attacks from the UE, we proceed with the traces captured at the UE, for both NAS and RRC layers. 
%\noindent \textbf{Decoding Packets. } 
We decode the packets using tshark~\cite{tshark}
% in Packet Details Markup Language (pdml), an XML-based format for the details of a decoded packet, and save them in an XML file. 
and use a Python script for protocol filtering, packet field extraction and packet field alignment
% , and handling missing values.
Lastly, we use \textit{scikit-learn}'s \textit{LabelEncoder} to encode the categorical fields 
% converting them 
into numerical representations.
\noindent \textbf{Data labeling.}
% We label the packets with predefined labels according to their reason for generation. 
We 
% store the predefined attack labels in a dictionary and 
assign each packet a label according to the reason for its generation. The NAS layer packets (fewer in number) are labeled manually
% with a script that 
by checking each packet and assigning a label according to the reason for its generation.
% the dictionary. 
Labeling all the NAS layer packets takes approximately 2 hours of one-time manual effort.
% , which we defined in the dataset creation experiments. 
An automated script is used to label the RRC layer packets (which are much more numerous) in batches of intervals by detecting attack intervals in the NAS layer traces.

\noindent \textbf{Dataset distribution per attacker level.}
Our \totalsizeofdata GB dataset is structured to represent five attacker levels, with six FBS experiments conducted per level. Among them level 0, 1, and 2 are for FBS only and level 3 and 4 are for FBS and MSA. The combined size for Levels 0, 1, and 2, which includes FBS experiments only, is 6.6 GB, divided as 2.5 GB for Level 0, 2.1 GB for Level 1, and 2.0 GB for Level 2. The remaining 2.6 GB belongs to Levels 3 and 4, where both FBS and  MSAs are included, with 1.4 GB for Level 3 and 1.2 GB for Level 4. 
%Level 0 features naive FBS setups with high signal strength, serving as the simplest baseline. Level 1 involves FBS configurations with optimal signal strength to induce handovers, challenging detection thresholds. Level 2 includes attacks cloning parameters such as MCC, MNC, TAC, and RF configurations, essential for evaluating impersonation detection. Level 3 comprises Multi-Step Attacks (MSAs), representing complex attack sequences. Finally, Level 4 contains highly sophisticated attacks designed to evade defenses. 
% By running multiple instances of each attack and experiment, we ensure consistent coverage of real-world scenarios across all sophistication levels, enabling robust evaluation of detection mechanisms.

\subsection{FBS Detection}
\noindent \textbf{Train-test split.} 
\looseness=-1
% Unlike the standard libraries, 
We use a custom script to split our 
% experimental 
dataset for training and testing, which does roughly $80-20$ split, preserving the sequence of packets
% For example, if we have $10$ instances of the FBS experiment, we append $8$ experiment's data to the train set and $2$ to the test set.
% This way,
% We preserve the packet sequence 
while ensuring the standard split; also, no experimental trace is cut in between. 
% We have the whole trace of each experiment as it occurred and was captured. 

\noindent \textbf{Stateful LSTM with attention model.} We utilize the TensorFlow functional API and use stateful LSTM and LSTM with attention in parallel layers. Their outputs are merged and fed forward to a common dense layer. Each network side is trained according to its architecture in this model.
The Stateful LSTM model is architecturally identical to the vanilla LSTM; however, the learning algorithm has been altered to maintain the states. Both return sequences and maintain state parameters are set to true.
For the LSTM with Attention model, the time-distributed dense layer is replaced by an attention layer. Return sequences are set to true, enabling the complete hidden layer sequences to be sent forward to the attention layer, where they are processed similarly to the encoder/decoder and vanilla LSTM models.

\subsection{MSA Recognition}
%As discussed in Subsection~\ref{subsubsec:Multi-Step-Attack-Recognition-using-Graph}, we use the Graph Learning approach to detect Multi-Step Attacks.
%\noindent \textbf{Graph Construction. } 
% \imtiaz{There should not be any package name in the paper, put all in Table 2.}
% We use the Python \textit{networkx}~\cite{networkx} library to create the directed graphs using the \textit{DiGraph} function.
For NAS layer packets, we create a node for every unique value of the \PacketState{nas\_eps\_nas\_msg\_emm\_type\_value} field, the packet name for NAS layer packets. 
Similarly, for the RRC layer packets, we create a node for every unique value of the \PacketState{lte-rrc\_c1\_showname} field, which is the packet name for the RRC layer packets. Every packet maps to a node in the graph corresponding to its packet name.
We add an edge from the node representing one packet to the node representing the next packet in the sequence and label that edge with the same label we labeled the next packet. This denotes if the transition was benign or due to an attack.
%The containers and libraries to create the directed graphs and graph learning are summarized in Table~\ref{tab:imp-efforts}.
% A graph created from an MSA, namely paging channel hijacking attack, depicted in Figure~\ref{fig:paging-channel-hijacking-attack-diagram}, can be seen in Figure~\ref{fig:paging_channel_hijacking_attack}.
%
% \imtiaz{Add example for Figure 3 and Figure 4.}
%
% For graph learning, we use pytorch container~\textit{torch\_geometric.nn}~\cite{pytorch-geometric}. 
%
%\noindent \textbf{Graph Learning. }
%On the constructed graphs we train \textit{Graph Convolutional Network (GCN)}, Graph Attention Network(GAT), GraphSAGE~\cite{GraphSAGE}, and GraphTransformer. All these Graph Neural Networks are implemented and available in pytorch container~\textit{torch\_geometric.nn}~\cite{pytorch-geometric}. 
%The models learn the graph structures and which transition is benign and which transition is due to a specific attack. Based on the learning, one can recognize different multi-step attacks when seeing such transitions in the graphs. The attack signatures are very evident in the constructed graphs and are learned very effectively by the Graph Neural Networks.
%
%
%
%
\subsection{Deployment and Integration} 
To deploy \system, we use Mobileinsight~\cite{Mobileinsight} to parse the baseband traces in the mobile phones, \textit{TensorflowLite}~\cite{tflite} to run the ML models, and \textit{Flutter}~\cite{flutter} 
to build the app. 
\section{Evaluation}
%In order to compare \system’s effectiveness in detecting FBSes and recognizing multi-step attacks, we investigate the following research question:
%In order to evaluate the effectiveness of \system we aim to answer the following research questions:
% To identify the characteristics and gain insights about the dataset, we show different distributions of the packets in our dataset in Appendix 
% \elisa{Please remember to add the reference to the appendix here.}
% Section~\ref{apndx-subsec:dataset}. 
We evaluate the effectiveness of \system based on the following research questions:
%
%\begin{description}[noitemsep,nolistsep]
    % to identify FBSes and recognize MSAs using ML algorithms effectively?
    \textbf{RQ1.}
    % Why are both packet-level classification and trace-level classification required for successful FBS detection? 
    % Why are statefulness and attention essential for the packet-level classification model?
    What is the performance of each step in the FBS detection framework, namely the packet classification and trace classification?
    What is the performance improvement of using stateful LSTM with attention in packet-level classification?
    How does combining predictions of NAS and RRC trace classification further improve performance? How does graph learning improve MSA recognition performance? Why does the simple heuristic-based detection not work?
    \textbf{RQ2.}
    What is the memory and power consumption of the detection framework?
    % when deployed in a real system? 
    How much time does the inference take?
    \textbf{RQ3.}
    Was \system deployed and tested in a real-world setup? How does it contrast with existing FBS detection solutions?
    \textbf{RQ4.}
    Is \system robust and generalizable against 
    % different types of FBS and MSA, especially with sophisticated attackers who can reshape the packets and change attack patterns? What happens in the event of such 
    reshaping and unseen attacks? Also can it detect Overshadowing attacks?
    %\item \textbf{RQ5.}
    %Why is the trace-level classification necessary along with the packet-level classification? 
    
%\end{description}
\noindent \textbf{Experimental Setup.}
% \subsection{Training Hardware.} 
For training the models, we utilized a Lenovo ThinkPad T480 equipped with 32GB of memory and an Intel Core i7-8650U CPU @ 1.90GHz × 8. The operating system was Ubuntu 22.04.1 LTS (64-bit), running GNOME Version 42.2.
We provide detailed information about the 
% experimental setup, including the training hardware specifications and 
model hyperparameters, in Appendix Section \ref{apndx-sec:experimental-setup}.
% \imtiaz{Include experimental setup here. Keep hyperparameters for appendix.}

In what follows, we delve into the details and answers to those research questions.
%\vspace{-0.3cm}

\subsection{RQ1. Packet and trace level classification} %, stateful-LSTM with attention, graph learning, predictions combining}
\label{subsec:fbs-det-eval}

\noindent\textbf{Packet and trace level classification.}
%Packet-level classification performance is shown in Table~\ref{tab:fbs-detection-performance-nas}. A subsequent trace-level classification performance on these labeled packets is shown in Table~\ref{tab:trace-level-fbs-classification-scores}. 
Our stateful-LSTM model with attention performs substantially better compared to the other packet-level models (shown in Table~\ref{tab:fbs-detection-performance-nas}). Meanwhile, for trace-level classification, all the classical ML models perform similarly (shown in Table~\ref{tab:trace-level-fbs-classification-scores}). This is expected because the heavy lifting of \system is done in the packet-level classification phase.
%\imtiaz{Please see the text above and revise if needed.}

\noindent\textbf{Performance improvement using Stateful-LSTM with attention. } 
%better detect FBSes.}
% Like legitimate BSes, FBSes generate sequences of packets.
% Figures~\ref{fig:fbs-seq-len-distr-nas} and \ref{fig:fbs-seq-len-distr-rrc} show the distribution of 
% \elisa{do you mean the distribution of the length of such sequences?}
% the length of the sequences. 
% \elisa{I'm not the text below is needed here; you have already said this in the design and implementation section. I have thus commented out your text and used a different text. }
%Therefore, detecting a single packet without context is often erroneous. 
% As mentioned in Section~\ref{sec:detailed_design}, 
%To take into account the context for each packet, we use 
%Motivated by this observation we design 
%a stateful-LSTM with attention model.
%and trains on these chunks of sequences. 
% Our experimental results, reported in Table~\ref{tab:fbs-detection-performance-nas} show that this approach performs better than any other model. For the detection in NAS layer packets, Sequential-LSTM beats its closest competitor XGBoost with a 6\% increase in recall, 2\% increase in F1-Score and 2\% increase in accuracy. For RRC layer packets the improvement over XGBoost is 1\% in precision, 10\% in recall, 6\% in F1-Score and 7\% in accuracy. 
%The experimental results reported in Table~\ref{tab:fbs-detection-performance-nas} demonstrate the superior performance of this approach over other models. 
Stateful-LSTM with attention 
% outperforms its closest competitor XGBoost, in detecting NAS layer packets with a improvement of 6\% in accuracy, 
improves the performance of the vanilla LSTM model by $6\%$ in NAS layer packet classification and $3\%$ in RRC layer packet classification (shown in Table~\ref{tab:fbs-detection-performance-nas}).
Substantial enhancements are also observed in precision, recall and f1-score. Improving recall and accuracy in malicious traffic classification means the model is better at capturing a larger proportion of actual threats, reducing the chances of missing malicious activity. This is crucial for minimizing false negatives and enhancing overall detection effectiveness.
% For RRC layer packets, the improvement over XGBoost though small in precision (1\%), significant improvement is seen in recall (10\%) in F1-Score (6\%), and accuracy (7\%).
% 10\% in recall, 6\% in F1-Score, and 7\% in accuracy. 
% crucial metrics for traffic classification tasks emphasizing low false positive rates.
%which can be seen in Table \ref{tab:fbs-detection-performance-nas} and \ref{tab:fbs-detection-performance-rrc}. 
%Tuning the sequence-length hyperparameter properly is very important because the detection performance depend largely on it. 
% More significant improvement is seen in precision and recall, which are more desirable performance metrics in traffic classification tasks where low false positive rates are a must.
Figures~\ref{fig:fbs-seq-len-distr-nas} and \ref{fig:fbs-seq-len-distr-rrc} show the distribution of the length of the FBS generated packet sequences and
Figures~\ref{fig:acc-vs-seq-len-nas} and \ref{fig:acc-vs-seq-len-rrc} show the impact of sequence length on the detection performance. The LSTM model performs better when the input sequence length is between $9-15$ for NAS layer packets and $80-120$ for RRC layer packets. 
% \elisa{This result is very interesting. Is there an explanation for the different lengths for RRC and NAS?}
The reason is that in an FBS session, based on the communication process of the FBSes, packets exchanged between the FBSes and the UEs in the NAS and RRC layer follow a specific distribution. These distributions are captured better when the sequence lengths are set in the range that can accommodate the distribution completely, and models trained on these segments of packets that contain these patterns can sufficiently learn better about the attack.  
% This fully agrees with the distribution of the sequence lengths we have observed previously.

\begin{table}[]
    \centering
    \renewcommand{\arraystretch}{1}
    \fontsize{6}{6}\selectfont
    \setlength{\tabcolsep}{0.5pt} 
    \begin{tabular}{l|cccc|cccc}
    \hline
        Model & \multicolumn{4}{c}{NAS Layer Packets} & \multicolumn{4}{c}{RRC Layer Packets} \\
        \cline{2-9}
         & Precision & Recall & F1-Score & Accuracy & Precision & Recall & F1-Score & Accuracy \\
        \hline
        Random Forest            & 0.85          & 0.87          & 0.82           & 0.84        & 0.68          & 0.81           & 0.16          & 0.69 \\
        SVM                      & 0.81          & 0.21          & 0.70           & 0.59        & 0.66          & 0.79           & 0.00          & 0.66\\
        Decision Tree            & 0.86          & 0.89          & 0.81           & 0.82        & 0.75          & 0.84           & 0.53          & 0.76\\
        XGBoost                  & 0.89          & 0.89          & 0.84           & 0.84        & 0.83          & 0.87           & 0.79          & 0.84\\
        k-NN                     & 0.86          & 0.81          & 0.80           & 0.79        & 0.86          & 0.89           & 0.82          & 0.83\\
        Na\"ive Bayes            & 0.53          & 0.87          & 0.35           & 0.58        & 0.80          & 0.08           & 0.52          & 0.37\\
        Logistic Regression      & 0.50          & 0.68          & 0.69           & 0.53        & 0.74          & 0.85           & 0.50          & 0.71 \\
        CNN                      & 0.86          & 0.39          & 0.57           & 0.52        & 0.84          & 0.67           & 0.78          & 0.66 \\
        FNN                      & 0.79          & 0.85          & 0.68           & 0.73        & 0.80          & 0.88           & 0.84          & 0.78\\
        LSTM                     & 0.86          & 0.85          & 0.82           & 0.89        & 0.89          & 0.86           & 0.81          & 0.89\\
        \textbf{Stateful-LSTM w/ attention} & \textbf{0.91} & \textbf{0.97} & \textbf{0.86}  & \textbf{0.95} & \textbf{0.94 }         & \textbf{0.97}  & \textbf{0.95} & \textbf{0.92} \\
        \hline
    \end{tabular}
    \caption{Performance of packet level classification for FBS detection}
    \vspace{-0.3cm}
    \label{tab:fbs-detection-performance-nas}
\end{table}

\begin{table}[ht]
\centering
\renewcommand{\arraystretch}{1}
\fontsize{6}{6}\selectfont
\setlength{\tabcolsep}{0.5pt} 
\begin{tabular}{l|cccc|cccc}
\hline
Model & \multicolumn{4}{c}{NAS Layer Trace} & \multicolumn{4}{c}{RRC Layer Trace} \\
\cline{2-9}
 & Precision & Recall & F1-Score & Accuracy & Precision & Recall & F1-Score & Accuracy \\
\hline
Logistic Regression & 0.95 & 0.94 & 0.94 & 0.94 & 0.92 & 0.91 & 0.91 & 0.91 \\
K-Nearest Neighbors & 0.95 & 0.94 & 0.94 & 0.94 & 0.92 & 0.91 & 0.91 & 0.91 \\
Decision Tree & 0.95 & 0.94 & 0.94 & 0.94 & 0.92 & 0.91 & 0.91 & 0.91 \\
Random Forest & 0.95 & 0.94 & 0.94 & 0.94 & 0.92 & 0.91 & 0.91 & 0.91 \\
Gradient Boosting & 0.95 & 0.94 & 0.94 & 0.94 & 0.92 & 0.91 & 0.91 & 0.91 \\
\textbf{Support Vector Machine} & \textbf{0.96} & \textbf{0.95} & \textbf{0.95} & \textbf{0.95} & \textbf{0.93} & \textbf{0.92} & \textbf{0.92} & \textbf{0.92} \\
\hline
\end{tabular}
\caption{Performance of trace level classification}
\label{tab:trace-level-fbs-classification-scores}
\vspace{-0.5cm}
\end{table}

\begin{table}[]
    \centering
    \renewcommand{\arraystretch}{1}
    \fontsize{8}{8}\selectfont
    \setlength{\tabcolsep}{2pt} 
    \begin{tabular}{l|ccc|c}
        \hline
        Model                               & Precision        & Recall                & F1-Score       & Accuracy      \\
        \hline
        Trace Level FBS Classifier (NAS)    & 0.96              & 0.95                  & 0.95          & 0.95 \\
        Trace Level FBS Classifier (RRC)    & 0.93             & 0.92                  & 0.92           & 0.92          \\
        \textbf{Combined Trace Level Prediction}        & \textbf{0.96}    & \textbf{0.95 }        & \textbf{0.96}  & \textbf{0.96} \\
        \hline
    \end{tabular}
    \caption{Performance of combined trace level classification for FBS detection }
    \vspace{-0.5cm}
    \label{tab:fbs-detection-performance-combined}
    \vspace{-0.5cm}
\end{table}

% \subsection{RQ3. Graph Learning, MSA Recognition, Unseen Attack Detection}
% \imtiaz{Add a topic sentence first}
% \noindent\textbf{Graph Learning in MSA Recognition.}
\noindent\textbf{Graph learning. }
%MSAs follow specific patterns (shown in Figure~\ref{fig:paging_channel_hijacking_attack}).  
% \elisa{We have said this before. Can you remove this text?}
% Following this intuition, we use Graph Learning approaches. 
%We train graph learning models to learn these patterns, and Table~\ref{tab:multi-step-attack-recognition-performance-nas-rrc} shows that by learning the patterns, the graph learning models perform better than any other models in recognizing MSAs. 
%GraphSAGE performs the best among all. So 
%
%
We use GraphSAGE as our MSA recognition model since, among the graph learning models, it performs better than any other models (shown in Table~\ref{tab:multi-step-attack-recognition-performance-nas-rrc}). Though the improvement seen in accuracy might seem small (1\%), a significant improvement is seen in precision (4-8\%), recall (10-12\%) and f1-score (8-10\%).
% by improving the performance metrics by 5-10\% over the classical ML models. 
% It is worth noting that Sequential-LSTM also achieves competitive performance compared to the other ML models.
% \imtiaz{Revise}
% Ensemble learning further improves this performance (shown in Table \ref{tab:multi-step-attack-recognition-performance-combined}).
% \elisa{This Table appears at the end of the paper. Could we make sure that appears here?}

\begin{table}
    \centering
    \renewcommand{\arraystretch}{1}
    \fontsize{6}{6}\selectfont
    \setlength{\tabcolsep}{0.5pt} 
    \begin{tabular}{l|cccc|cccc}
        \hline
        \multirow{2}{*}{Model} & \multicolumn{4}{c}{NAS} & \multicolumn{4}{c}{RRC} \\
        \cline{2-9}
        & Precision & Recall & F1-Score & Accuracy & Precision & Recall & F1-Score & Accuracy\\
        \hline
        Random Forest               & 0.617 & 0.628 & 0.454 & 0.76 & 0.343 & 0.278 & 0.254 & 0.42 \\
        SVM                         & 0.088 & 0.200 & 0.122 & 0.44 & 0.076 & 0.200 & 0.110 & 0.38 \\
        Decision Tree               & 0.342 & 0.402 & 0.356 & 0.66 & 0.076 & 0.200 & 0.110 & 0.38 \\
        XGBoost                     & 0.794 & 0.573 & 0.786 & 0.84 & 0.586 & 0.530 & 0.548 & 0.47 \\
        k-NN                        & 0.734 & 0.702 & 0.706 & 0.81 & 0.484 & 0.454 & 0.392 & 0.47 \\
        Naive Bayes                 & 0.440 & 0.384 & 0.274 & 0.29 & 0.440 & 0.258 & 0.178 & 0.11 \\
        Logistic Regression         & 0.284 & 0.420 & 0.142 & 0.46 & 0.412 & 0.318 & 0.300 & 0.49 \\
        CNN                         & 0.132 & 0.274 & 0.114 & 0.22 & 0.072 & 0.259 & 0.106 & 0.36 \\
        FNN                         & 0.128 & 0.282 & 0.222 & 0.24 & 0.274 & 0.220 & 0.144 & 0.40 \\
        LSTM                       & 0.150 & 0.320 & 0.200 & 0.49 & 0.128 & 0.164 & 0.130 & 0.30 \\
        % Sequential-LSTM   \imtiaz{Revise}           & 0.744 & 0.726 & 0.784 & 0.85 & 0.408 & 0.342 & 0.348 & 0.53 \\
        Graph Attention Network     & 0.868 & 0.672 & 0.865 & 0.84 & 0.316 & 0.338 & 0.298 & 0.36 \\
        Graph Attention Network v2  & 0.842 & 0.864 & 0.857 & 0.83 & 0.504 & 0.411 & 0.421 & 0.42 \\
        Graph Convolutional Network & 0.832 & 0.754 & 0.818 & 0.80 & 0.306 & 0.464 & 0.352 & 0.40 \\
        Graph Transformer           & 0.836 & 0.882 & 0.841 & 0.81 & 0.436 & 0.495 & 0.430 & 0.41 \\
        \textbf{GraphSAGE}          & \textbf{0.872} & \textbf{0.924} & \textbf{0.883} & \textbf{0.85}  & \textbf{0.633} & \textbf{0.561} & \textbf{0.557} & \textbf{0.59} \\
        \hline
    \end{tabular}
    \caption{Performance of MSA recognition from NAS and RRC layer packets}
    \vspace{-0.3cm}
    \label{tab:multi-step-attack-recognition-performance-nas-rrc}
\end{table}

\begin{table}[]
    \centering
    \renewcommand{\arraystretch}{1}
    \fontsize{8}{8}\selectfont
    \setlength{\tabcolsep}{3pt} 
    \begin{tabular}{l|cccc}
        \hline
        Model & Precision & Recall & F1-Score & Accuracy \\
        \hline
        GraphSAGE (NAS) & 0.867 & 0.879 & 0.856 & 0.850 \\
        GraphSAGE (RRC) & 0.611 & 0.480 & 0.52 & 0.590 \\
        \textbf{Combined Prediction} & \textbf{0.875} & \textbf{0.901} & \textbf{0.874} & \textbf{0.865}\\
        \hline
    \end{tabular}
    \caption{Performance of MSA recognition after combining predictions}
    \vspace{-0.3cm}
    \label{tab:multi-step-attack-recognition-performance-combined}
\end{table}

% \noindent\textbf{Weighted Confidence-based Fusion Method for Combining Predictions.}
\noindent\textbf{Predictions combining.}
% Ensemble techniques offer a robust means to blend the outputs of NAS and RRC layer models, facilitating the creation of a unified and accurate predictive framework that captures the strengths of each layer. 
% We take the best-performing models from each layer and create an ensemble model that combines the predictions of each layer. 
The performance after combining the predictions of NAS and RRC trace classification using the weighted confidence-based fusion method
% ensemble models
for FBS detection and MSA recognition is shown in Table~\ref{tab:fbs-detection-performance-combined} and Table ~\ref{tab:multi-step-attack-recognition-performance-combined}, respectively. 
By leveraging knowledge from both layers, 
% this method gives a unified prediction and performs better than any individual detection model. 
this method improves on all the performance metrics from the best trace-level classification model among the two and makes the system robust, especially by improving recall.

\looseness = - 1
\noindent \textbf{Necessity of Trace Level Classification.} 
%Packet-level classification often lacks the broader context required to identify complex malicious activities. Malicious behaviors often unfold over a sequence of packets and exhibit patterns that cannot be detected in isolation. 
%There can be false positives in packet level classification while classifying single packets.
%Trace-level classification aggregates information from multiple packets to analyze these patterns, providing a contextual understanding of the traffic, helping reduce false positives by distinguishing between isolated anomalies and those indicative of actual attacks. 
To evaluate the efficacy of the trace-level classification model 
we conduct a performance comparison between packet-level classification only and the combined approach with trace-level classification. For the packet-level classification-only approach, we assume that if at least one packet is flagged as malicious, we consider the whole trace as malicious. In Table~\ref{tab:pkt-level-vs-trace-level-and-zero-shot-det}, we see by enriching the analysis with aggregated data, trace-level classification enhances detection accuracy and ensures a more comprehensive approach to identifying threats.

\begin{table}[]
    \centering
    \renewcommand{\arraystretch}{1}
    \fontsize{8}{8}\selectfont
    \setlength{\tabcolsep}{2pt} 
    \begin{tabular}{l|cccc}
        \hline                     
        Task & Prec & Rec & F1 & Acc \\
        \hline
        Packet Level Classification Only & 0.91 & 0.90 & 0.91 & 0.90 \\
        Combined w/ Trace Level Classification & \textbf{0.96} & \textbf{0.95} & \textbf{0.96} & \textbf{0.96} \\
        \hline
        \hline
        Overshadow attack~\cite{yang2019hiding} detection (Zero Shot) & 0.84 & 0.82 & 0.83 & 0.86 \\
        %aLTEr~\cite{ALTER} attack detection (Zero Shot) & 0.8 & 0.83 & 0.81 & 0.81 \\
        \hline
    \end{tabular}
    \caption{Performance of combined packet and trace level classification for FBS detection and a zero shot detection evaluation of overshadowing attack}
    \vspace{-0.5cm}
    \label{tab:pkt-level-vs-trace-level-and-zero-shot-det}
    \vspace{-0.5cm}
\end{table}

\subsection{RQ2. Overhead Analysis} %: Time Consumption, Memory Consumption, Power Consumption}
\begin{figure*}[t]
     \centering
     \begin{subfigure}[b]{0.23\textwidth}
        \centering
        \includegraphics[width=1.0\linewidth]{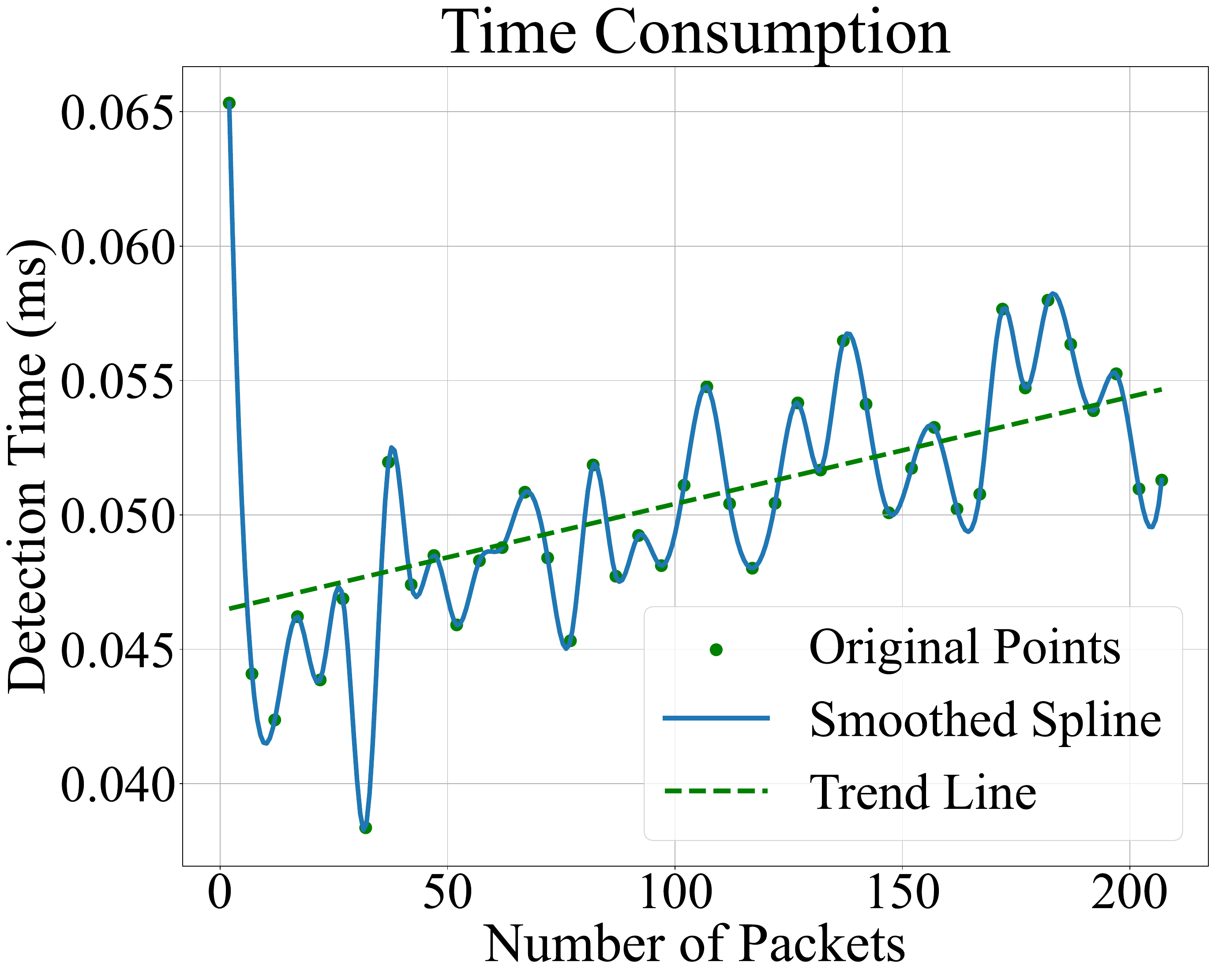}
        \caption{Time consumption}
        \label{fig:time-consumption}
     \end{subfigure}
     \begin{subfigure}[b]{0.23\textwidth}
        \centering
        \includegraphics[width=1.0\linewidth]{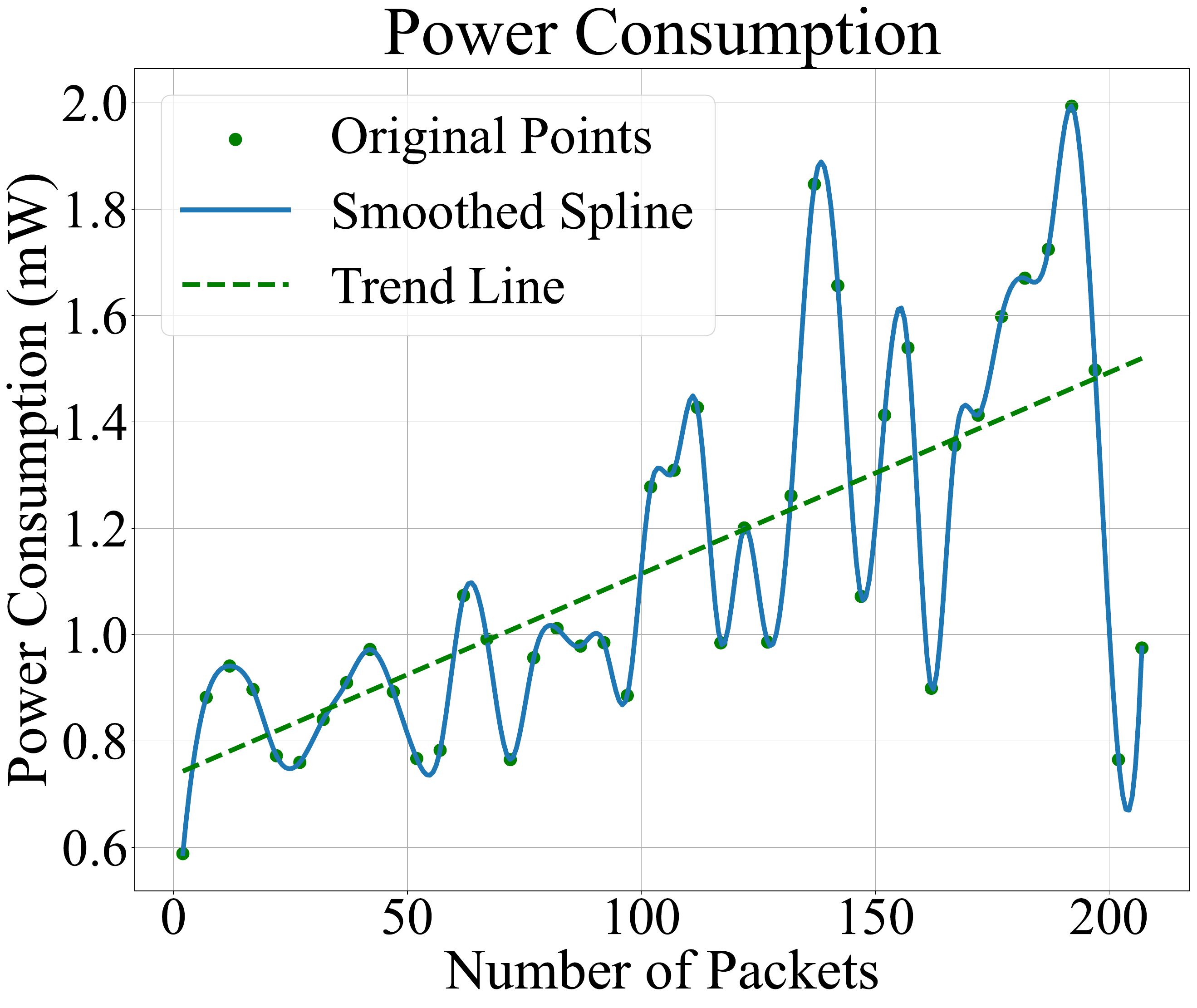}
        \caption{Power consumption}
        \label{fig:power-consumption}
     \end{subfigure}
     \begin{subfigure}[b]{0.23\textwidth}
        \centering
        \includegraphics[width=1.0\linewidth]{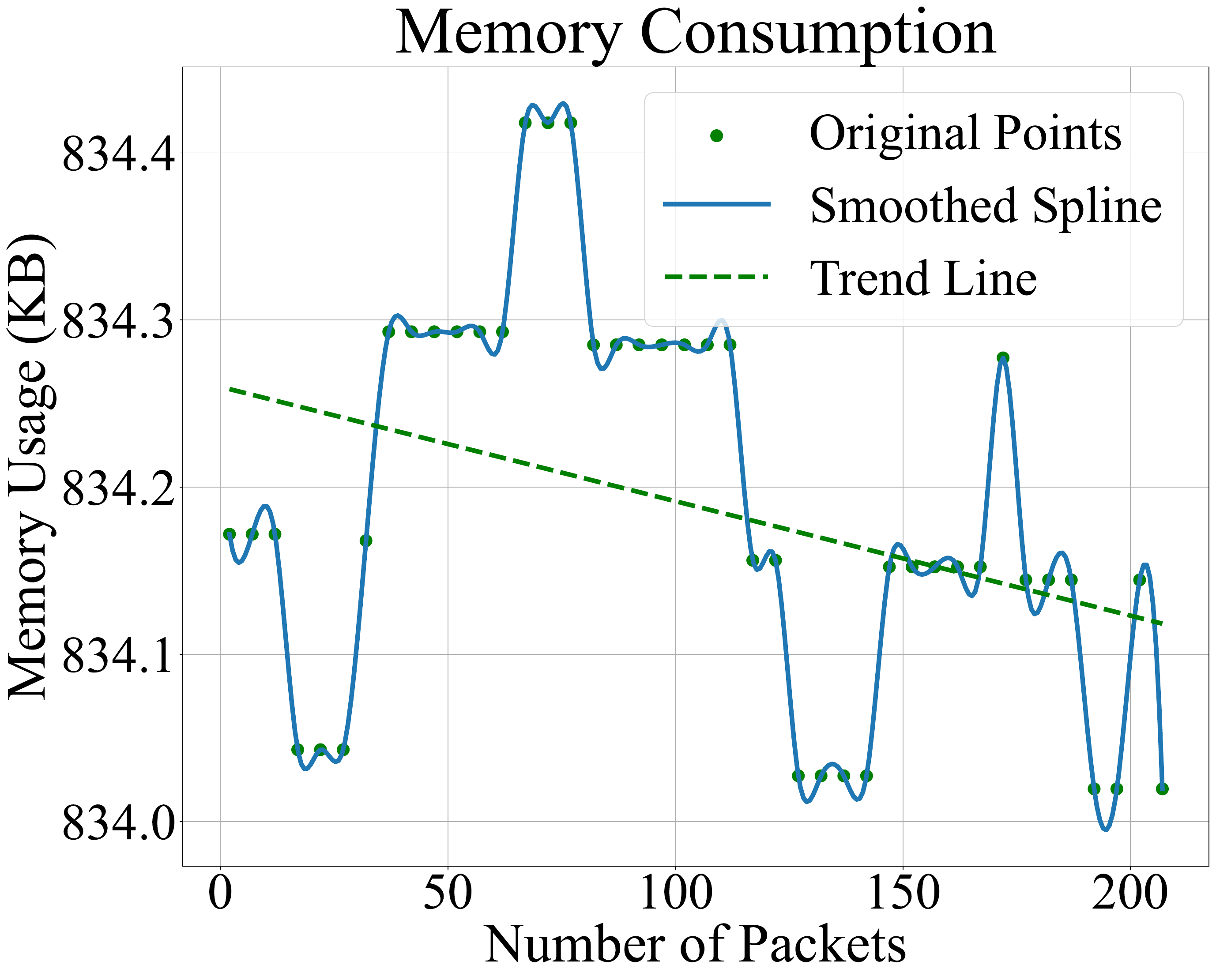}
        \caption{Memory consumption}
        \label{fig:memory-consumption}
     \end{subfigure}
     \begin{subfigure}[b]{0.23\textwidth}
        \centering
        \includegraphics[width=1.0\linewidth]{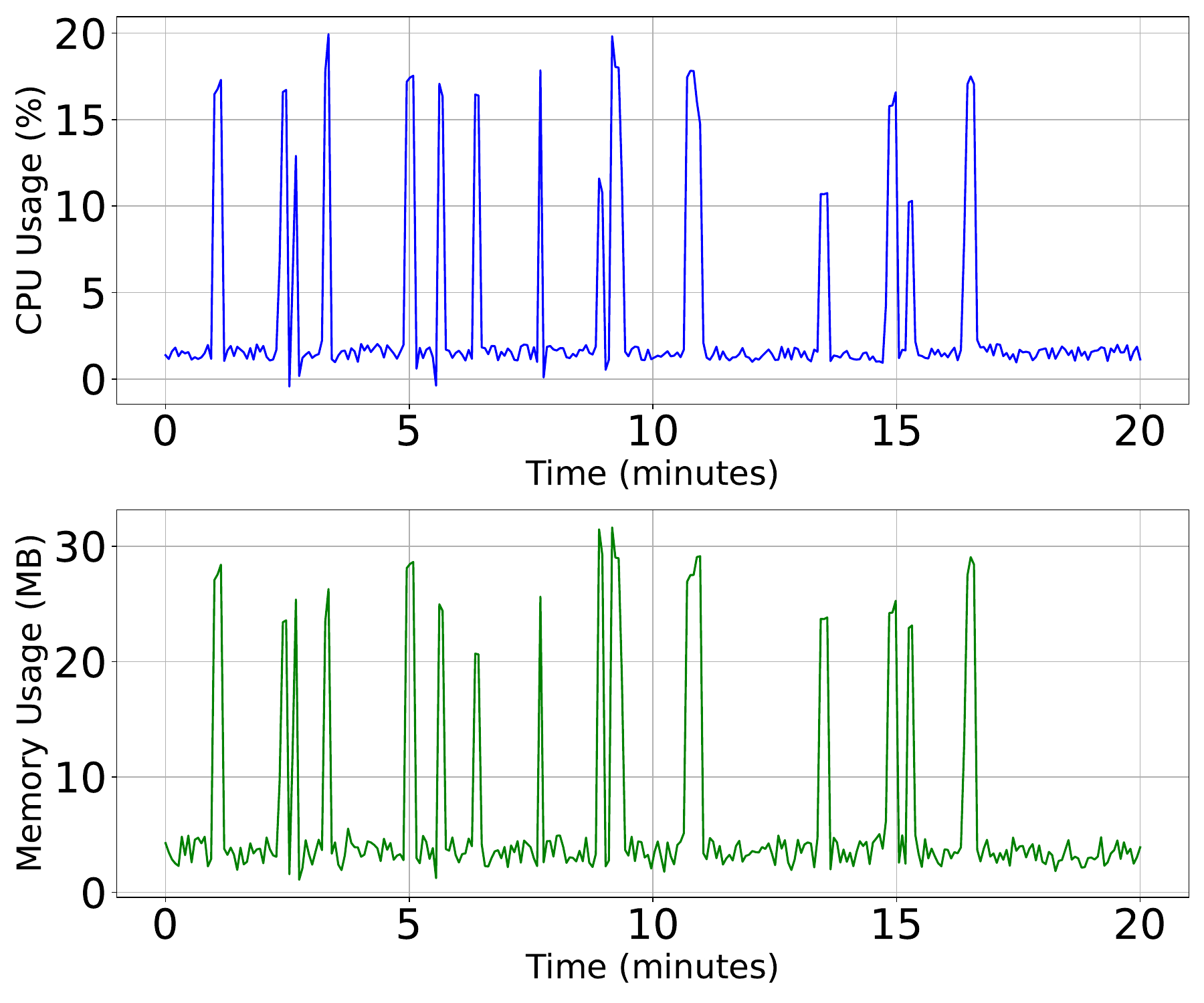}
            \caption{Load testing of app}
        \label{fig:cpu-usage}
     \end{subfigure}
    \caption{Performance overhead of \system}
    \vspace{-0.4cm}
    \label{fig:perf-overhead}
    \vspace{-0.2cm}
\end{figure*}

%When deployed in real devices, \system adds only a small overhead to the system. 
\noindent \textbf{Overhead Analysis of ML Models.} 
We evaluate the overhead of the ML models used in \system based on several criteria.
Figure~\ref{fig:time-consumption} shows the time required to predict packets, which increases linearly with the number of packets. The slope of the increase is minimal, which means that the solution can scale to high throughput applications. Figure~\ref{fig:memory-consumption} shows the memory consumption and Figure~\ref{fig:power-consumption} shows the power consumption of \system, consumed in packet processing and running the ML model on the processed packets to generate inferences. The trend of power consumption also linearly increases, with a small slope and the trend of memory consumption decreases, which can result from multiple hardware-level optimizations by the operating system. Compared to a recent approach~\cite{PHOENIX}, which uses an average of 4 mW power, \system uses less than 2 mW of power to detect a FBS. These results show that \system is a promising solution for deployment in real-world systems, having negligible overhead. The load testing (CPU and memory usage under different loads) for the mobile app is shown in Figure~\ref{fig:cpu-usage} and discussed in detail in Appendix Section~\ref{appndx:apploadtesting}.
% \imtiaz{Need comparison. Bring for Phoenix, there's is around 4mW ours is less than 2}
% \imtiaz{Fix.}

\subsection{RQ3. Validation and Comparison}\label{subsec:rq3}

\begin{table}[]
    \centering
    \renewcommand{\arraystretch}{1}
    \fontsize{7}{7}\selectfont
    \setlength{\tabcolsep}{1pt} 
    \begin{tabular}{l|C{1cm}|C{1.2cm}|C{1.2cm}|C{1.2cm}||C{0.8cm}}
    \hline
         Solution & Supports 4G & No Change in Protocol & No Additional Hardware Required & Source Available & Detects FBS\\
         \hline
         Crocodile Hunter\cite{Cooper2021} & \ding{51} & \ding{51} & \ding{55} & \ding{55} & -\\
         Darshak~\cite{Darshak} & \ding{55} & - & - & - & -\\
         Baron~\cite{BARON} & \ding{51} & \ding{55} & - & - & -\\
         Phoenix~\cite{PHOENIX} & \ding{51} & \ding{51} & \ding{51} & \ding{55} & -\\
         Android IMSI Catcher~\cite{AIMSICD} & \ding{51} & \ding{51} & \ding{51} & \ding{51} & \ding{55}\\
         SnoopSnitch~\cite{SnoopSnitch} & \ding{51} & \ding{51} & \ding{51} & \ding{51} & \ding{55}\\
         \system & \ding{51} & \ding{51} & \ding{51} & \ding{51} & \ding{51}\\
         \hline
    \end{tabular}
    \caption{Comparison of \system with existing solutions.}
     \vspace{-0.5cm}
    \label{tab:comp-prev-works}
    \vspace{-0.2cm}
\end{table}

% \imtiaz{I revised.}
\noindent \textbf{Real-world App validation.} 
To validate \system app's performance against threats in the wild in real environments, we perform tests in our controlled lab environment. This is because it is illegal to deploy FBS in public places. 

\noindent \emph{Lab environment:} For the controlled lab environment, we create a testbed using (1) two USRP B210~\cite{USRP-B210}, (2) two engineering laptops and (3) a smartphone with a Google Fi sim card with the \system app installed. We used Open5GS~\cite{Open5GS} for the core network and srsRAN~\cite{srsRAN} and OAI~\cite{OpenAirInterface} for the BS. Following the standard approaches, we create and spawn an FBS using the laptop as the core network and the USRP B210 SDR as the BS. To test the FBS and MSA detection in different setups, we create the following scenarios: (1) lab 4G network with our own SIM card as legitimate; (2) commercial network with a Google Fi SIM card; (3) varying distance between FBS and device; (4) limited mobility in the lab. The real-world setup is shown in Figure~\ref{fig:real-world-test}. For a thorough evaluation, we run the app for 24 hours with the Google Fi SIM card, create an FBS and run all 21 MSAs each for 5 times. The results are shown in Table~\ref{tab:real-world-msa-attacks-evaluation}. For general FBS, all five times are detected. For the MSA, in a total of 105 attacks, 88 are detected as True Positive, 5 are misclassified as benign and 17 MSAs are classified into other attacks. For the whole 24-hour period, we got 6 false positives, where no attack was conducted, but the app still showed a notification.

\noindent \emph{\underline{Long Term Evaluation.}} 
% \imtiaz{Revise.}
For long-term evaluation with the \system app we ran the app for total of 7 days with different use cases such as web browsing, video streaming, calling, idle time, maps and navigation. Within this time, we ran the app in different areas with varying population densities, such as metropolitan cities and high-population events. Lastly, we ran the app in 2 different countries with local 4G connectivity providers. 
For stress testing, we conducted all $21$ attacks $5$ times within $24$ hours. For the extensive longer tests, we just ran the app for $7$ days with commercial SIM cards. From the packets that we saved in that period, for the stress test, it was $105,561$ (NAS and RRC combined) packets within $24$ hours, whereas for the longer tests, it was $326,385$ (NAS and RRC combined) packets in $7$ days.
On the whole we found 2 alerts during the longer tests. Since we do not have any ground truth data we can not certainly discuss false positives and false negatives. However, even if we consider the 2 alerts as false-positives, compared to previous stress testing, the performance is significantly better. Therefore, we can argue in real-world usage \system would perform better in terms of False Positives.

\looseness = -1
\noindent \textbf{Existing solutions comparison.} We compare \system with different real-world solutions. The criteria for the comparison are: (1) The solution must not require any changes in the protocol. (2) The source of the solution must be available and maintained. (3) No additional hardware is required for operation. The comparison summary in Table~\ref{tab:comp-prev-works} shows that only Android IMSI Catcher (AIMSICD)~\cite{AIMSICD} and SnoopSnitch~\cite{SnoopSnitch} satisfy all comparison criteria.
% \imtiaz{Why is Table 3, 2 pages up?}
Therefore, we test them in our testbed by creating an FBS in our controlled lab environment and spawning it near a mobile device with AIMSICD and SnoopSnitch installed and running. We ran the experiment several times, but neither of the solutions could detect the FBS, whereas \system detected the FBS every time. Note that, the comparison with AIMSICD and SnoopSnitch was conducted by downloading and testing them in same controlled setup used for testing our app. These apps are closest compared to our app, are developed by open-source communities, are not well maintained, and we do not have many details about their inner workings. We have contacted the developers but, at the time of the write-up, have not heard back.

\noindent \emph{\underline{Comparison with Phoenix.}} For a more comprehensive comparison of MSA detection with existing solutions, we find Phoenix~\cite{PHOENIX} the most appropriate according to Table~\ref{tab:comp-prev-works}. To compare \system with Phoenix, we first implement a simple implementation of Phoenix in Python as the implementation is not publicly available. Phoenix uses three different signature representations, (1) Deterministic Finite Automata (DFA); (2) Mealy machine (MM)~\cite{mealymachine}; (3) propositional, past linear temporal (PLTL)~\cite{PLTL}. We create and run Python scripts for all three signature representations on our attack traces (level 0-4) that Phoenix can detect. From the results shown in Table~\ref{tab:comp-with-phoenix}, we can see that \system performs significantly better than Phoenix for all the attacks.

\noindent \emph{\underline{Why heuristic/signature-based approaches fail.}} For further evaluating signature-based detection approaches with \system we evaluate with traces from an adaptive adversary.
The adaptive adversary reshapes attack to evade detection and active employs the two techniques discussed in Section~\ref{subsec:adaptive}) level 4 attacker ability. We conduct this evaluation with Phoenix~\cite{PHOENIX} as well. We chose Phoenix for several reasons: (i) similar to \system it also deploys a device-centric attack detection mechanism; (ii) it deploys sophisticated signature based schemes for MSA detection. From Table~\ref{tab:cross_validation_attacks}, we see that signature-based detection techniques used in Phoenix~\cite{PHOENIX} with PLTL (the best performing signature) struggle to detect reshaped attacks like Attach Reject, IMSI Catching, and Service Reject due to their reliance on rigid, predefined rules and patterns. Attackers can reshape attacks by subtly modifying the attack behavior to avoid violating these rules, leading to misclassification or missed detections.
For instance, for Attach Reject, Phoenix's signature is to detect an attack as soon as it receives an \M{AttachReject} message. Therefore, sending an out-of-sequence \M{AttachReject} with a different cause field misclassifies all the attacks to this category. Similar results are for IMSI catching, Service Reject, and TAU Reject, where the signature detects attacks based on only the specific message type. Similarly, signatures where temporal orderings are included are broken by changing the sequence. For instance, Phoenix detects a Numb Attack when \M{AuthenticationReject} message is received without the previous NAS message being \M{AuthenticationResponse}. This is broken by sending an \M{AuthenticationRequest} message before sending the reject message.
In contrast, our machine learning-based approach excels by learning complex patterns and relationships in data, enabling it to generalize across diverse attack variations. %By leveraging high-dimensional features, contextual protocol data, and adversarial training, ML models can accurately identify anomalies and correctly classify reshaped attacks, even when they mimic legitimate behaviors. This adaptability makes ML-based detection significantly more effective in identifying dynamic, evolving threats compared to static heuristic methods.

\begin{figure}
    \centering
    \includegraphics[width=\linewidth]{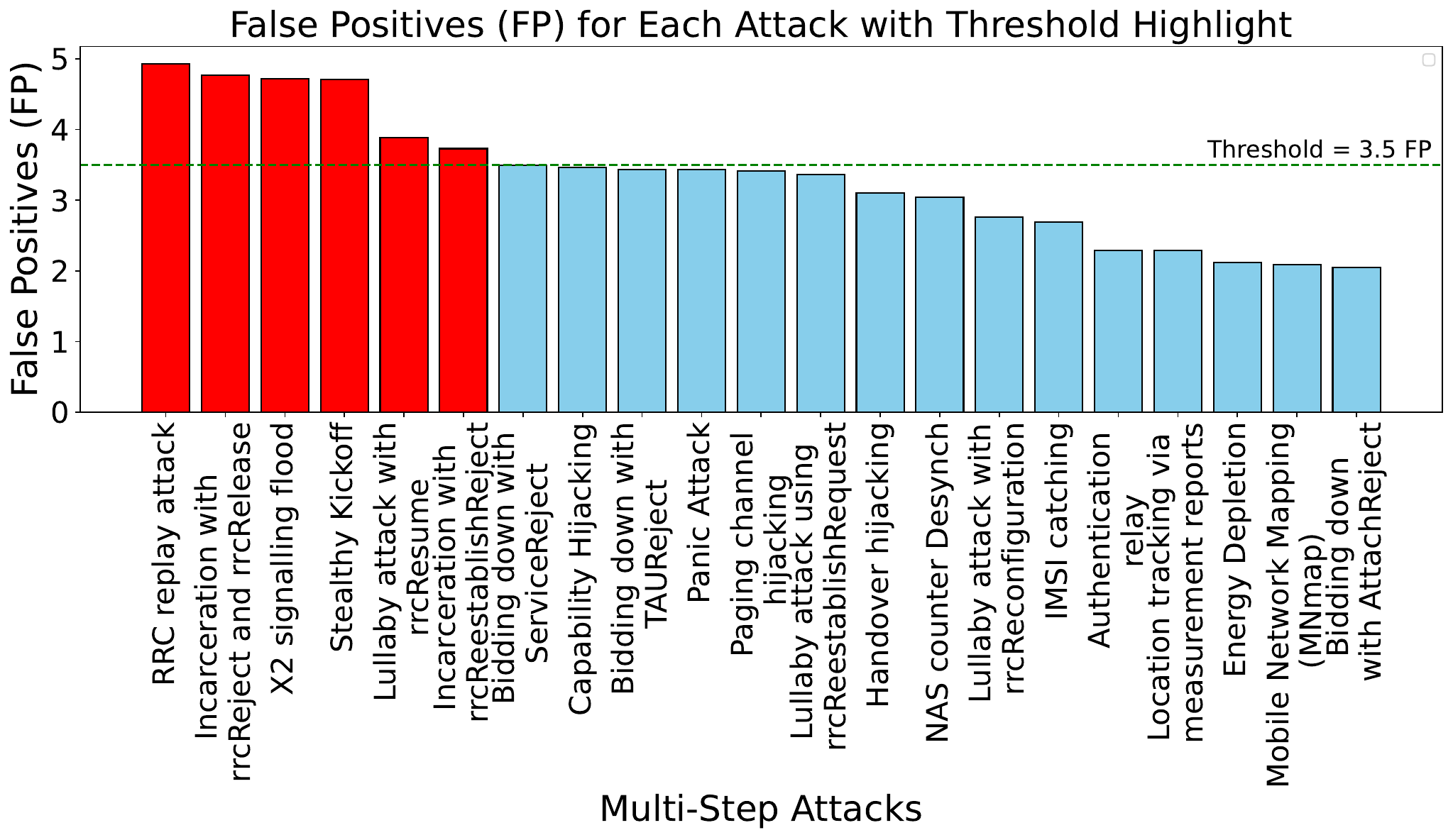}
    \caption{MSA detection FP breakdown}
    \vspace{-0.5cm}
    \label{fig:msa-det-perf-breakdown}
    \vspace{-0.2cm}
\end{figure}

\noindent \textbf{Error Analysis.}
% \imtiaz{Discuss FN as well.}
For FBS detection, level 0 and level 1 attacks both have low  False Positive (FP) and False Negative (FN) rates (shown in Table~\ref{tab:attacker-level-fpr-breakdown}). However, the FP and FN increase in levels 2-4, where an attacker clones all the parameters of the legitimate BS, and it becomes harder to detect the FBS. This is mostly due to the lack of features when an attacker clones all the parameters, and the behavior completely resembles a legitimate BS.
For MSA, the analysis in Table~\ref{tab:msa-fpr-breakdown} highlights that the RRC replay attack, Incarceration with \M{rrcReject} and \M{rrcRelease}, X2 signaling flood, and Stealthy Kickoff Attack are the top contributors to FP, accounting for $4.93\%$, $4.77\%$, $4.72\%$, and $4.71\%$ of the total FP, respectively. Approximately $20\%$ of the attacks are responsible for majority of the FP, which means that the overall false positives of the system will be lower on average (see Figure~\ref{fig:msa-det-perf-breakdown}). 
For all these attacks, from the UE perspective, the attack behavior is precisely the same as the benign behavior of the network. For instance, for the Stealth Kick-off Attack, the attacker clones the paging message and includes the UEs IMSI. From the Layer-3 packets and features, it is difficult for the ML model to detect it and cause FP.
The same goes for FN behavior, where the attack behavior closely resembles legitimate network activities, making detecting challenging due to overlapping features or insufficient distinction between benign and malicious patterns.
The attacks contributing most to FN in the system include Paging Channel Hijacking and Lullaby Attack using \M{rrcReestablishRequest}, among others.

%The highest FP and FN attacks are on the RRC layer MSAs, and the high FP rates may stem from the (i) high number of RRC packets and (ii) low complexity of these attacks, as their behavior closely resembles legitimate network activities, making detection challenging. This could be due to overlapping features or insufficient distinction between benign and malicious patterns.

%These high FNs suggest that, similarly to the false positives, detection model struggles with certain attack patterns due to insufficient features, stealthy attack designs, and overlapping behavior with legitimate network traffic. 

\noindent \textbf{Zero shot detection of Overshadow attack.}
\label{subsec:overshadow}
% \imtiaz{How overshadow and alter dataset was generated?}
%We generate Overshadowing~\cite{yang2019hiding} and aLTEr~\cite{ALTER} attack data in our controlled lab environment as per our discussion in Section~\ref{subsec:dep-scope} as generating these sophisticated attacks that require precise control over the location of the network components is not feasible in POWDER. 
To generate overshadowing attack data in our controlled lab environment, we start by configuring the legitimate eNodeB with specific LTE parameters and connecting a UE to establish a baseline connection. A second SDR is then set up as a malicious transmitter. This transmitter synchronizes with the legitimate LTE network to ensure proper timing and frequency alignment. The malicious transmitter generates high-power LTE signals with the same Cell ID to overshadow legitimate signals. The success of the attack is validated by confirming that the UE receives the malicious signal instead of the legitimate one.
%
%The aLTEr attack requires a different setup involving a malicious relay positioned between the UE and eNodeB by impersonating a UE towards to the network and the eNodeB towards the user. 
%traffic. %During the attack user-plane packets are manipulated in real-time. For example, DNS responses are altered to redirect traffic or disrupt applications. Data is logged and inspected for modifications at the application level, ensuring the attack's success is evident from the captured logs. 
Table~\ref{tab:pkt-level-vs-trace-level-and-zero-shot-det} shows the zero shot detection capability of \system for Overshadowing attack. The attack is detected with 86\% accuracy.

\subsection{RQ4. Unseen and Reshaped Evaluation}
% \elisa{It would have been good to do some experiments about this aspect and report the results in section 7. I guess that now we do not have time.}
To establish a benchmark and prove the robustness and generalizability of \system, 
we evaluate its capability to detect unseen and reshaped attacks.

\noindent \textbf{Unseen attacks.}
\system can detect unseen attacks by leveraging anomalies in behavior that deviate from benign patterns. In cases where it encounters an attack it has not seen before, there will be a misclassification into an existing class. However, the attack would not go undetected, as the deviation from established benign behavior will still trigger an alert, ensuring that all anomalies are identified and addressed.
% , even if the specific attack type is not previously encountered. 
%This approach also works if the attackers reshape the packets or change the attack signature, as this deviates their behavior from the benign behavior, making the system robust and generalizable against different types of FBS and MSA, especially those with more sophisticated capabilities. 
We validate this using k-fold cross-validation. We keep one attack aside while training and then test on it. 
% by keeping the stealthy kickoff attack data unseen during training. 
Table~\ref{tab:cross-val-unseen-attack} shows that all the unseen attacks are classified as another type of attack, proving that attacks will not go undetected. This shows the capability and robustness of \system to detect unseen attacks and to generalize.

\noindent \textbf{Reshaped attacks.}
% \imtiaz{Revised.}
In case of attack reshaping, especially for attackers with more sophisticated capabilities, even if the attacker is aware of the presence of \system and reshapes the attack pattern completely to evade detection, \system can still detect it. This reshaped behavior deviates from benign behavior and overlaps with the original attack that was reshaped, and \system can detect it from this deviation and overlap. To test \system's capability to detect reshaped attacks properly, we create additional reshaped data following level 4 of attacker capability (section~\ref{subsec:attacker_ability}) and evaluate \system's performance.
% We test this by
%we reshape each attack by (i) changing different fields in different packets, (ii) changing the packet intervals, and (iii) sending additional out-of-sequence packets to change the pattern.
% . We achieve this 
%changing different fields in different packets sent during the attack and changing the sending intervals. We also send some additional and out-of-sequence packets to reshape the attack further. 
%For instance, we reshape the paging channel hijacking attack in Figure~\ref{fig:paging_channel_hijacking_attack} by including additional \M{IdentityRequest} messages before sending the \M{Paging(empty)} message.
Table~\ref{tab:reshaped-attack} shows that all of the reshaped attack packets are classified as the original attack. %This proves the capability of \system to detect reshaped attacks.

\section{Related Work}
Several approaches have been proposed to address the challenge of detecting FBSes. %We now present an overview of those approaches and highlight their key contributions and limitations.
Recent efforts have introduced certificate-based solutions and digital signatures~\cite{BARON, Insecure-Connection-Root-of-All-Evil, Look-Before-You-Leap, Singla2020ProtectingT4, Trustin5G, FBSSpecifications} for BS authentication. %However, these solutions are not scalable as they are network-based and need to be added to existing network monitoring infrastructures. 
However, these techniques require modifications to specifications, require huge infrastructure changes, add overhead, and are not able to defend the billions of devices currently in the market. This makes \system highly suitable for defending a wide variety of attacks. The other approaches are based on several simple heuristics and signatures~\cite{5G-Spector:NDSS24, Darshak, AIMSICD, SnoopSnitch, PHOENIX}. As shown by the extensive experiments in this paper, heuristic based approaches are not well suited to defend against an adaptive adversary. 
The approaches proposed in~\cite{OVERWATCH, SeaGlass} require installing expensive hardware and in most cases they are proprietary. The techniques in ~\cite{Li2017FBSRadarUF} depends on crowd-sourced data, which is not a practical solution for scaling up.
ML based efforts are effective in detecting attacks and anomalies from traces~\cite{NakarmiML2022}. 
% \imtiaz{Add a sentence to show why they are still not good. For instance, they train models with only simulated data.}
However, they work on small datasets generated in simulated environments that lack diversity; also because they work only in the data plane, they cannot detect MSAs.
Recently, researchers have used simulation models to analyze fake base station attacks on a large scale~\cite{Rupprecht2024}, but they fail to capture different real-world scenarios.
Previous works have shown that information related to the connection between a UE and a BS can be used to reason about the authenticity of the BS~\cite{Arslan2019, Hamad2019ICCAIS, Dabrowski2016TheMS, CatchMeIfYouCan, Huang2018IdentifyingTF, Murat, Karaçay2021}. 
\section{Discussion}
\label{section:discussion}

\noindent \textbf{Applicability to 5G.}
To the best of our knowledge, no open-source protocol stack for the standalone 5G core network supports handover, which is a prerequisite for creating a real-world FBS and MSA dataset. Therefore, we leave the detection of FBSes and MSAs in 5G cellular networks with \system as a future work. However, we believe that,  based on 4G, our approach is equally applicable to 5G because most of the layer 3 procedures are unchanged from 4G. Thus, the ML model designs will remain the same when porting \system to 5G.

\noindent \textbf{Deployment.}
If  \system were deployed in a real-world setting, we envision the model will be periodically retrained to incorporate new attacks, with updates pushed promptly to user devices via app updates. Since we collect NAS and RRC traces, which can contain sensitive information, data collection is enabled via user consent, ensuring transparency and privacy compliance. The \system app is built on top of MobileInsight~\cite{Mobileinsight}. Therefore, the requirements for running MobileInsight apply to our solution as well, which includes rooting the phone for most smartphone models (for more details, we refer to the MobileInsight website). Apart from this deployment scenario recently, we have been in discussion with a commercial connectivity vendor about applying \system on top of the baseband directly, without requiring the phone to be rooted.

\noindent \textbf{Defense against the detected FBSes.}
For defense in our mobile app, the user is notified immediately for an FBS upon detection. Additionally the user has the capability to switch to another cell, and add the current cell to a temporary block list. For instance, if the \system app detects an IMSI-catcher after receiving an \M{IdentityRequest} in the sequence, it turns the radio off to stop leaking sensitive information (with permission from the user). 
%
%
%we can implement the option for users to disconnect from the FBS immediately upon detection. Additionally, 
%Elisa: To what  does "They" refer to? To the users? Also once the malicious FBS is blacklisted, how does our app prevents the UE to reconnect to the malicious FBS? Also can a FBS assume the identify of a legitimate one (perhaps a legitimate one in a different tracking area); then the phone will not connect to the legitimate one when moving to the different trackin area.
%the users can have the capability to blacklist the identified FBS for a certain period, preventing future unauthorized connections. Moreover, our app can facilitate a seamless transition to a new, legitimate base station upon user request, ensuring uninterrupted connectivity and safety. 
Another advanced solution can be to design and integrate a learning-based (e.g., reinforcement learning) decision-making agent directly into the baseband that can not only detect but also recover from the attack in an automated way. %We leave this open research problem to design a defense against the attacks after detection as future work. 

\noindent \textbf{UE vs network side defense.} 
Defense against FBS can be deployed at the UE or network sides. These network side solutions are designed for traditional cellular architecture~\cite{Murat, NakarmiML2022} and for the emerging O-RAN architecture~\cite{Trustin5G,5G-Spector:NDSS24}. %However, with O-RAN gaining popularity, a network-level detection approach might become more feasible. 
There are two critical limitations with network-based deployments: (i) 
a network-level solution might be able to detect that a cell is affected by an FBS but would be unable to take any necessary action to protect user privacy and prevent attacks. For instance, \system prevents sensitive information leakage upon detecting attacks by turning off the radio; (ii) certain MSAs necessarily cannot be observed by the network operators, which is observable only from the device vantage point. For instance, after an FBS has been connected to the device, it is not possible for the network operator to uncover the type of attack. Because of these reasons \system opts for an UE-centric solution. However, device-centric solutions also have limitations, such as requiring root access or sensitive permissions, especially on Apple devices. In the future, this can be resolved by deploying the UE-centric solutions on top of the baseband directly without requiring the phone to be rooted. Overall, we conclude that both network and UE side defense and detection mechanisms would ultimately be needed to defend against FBS attacks and create a robust ecosystem to prevent attacks altogether.

%\noindent \textbf{Further real-world validations.} Since it is not legal to create FBS in public places, for validating \system performance in detecting FBS and MSA's, we perform experiments in a lab environment. However, in the future, we plan to do a large-scale measurement study with the Android app, especially in crowded and important places, to detect the presence of FBS and MSAs. We leave it as a future work.

%However, further experiments could be carried out done using different commercial sim cards and more complex movements and scenarios. We consider this as a future work.

%\imtiaz{Large usability study in future in different areas and countries.}

\looseness = -1
\noindent \textbf{Implication of FP and FN.}
%\imtiaz{Revise.}
\system detects FBSes with 96\% accuracy. %Approximately 20\% are of the attacks responsible for the majority of the FP and FN. This means the overall FP of the system will be lower in average. 
However, FBS detection is a hard problem not because of the difficulty of detecting attacks but because it is hard to prevent false positives, given that attacks are rare. Therefore, a system with high FP might not be suitable for general use.
\system has a 2.96\% FP rate, and in the \system app stress testing, we found 6 FP instances out of 110 instances; in longer tests, we found 2 alerts. The ambition of \system is to bring FBS detection to the masses; however, currently, it is more suitable for safety-critical UEs and communication for people with sensitive information, where security is prioritized. In the future, we plan to improve the FP rate further by incorporating lower-layer features and pushing direct app updates.
\section{Conclusion And Future Work}
In this paper, we present \system, an ML-based FBS and MSA detection system for cellular networks, 
% \elisa{I have made some minor changes here.}
which leverages network traces at Layer 3. To train the ML models, we have created \fbsdataset and \msadataset, the \emph{first-ever} large-scale real-world datasets.
% of \totalsizeofdata GB.
% combined.
% , containing \totalnumberofpackets packets. %Our novel models, designed specially to detect FBSes and MSAs, perform better than any state-of-the-art model and detect FBSes with a high accuracy of \fbsdetectionaccuracy and a low false positive rate of \fbsdetectionfpr, and MSAs with \msadetectionaccuracy accuracy and \msadetectionfpr false positive rate. 
We deploy \system on a mobile app that effectively detects FBSes and MSAs in all the tested real-world scenarios.

\noindent \textbf{Future work.}
In the future, we will port \system to 5G, support overshadow attacks and focus on detecting FBSes within the emerging Open Radio Access Network (ORAN) environment, employing advanced machine learning algorithms through xApps. We will also investigate different defense mechanisms to effectively stop an attack once detected by \system.
%\newpage

\section{Acknowledgements}
This work is supported by NSF Grant No. 2112471. We thank the POWDER team for their support in utilizing the platform and Mir Imtiaz Mostafiz for his feedback on the paper.

%-------------------------------------------------------------------------------
\section*{Ethics considerations}
%-------------------------------------------------------------------------------
All the experiments in this paper followed the ethics consideration policies. The experiments 
were done in an isolated lab setup where all the victim UE's belong to us. Furthermore, we use a shielding box to prevent our USRPs from interfering with the commercial networks' licensed spectrum, following ethical guidelines. We also use logical precautions to make sure only our UE IMSIs are connected to the attack setup and we do not affect any other UEs. These ethical steps and precautions are in line with all the previous research on cellular network security.

%-------------------------------------------------------------------------------
\section*{Compliance with the open science policy}
%-------------------------------------------------------------------------------
% In compliance with the open science policy, we will release the datasets, the models and the Android App for general users to use and foster further research. Furthermore, we will participate in the artifact evaluation and will aim for an artifact-reproduced badge.

In compliance with the open science policy, we release the datasets, the models and the Android App for general users to use and foster further research\footnote{Datasets, codebase and models for \system are publicly available at \href{https://zenodo.org/records/14720824}{https://zenodo.org/records/14720824}}.

% Conference papers do not normally have an appendix

%\section*{Acknowledgment}


\bibliographystyle{unsrt}
\bibliography{ref}


% \end{thebibliography}

\appendix
\newpage
\section{Algorithms for FBS and MSA Detection}
% \subsection{FBS Detection Algorithm}
The FBS detection algorithm is described in Algorithm~\ref{alg:fbs-det-algo}, and the MSA recognition algorithm is described in Algorithm~\ref{alg:msa-det-algo}.

\begin{algorithm}
\caption{Stateful LSTM w/ Attention}
\label{alg:fbs-det-algo}
\begin{algorithmic}[1]
\State \textbf{Input:} Labeled dataset: \fbsdataset, hyperparameter: $len_{seq}$
\State \textbf{Output:} Classified traces indicating FBS activity
\Procedure{StatefulLSTM}{$x_t$}
\State Initialize LSTM parameters $\theta_s$
\State Set $\text{stateful} = \text{true}$, $\text{return\_sequences} = \text{true}$
\For{each timestep $t$}
    \State $h_t, c_t \gets \text{LSTM}(x_t, h_{t-1}, c_{t-1}; \theta_s)$
\EndFor
\State \textbf{return} $h_t$
\EndProcedure

\Procedure{LSTMwithAttention}{$x_t$}
\State Initialize LSTM parameters $\theta_a$
\State Set $\text{return\_sequences} = \text{true}$
\State $H \gets \text{LSTM}(x_t; \theta_a)$
\State $c_t \gets \text{context vector from attention mechanism over } H$
\State $h'_t \gets \text{tanh}(W_c[c_t; H_t] + b_c)$
\State \textbf{return} $h'_t$
\EndProcedure

\State $x_t \gets \text{input(sequence\_length = $len_{seq}$)}$
\State $h_t^{\text{stateful}} \leftarrow \Call{StatefulLSTM}{x_t}$
\State $h_t^{\text{attention}} \leftarrow \Call{LSTMwithAttention}{x_t}$
\State $y_t = \text{Dense}(\text{concat}(h_t^{\text{stateful}}, h_t^{\text{attention}}))$
\State Train model on loss $\mathcal{L}(y, \hat{y})$, propagate back
\end{algorithmic}
\end{algorithm}

% \subsection{MSA Detection Algorithm}
\begin{algorithm}
\caption{MSA Recognition Model}
\label{alg:msa-det-algo}
\begin{algorithmic}[1]
\State \textbf{Input:}
Labeled dataset: \msadataset
\State \textbf{Output:} Graph Model (GM)
%\State \textbf{Variables:} Seq-LSTM, GM

\Procedure{Graph Learning}{}
\State \textbf{Variables:} Graph $G(V,E)$
% \For{each unique packet $p$ in $D_{MSA}$}
%     \State Create a node $N_p$
% \EndFor
\State Create start node $V_1$ with the first packet $p_1$
\For{each subsequent packet $p_i$ in \msadataset}
    \If{$V_p$ not in $G$}
        \State Create a node $V_p$
    \EndIf
    \State Add an incoming edge $E_p$ from $V_{p-1}$ to $V_p$
    \State Label $E_p$ with $L_p$, the Label for Packet $p$
\EndFor
\State $GM = \text{train}(G)$
\EndProcedure
\State \textbf{return} GM
\end{algorithmic}
\end{algorithm}

\section{Experimental Setup}\label{apndx-sec:experimental-setup}
\subsection{Model Hyperparameters.}
\par ~ \par
\noindent \textbf{Stateful LSTM w/ attention.}
The Stateful LSTM w/ attention consists of two parallel \textit{LSTM} layers with $64$ units and a \textit{sigmoid} activation function configured to return sequences. One additional attention layer is added to the LSTM w/ attention and the \textit{stateful} hyperparameter is set to true for the stateful LSTM. The subsequent layer after these parallel layers are concatenated is a dense layer with a single unit and \textit{sigmoid} activation. For optimization, the \textit{stochastic gradient descent (sgd)} was chosen, with a \textit{mean squared error (mse)} loss function. The model's performance is assessed using \textit{accuracy} as the main metric, complemented by the inclusion of a custom metric, \textit{false positives}, to evaluate its classification capabilities further.

\noindent \textbf{GraphSAGE.}
The graph model features a single \textit{SAGEConv} layer. The \textit{SAGEConv} layer utilizes $2$ attention heads to capture graph-based relationships. The model's architecture is encapsulated within a \textit{PyTorch} Module, with the forward function defining the flow of information through the single layer. Logarithmic \textit{softmax} is employed for the final layer's output activation.

\noindent \textbf{Other graph models.}
We used the same configurations as the \textit{GraphSAGE} model for the other graph models with their own convolutional layer. For example, for the Graph Attention Network, we used \textit{GATConv}, and for the Graph Convolutional Network, we used \textit{GCNConv}.

\noindent \textbf{Other classification models.}
The Random Forest Classifier and the Decision Tree Classifier were configured with a Gini criterion, a maximum depth of 3. For the XGBoost Classifier, Support Vector Classifier (SVC), K-Nearest-Neighbors (KNN) Classifier, Gaussian Naive Bayes and Logistic Regression we adopted default configurations.

\noindent \textbf{Convolutional Neural Network (CNN).}
The CNN architecture comprises two \textit{Conv1D} layers with $32$ and $64$ filters, respectively, followed by a \textit{ReLU} activation function. \textit{MaxPooling1D} layers with a pool size of $2$ were inserted after each convolutional layer to downsample the spatial dimensions. A \textit{GlobalAveragePooling1D} layer was then employed to aggregate the spatial information across the entire sequence. Subsequently, two Dense layers were added with $64$ units and \textit{ReLU} activation in the first, and a single unit with a \textit{sigmoid} activation in the final layer for classification. The model was compiled using the Adam optimizer, binary cross-entropy loss function, and accuracy as the metric for performance evaluation.

\noindent \textbf{Feedforward Neural Network (FNN).}
The FNN architecture consists of three Dense layers, with the first two layers containing $64$ units each and utilizing the \textit{ReLU} activation function. 
The final dense layer, with a single unit and a \textit{sigmoid} activation function, is employed for classification. The model was compiled using the \textit{Adam} optimizer, \textit{binary cross-entropy} as the loss function, and \textit{accuracy} as the metric for assessing its performance.

\subsection{Impact of sequence length}
\begin{figure*}[t]
     \centering
     \begin{subfigure}[b]{0.23\textwidth}
         \centering
         \includegraphics[width=\linewidth]{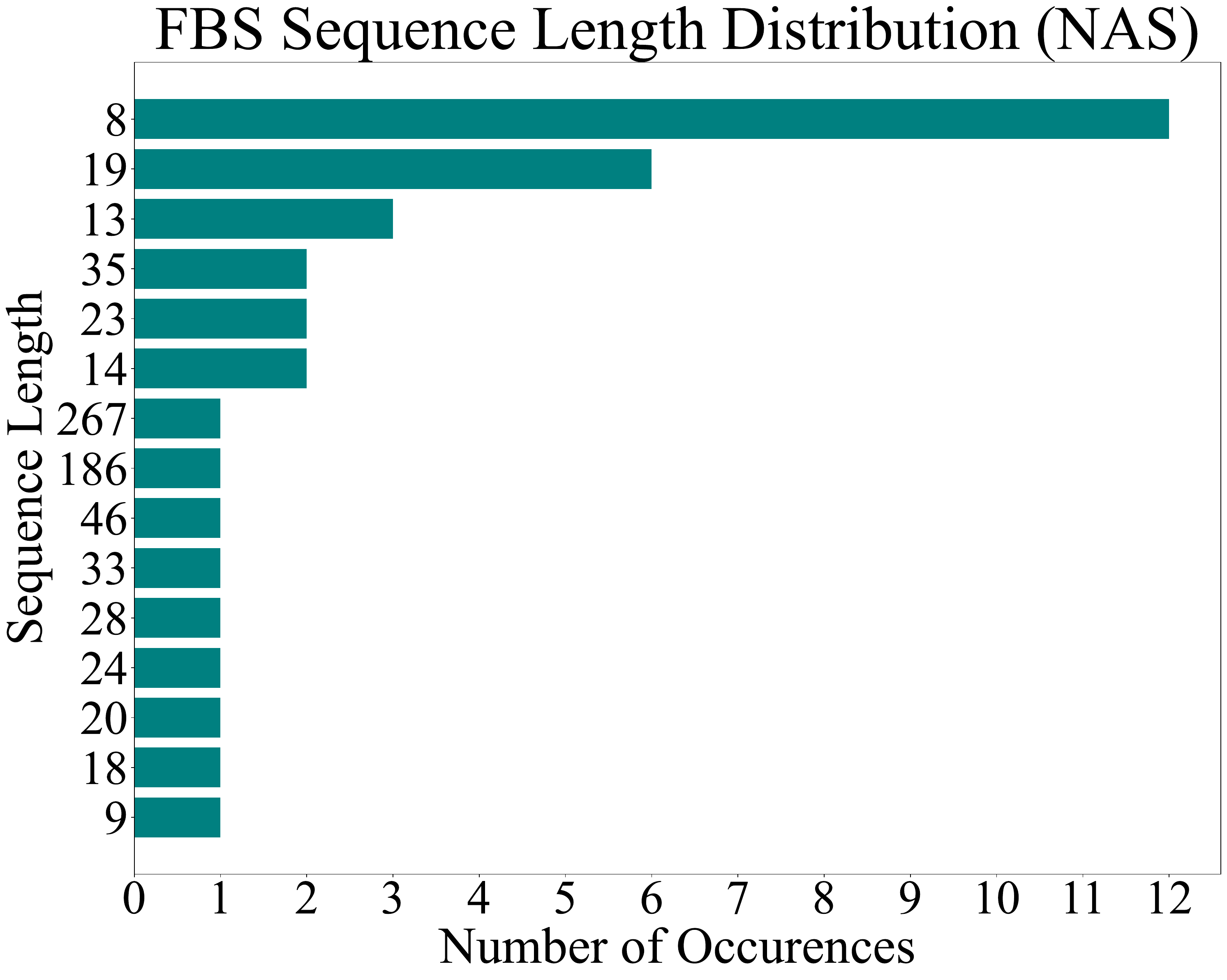}
         \caption{NAS sequence length}
         \label{fig:fbs-seq-len-distr-nas}
     \end{subfigure}
     % \hfill
     \begin{subfigure}[b]{0.23\textwidth}
         \centering
         \includegraphics[width=\linewidth]{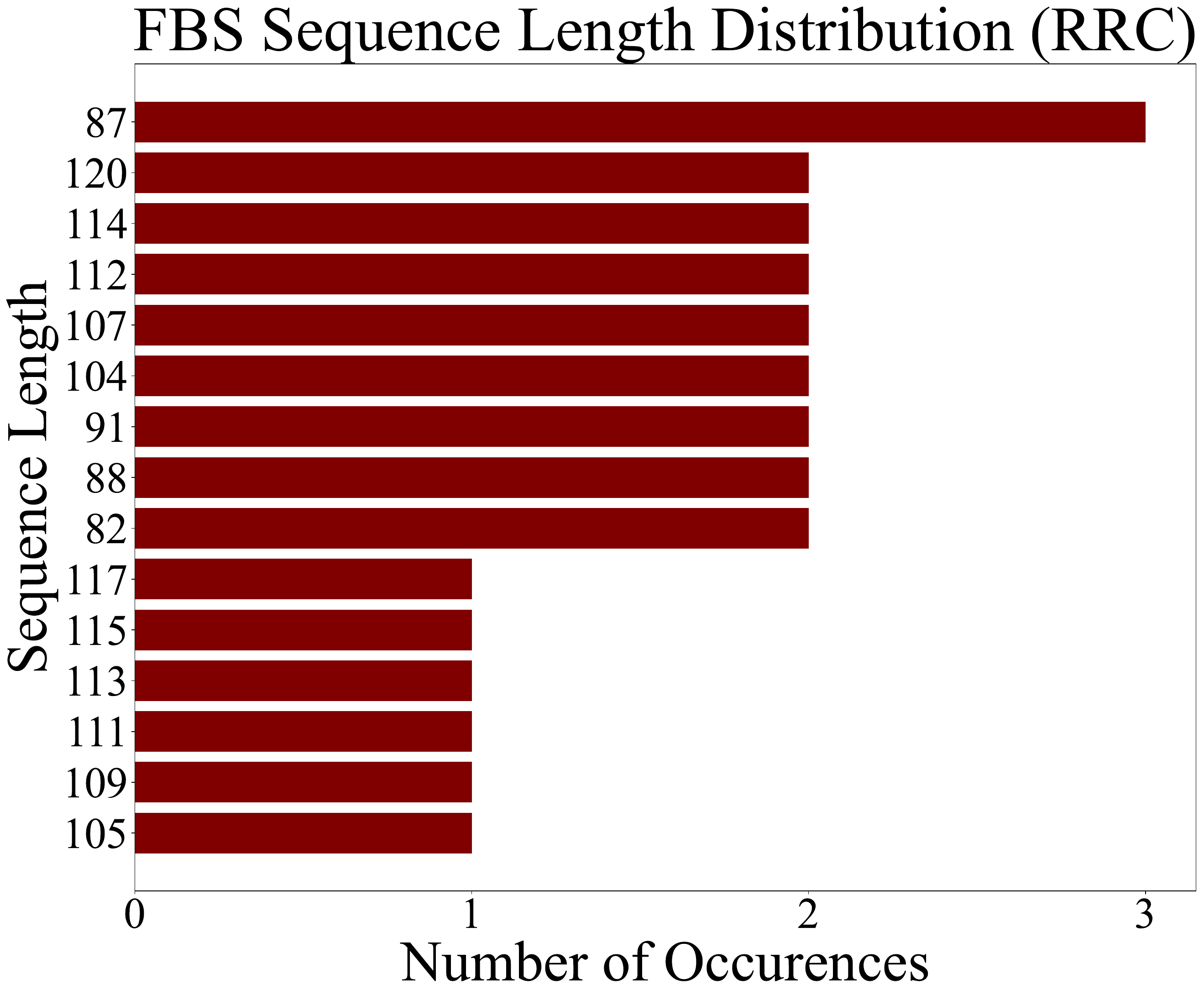}
         \caption{RRC sequence length}
         \label{fig:fbs-seq-len-distr-rrc}
     \end{subfigure}
     % \hfill
     \begin{subfigure}[b]{0.25\textwidth}
         \centering
        \includegraphics[width=1.0\linewidth]{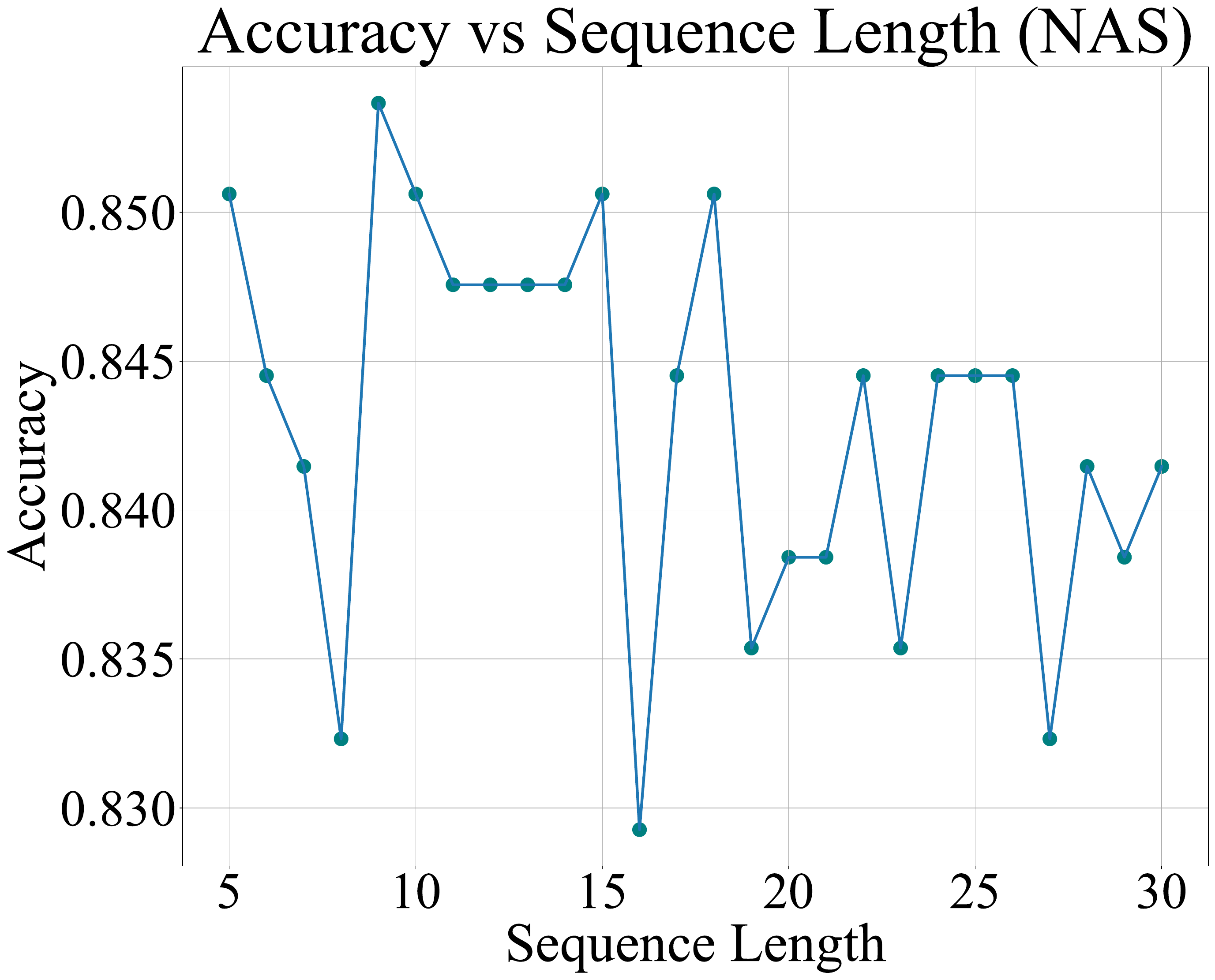}
        \caption{Accuracy in NAS}
        \label{fig:acc-vs-seq-len-nas}
     \end{subfigure}
     % \hfill
     \begin{subfigure}[b]{0.25\textwidth}
         \centering
        \includegraphics[width=1.0\linewidth]{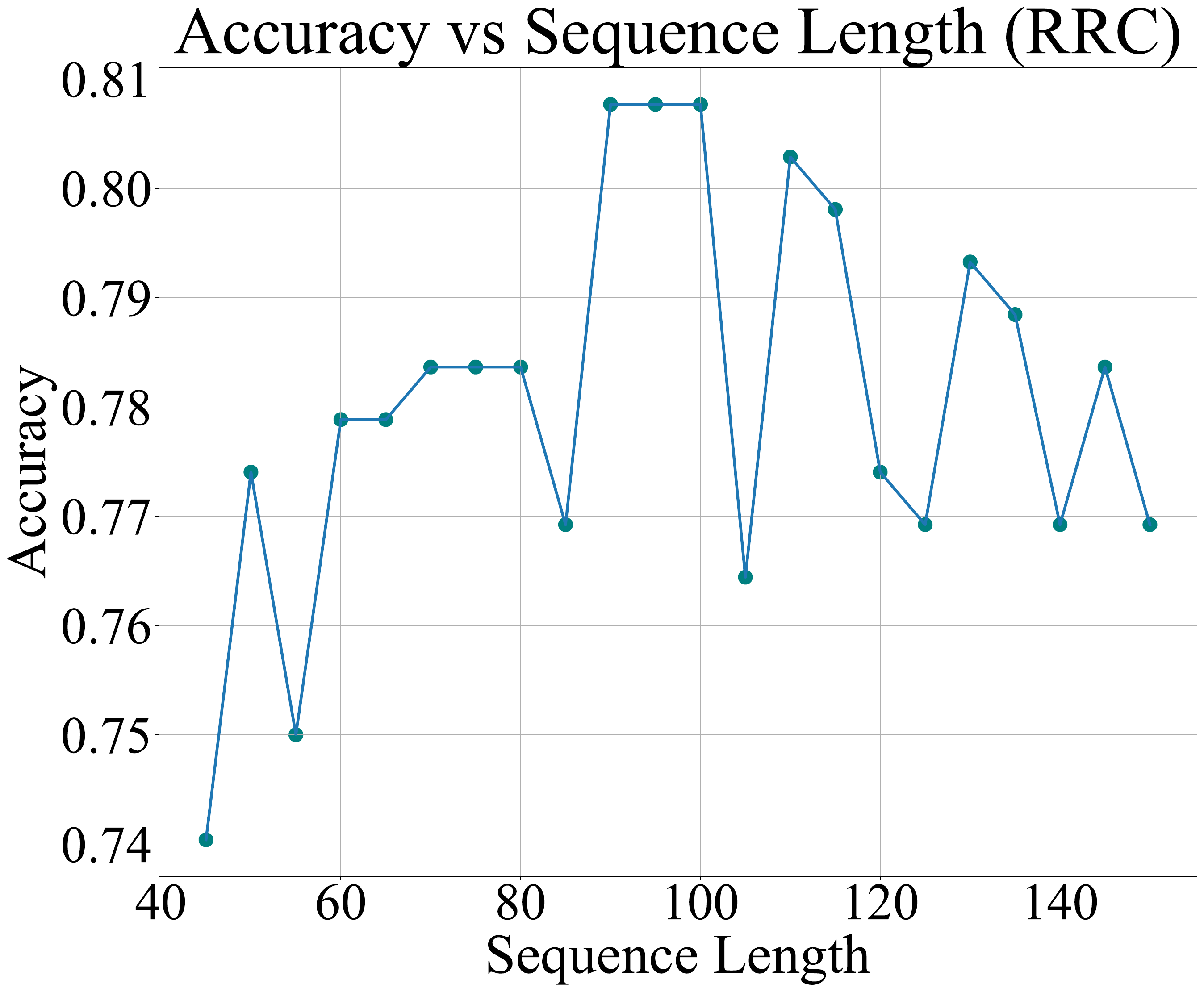}
        \caption{Accuracy in RRC}
        \label{fig:acc-vs-seq-len-rrc}
     \end{subfigure}
        \caption{Distribution and impact of NAS and RRC sequence length in FBS detection}
        \label{fig:fbs-seq-dst-and-det-acc}
        \vspace{-0.5cm}
\end{figure*}

The distribution of the length of the FBS generated packet sequences and the impact of sequence length on the detection performance is shown in Figure~\ref{fig:fbs-seq-dst-and-det-acc}

\section{POWDER}\label{apndx-sec:powder}
% \imtiaz{All which challenges? Revise.}

\begin{figure}
    \centering
    \includegraphics[scale=0.4]{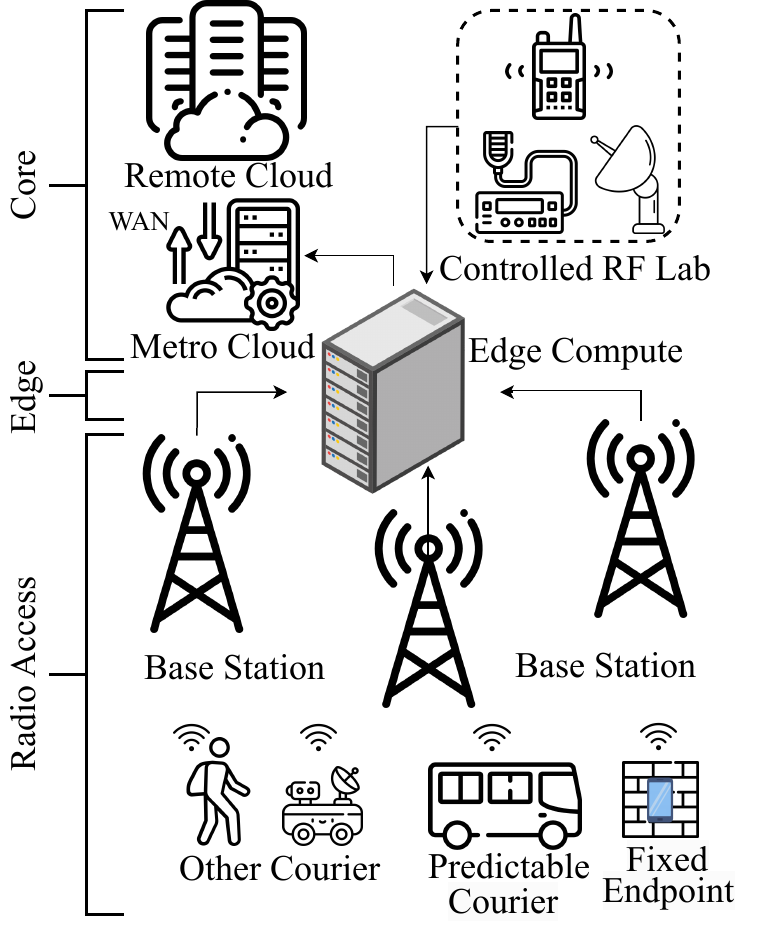}
    \caption{POWDER overview}
    \label{fig:powder-overview}
\end{figure}

POWDER (\textbf{P}latform for \textbf{O}pen \textbf{W}ireless \textbf{D}ata-driven \textbf{E}xperimental \textbf{R}esearch) resolves all these challenges and provides us with a way to generate high-quality, real-world, over-the-air datasets of FBSes with all the real-world scenarios involved, especially mobility. This invaluable resource enables us to develop and test FBS detection algorithms in a %controlled yet 
realistic environment. 
POWDER is a highly flexible city-scale scientific instrument that enables research at the forefront of the wireless revolution~\cite{powder} and is currently deployed in Salt Lake City, Utah. POWDER is a partnership between the University of Utah, Salt Lake City, and other public and private organizations (local, national, and global). %POWDER is one of the platforms being developed as part of the National Science Foundation (NSF) Platforms for Advanced Wireless Research (PAWR) program. The PAWR program is a public-private partnership between the NSF and an industry consortium of more than thirty organizations.

\subsection{Hardware and Mobility} 
% POWDER has deployed dozens of programmable radio nodes at fixed locations over an area of fourteen square kilometers, with approximately fifty mobile programmable radio nodes traveling through the area on couriers. This contiguous space covers three distinct environments: an urban downtown, a moderate-density residential area, and a hilly campus environment. 
The physical deployment of POWDER encompasses three key components: (1) \textbf{Core Component}: This includes the Controlled RF Lab and the Metro Cloud situated on-premise, connected to a remote cloud through a wide area network (WAN). (2) \textbf{Edge Component}: This involves edge computing devices to facilitate processing closer to data sources. (3) \textbf{Radio Access Component}: Comprising BSes, fixed, and mobile endpoints, this component establishes the fundamental wireless communication infrastructure. %These components collectively form the foundation of POWDER's infrastructure, enabling research across a spectrum of wireless networking domains.
The physical deployment offers a variety of configurable “coverage” scenarios, e.g., conventional macro-cell, small-cell (enabled by the campus “dense” deployment), or combinations thereof. Diversity in \emph{mobility} is provided by using mobile couriers that have relatively predictable movement patterns (e.g., buses), less predictable but bounded mobility (e.g., maintenance vehicles), and couriers that are “controllable” (e.g., backpacks/portable endpoints that can be moved by researchers that come on-site). Each deployed node consists of user-programmable software-defined radios (SDRs), off-the-shelf (OTS) radio equipment, RF front-ends, and antennas. Figure \ref{fig:powder-overview} gives an architectural overview of POWDER.

% Each node is also designed to support a modular “bring your own device” (BYOD) approach whereby experimenters can augment or “replace” functionality in the nodes. \imtiaz{Are we using this BYOD functionality? If not please remove the discussion.}
% All POWDER nodes have out-of-band access so that experimenters can remotely control, monitor, and collect data from their experiments. Nodes also have modest local compute and storage capabilities (i.e., edge compute with sub-ms latency), and the ability to access large amounts of cloud computing capacity both in the metro area (with a few milliseconds of latency) and across the country. 
% \imtiaz{The above discussion is not important to us. We are emphasizing more on what sort of tests can be done on POWDER and what testing capabilities it has. Resource capabilities are not useful for our discussion. }
% Fixed nodes deployed as base stations are connected with each other and compute resources via a dedicated fiber front-haul/back-haul network. \imtiaz{Remove discussion about fiber-connected networks. What we are after is only their over-the-air testing capability.}

\subsection{Software and Profiles}  
% \imtiaz{All this previous discussion can be removed. Remember we are not designing POWDER so their challenges are useless to us. Just say POWDER allows us to create profiles and using profiles we can access raw hardware to create different scenarios including FBSes without any legal issues.}
In the context of POWDER, profiles refer to configurable sets of parameters that define various aspects of the wireless network environment. These parameters encompass hardware, network configurations, signal propagation characteristics, mobility patterns, interference scenarios, and more. Profiles make it possible to replicate real-world conditions for experimentation, allowing one to create diverse scenarios, including those involving FBSes. Using profiles, one can manipulate the network environment within POWDER's controlled setting and conduct experiments that closely mirror real-world situations while avoiding legal and regulatory challenges. By creating different profiles, one can effectively create real-world conditions. 
%effectively and advance  research in FBS detection and related fields. 
This capability is a significant advantage, allowing one to conduct comprehensive experiments and analyses within a controlled environment.

% \imtiaz{On the whole this previous paragraph can be removed. For use cases, we only want to focus on what capabilities we need to use POWDER that's it, and what capabilities POWDER has for us. Other use cases are not needed.}

\noindent \textbf{POWDER feedback}
\looseness=-1
POWDER makes it possible to replicate real-world conditions for experimentation. 
One can manipulate the network environment within POWDER’s controlled setting and conduct experiments using POWDER profiles.
% that closely mirrors real-world situations while avoiding legal and regulatory challenges. 
However, only a limited number of profiles with some fixed sets of parameters are currently available at POWDER. 
Users can create their profiles according to their requirements. Still, a new user might find it challenging to create a profile because there is not enough documentation and tutorials on creating these profiles. The incorporation of more profiles with diverse scenarios, automation of the profile creation process, and extensive documentation and tutorials will help users utilize POWDER's capabilities to a great extent.

\iffalse
\section{Mobile App}\label{appendix-subsubsec:mobile-app-demo}
Different components of the \system Android app are demonstrated in Figure \ref{fig:app-demo}.

\begin{figure*}[]
     \centering
     \begin{subfigure}[b]{0.3\textwidth}
         \centering
         \includegraphics[width=0.7\linewidth]{figures/homescreen.png}
        \caption{Homepage of FBS Detector}
        \label{fig:homepage}
     \end{subfigure}
     % \hfill
     \begin{subfigure}[b]{0.3\textwidth}
         \centering
         \includegraphics[width=0.7\linewidth]{figures/nofbspage.png}
         \caption{No Fake Base Station detected}
         \label{fig:nofbspage}
     \end{subfigure}
     % \hfill
     \begin{subfigure}[b]{0.3\textwidth}
         \centering
         \includegraphics[width=0.7\linewidth]{figures/fbspage.png}
         \caption{Fake Base Station detected}
         \label{fig:fbspage}
     \end{subfigure}
    \caption{Android app for \system}
    \label{fig:app-demo}
\end{figure*}

\fi

\section{Load testing of App}\label{appndx:apploadtesting}
To evaluate our app's performance under load, we conducted load testing using Android Studio's profiler to measure CPU and memory usage. The profiling was carried out over a period of 20 minutes on a Google Pixel 3a device, which is equipped with a Qualcomm SDM670 Snapdragon 670 chipset (10 nm), an Octa-core CPU (2x2.0 GHz Kryo 360 Gold \& 6x1.7 GHz Kryo 360 Silver), an Adreno 615 GPU, and 4GB of RAM. Under load conditions, the CPU usage ranged between 18-20\%, while memory usage was between 20-30 MB. In contrast, CPU usage dropped to 2-3\% under no-load conditions, and memory usage was reduced to 4-5 MB. 
The load testing is shown in Figure~\ref{fig:cpu-usage}.
The peaks in the CPU and memory usage graphs correspond to the attacks we executed during the test. Specifically, from 0 to 5 minutes, we ran two FBS attacks, followed by ten Multi-Step Attacks (MSAs) from 5 to 20 minutes. 

\section{Comparison with existing solutions}
The comparison between \system and Phoenix~\cite{PHOENIX} is shown in Table~\ref{tab:comp-with-phoenix}.
\begin{table}[]
    \centering
    \renewcommand{\arraystretch}{1}
    \fontsize{6}{6}\selectfont
    \setlength{\tabcolsep}{1.6pt} 
    \begin{tabular}{l|ccc|ccc|ccc|ccc}
    \hline
    \multirow{3}{*}{Attack} & \multicolumn{9}{c|}{Phoenix} & \multicolumn{3}{c}{\system} \\
     \cline{2-13}
      & \multicolumn{3}{c}{DFA} & \multicolumn{3}{c}{MM} & \multicolumn{3}{c|}{PLTL} & \multirow{2}{*}{Acc} & \multirow{2}{*}{Prec} & \multirow{2}{*}{Rec} \\
      \cline{2-10}
       & Acc & Prec & rec & Acc & Prec & Rec & Acc & Prec & Rec & & & \\
     \hline
       Attach Reject            & 0.487 & 0.35  & \underline{0.799} & \underline{0.89}  & \underline{0.86}  & 0.79  & 0.868 & 0.70  & 0.767 & \textbf{0.95} & \textbf{0.97} & \textbf{0.95} \\
       IMSI Catching            & 0.667 & 0.538 & \underline{0.876} & 0.785 & 0.79  & 0.858 & \underline{0.798} & \underline{0.81}  & 0.797 & \textbf{0.98} & \textbf{0.94} & \textbf{0.97} \\
       Service Reject           & 0.712 & 0.704 & 0.721 & 0.797 & 0.725 & 0.753 & \underline{0.871} & \underline{0.81}  & \underline{0.844} & \textbf{0.95} & \textbf{0.96} & \textbf{0.93} \\
       TAU Reject               & 0.627 & \underline{0.95}  & \underline{0.756} & 0.763 & 0.865 & 0.715 & \underline{0.789} & 0.803 & 0.751 & \textbf{0.94} & \textbf{0.95} & \textbf{0.92} \\
       Measurement Report       & 0.445 & 0.434 & 0.456 & 0.766 & 0.766 & 0.845 & \underline{0.878} & \underline{0.864} & \underline{0.871} & \textbf{0.97} & \textbf{0.95} & \textbf{0.97} \\
       Paging with IMSI         & 0.574 & 0.634 & \underline{0.918} & 0.783 & 0.765 & 0.81  & \underline{0.84}  & \underline{0.822} & 0.786 & \textbf{0.94} & \textbf{0.96} & \textbf{0.95} \\
       Authentication Failure   & 0.802 & 0.671 & \underline{0.897} & 0.805 & \underline{0.79}  & 0.749 & \underline{0.849} & 0.788 & 0.863 & \textbf{0.98} & \textbf{0.96} & \textbf{0.95} \\
       Numb Attack              & 0.817 & 0.811 & 0.799 & \underline{0.846} & 0.711 & \underline{0.818} & 0.732 & \underline{0.833} & 0.722 & \textbf{0.97} & \textbf{0.99} & \textbf{0.95} \\
     \hline 
    \end{tabular}
    \caption{Comparison between \system and Phoenix}
    \label{tab:comp-with-phoenix}
\end{table}

\begin{table}[]
    \centering
    \renewcommand{\arraystretch}{1}
    \fontsize{6.5}{6.5}\selectfont
    \begin{tabular}{l|C{0.7cm}|C{0.8cm}|C{0.7cm}}
    \hline
        \multirow{2}{*}{Attack} & \multicolumn{3}{c}{Classified As}  \\
        \cline{2-4}
        & Attack (TP) & Benign (FN) & Other Attack \\
         \hline
         FBS &                                                              5 & 0 & 0\\
         Authentication relay attack &                                      4 & 1 & 0\\
         Bidding down with \MTable{AttachReject} &                          3 & 0 & 2\\ 
         Paging channel hijacking attack &                                  3 & 1 & 1 \\
         Location tracking via measurement reports &                        5 & 0 & 0\\
         Capability Hijacking &                                             5 & 0 & 0 \\
         Incarceration with \MTable{rrcReestablishReject} &                 4 & 0 & 1\\ 
         Lullaby attack using \MTable{rrcReestablishRequest} &              3 & 0 & 2\\ 
         Bidding down with \MTable{ServiceReject} &                         3 & 0 & 2\\
         Mobile Network Mapping (MNmap) &                                   5 & 0 & 0 \\
         Energy Depletion attack &                                          5 & 0 & 0\\
         Lullaby attack with \MTable{rrcResume} &                           3 & 0 & 2\\
         Stealthy Kickoff Attack &                                          4 & 0 & 1\\
         Incarceration with \MTable{rrcReject} and \MTable{rrcRelease} &    4 & 0 & 1\\
         IMSI catching &                                                    5 & 0 & 0\\
         NAS counter Desynch attack &                                       3 & 1 & 1\\
         X2 signalling flood &                                              5 & 0 & 0\\
         Handover hijacking &                                               4 & 1 & 0\\
         RRC replay attack &                                                5 & 0 & 0\\
         Lullaby attack with \MTable{rrcReconfiguration} &                  3 & 0 & 2\\
         Bidding down with \MTable{TAUreject} &                             3 & 0 & 2\\
         Panic Attack &                                                     4 & 1 & 0\\
        \hline
         Total &                                                            88 & 5 & 17\\
        \hline
        \hline
         No Attack (Benign) &                                              - & - & 6 (FP)\\
        \hline
    \end{tabular}
    \caption{Real world attack detection performance of \system}
    \label{tab:real-world-msa-attacks-evaluation}
\end{table}

\begin{table*}[ht]
\centering
\renewcommand{\arraystretch}{1}
\fontsize{6.5}{6.5}\selectfont
\setlength{\tabcolsep}{1.6pt}
\begin{tabular}{l|c|c|c|c|c|c|c|c|c|c}
\toprule
\textbf{Attack Type} & \textbf{System} & \textbf{Benign} &  \textbf{Attach Reject} & \textbf{IMSI Catching} & \textbf{Service Reject} & \textbf{TAU Reject} & \textbf{Measurement Report} & \textbf{Paging with IMSI} & \textbf{Auth Failure} & \textbf{Numb Attack} \\
\hline
\multirow{2}{*}{Attach Reject} & FBSDetector & 9.35 &  \textbf{67.27} & 1.24 & 6.23 & 6.20 & 0.81 & 2.71 & 4.64 & 1.56 \\
                               & Phoenix & 9.12 & 2.14 &  7.39 &  3.66 &  3.48 &  8.24 &  3.03 & 18.21 & \textcolor{red}{\textbf{44.72}}  \\
\hline                               
\multirow{2}{*}{IMSI Catching} & FBSDetector & 3.83 & 2.45 & \textbf{47.49} & 15.97 & 7.53 & 4.14 & 5.76 & 0.42 & 0.66  \\
                               & Phoenix & 4.16 & \textcolor{red}{\textbf{26.26}} & 5.71 & 16.16 & 21.74 & 6.62 & 3.88 & 14.32 & 1.15  \\ 
\hline
\multirow{2}{*}{Service Reject} & FBSDetector & 1.44  & 14.18 & 4.26 & \textbf{51.77} & 10.30 & 2.14 & 11.49 & 7.32 & 8.79 \\
                                     & Phoenix & 8.67 & 8.82 & 5.38 & 7.14 & 4.74 & 0.26 & \textcolor{red}{\textbf{51.95}} & 6.93 & 6.12 \\ 
\hline
\multirow{2}{*}{TAU Reject}          & FBSDetector & 1.35 & 2.88 &   8.37 &   1.13 &  \textbf{78.34} &   3.90 &   3.19 & 0.09 &   0.71  \\
                                     & Phoenix & 9.28 & 5.98 & 1.84  & \textcolor{red}{\textbf{60.22}}  & 5.52 & 0.15 & 1.61 & 3.89 & 11.49 \\ 
\hline
\multirow{2}{*}{Measurement Report}  & FBSDetector & 3.02 & 0.51 & 11.81 & 0.52 &  1.74 & \textbf{69.78} &  8.03 & 3.31 & 1.24  \\
                                     & Phoenix & 2.67 & 17.92 & 3.54 & 2.48 & 17.39 & 10.72 & 7.59 & 7.39 & \textcolor{red}{\textbf{30.29}} \\ 
\hline
\multirow{2}{*}{Paging with IMSI}    & FBSDetector & 6.55 & 4.69 & 10.32 & 1.06 &  9.54 &  3.76 & \textbf{50.92} & 3.94 &  9.18 \\
                                     & Phoenix & 4.43 & 1.68  & 0.71 & 0.52 & 4.76 & 4.23 & 23.14 & 11.5 & \textcolor{red}{\textbf{49.03}} \\
\hline
\multirow{2}{*}{Authentication Failure} & FBSDetector & 6.97 & 0.54 &  6.9 & 0.19 & 0.99 & 0.18 & 1.12 & \textbf{72.32} & 10.79 \\
                                     & Phoenix & 5.01 & \textcolor{red}{\textbf{63.20}} & 0.34 & 1.75 &  1.47 &  8.07 &  5.81 & 11.17 &  3.18  \\
\hline
\multirow{2}{*}{Numb Attack}         & FBSDetector & 9.5 & 9.13 & 10.17 &  2.07 & 13.07 &  6.26 &  0.5 &   5.13 & \textbf{44.16} \\
                                     & Phoenix & 7.21 & 0.63 &  9.45 &  6.46 & 12.63 & \textcolor{red}{\textbf{28.34}} & 4.1 & 14.23 & 16.95 \\ \bottomrule
\end{tabular}
\caption{Cross-validation comparison with Phoenix~\cite{PHOENIX} for an adaptive adversary. The numbers represent percentage of the attack packets being classified as benign and different other attack packets. \system accurately classifies majority of the packets to it original attack (numbers shown in bold) whereas Phoenix misclassifies to other attacks (numbers shown in red).}
\label{tab:cross_validation_attacks}
\end{table*}

\begin{table}[]
    \centering
    \renewcommand{\arraystretch}{1}
    \fontsize{7}{7}\selectfont
    \begin{tabular}{c|l|c|c|c|c}
    \hline
       Sl & Attack & TP & TN & FP & FN \\
        \hline
        1 & Authentication relay attack & 46.08 & 48.49 & 2.29 & 3.14  \\
        \hline
        2 & Bidding down with \MTable{AttachReject} & 52.2 & 42.56 & 2.05 & 3.19 \\ 
        \hline
        3 & Paging channel hijacking attack & 51.18 & 38.6 & 3.41 & 6.81 \\
        \hline
        4 & Location tracking via measurement reports & 51.93 & 41.52 & 2.29 & 4.26 \\
        \hline
        5 & Capability Hijacking & 50.62 & 40.58 & 3.46 & 5.34\\
        \hline
        6 & Incarceration with \MTable{rrcReestablishReject} & 49.65 & 42.81 & 3.73 & 3.81\\ 
        \hline
        7  & Lullaby attack using \MTable{rrcReestablishRequest} & 44.16 & 46.36 & 3.36 & 6.12\\ 
        \hline
        8 & Bidding down with \MTable{ServiceReject} & 48.1 & 44.57 & 3.49 & 3.84\\
        \hline
        9 & Mobile Network Mapping (MNmap) & 53.05 & 40.99 & 2.09 & 3.87\\
        \hline
        10 & Energy Depletion attack & 52.44 & 40.95 & 2.12 & 4.49 \\
        \hline
        11 & Lullaby attack with \MTable{rrcResume} & 44.79 & 46.85 & 3.89 & 4.47 \\
        \hline
        12 & Stealthy Kickoff Attack & 52.71 & 37.84 & 4.71 & 4.74\\
        \hline
        13 & Incarceration with \MTable{rrcReject} and \MTable{rrcRelease} & 48.01 & 42.84 & 4.77 & 4.38\\
        \hline
        14 & IMSI catching & 39.17 & 53.36 & 2.69 & 4.78\\
        \hline
        15 & NAS counter Desynch attack & 49.81 & 42.49 & 3.04 & 4.66\\
        \hline
        16 & X2 signalling flood & 42.75 & 49.46 & 4.72 & 3.07\\
        \hline
        17 & Handover hijacking & 40.2 & 51.8 & 3.1 & 4.9\\
        \hline
        18 & RRC replay attack & 44.16 & 46.89 & 4.93 & 4.02\\
        \hline
        19 & Lullaby attack with \MTable{rrcReconfiguration} & 47.24 & 45.9 & 2.76 & 4.1 \\
        \hline
        20 & Bidding down with \MTable{TAUReject} & 50.52 & 42.27 & 3.43 & 3.78\\
        \hline
        21 & Panic Attack & 50.52 & 42.27 & 3.43 & 3.78\\
        \hline
    \end{tabular}
    \caption{MSAs detection performance breakdown (in percentage)}
    \vspace{-0.7cm}
    \label{tab:msa-fpr-breakdown}
\end{table}

\begin{table}[]
    \centering
    \renewcommand{\arraystretch}{1}
    \fontsize{7}{7}\selectfont
    \begin{tabular}{c|c|c|c|c}
    \hline
       Attacker Level & TP & TN & FP & FN \\
        \hline
        Level 0 & 48.31 & 48.31 & 2.32 & 1.05\\
        \hline
        Level 1 & 43.20 & 52.76 & 2.93 & 1.11\\
        \hline
        Level 2 & 39.64 & 56.00 & 3.19 & 1.16 \\ 
        \hline
        Level 3 & 52.48 & 41.99 & 3.27 & 2.27 \\
        \hline
        Level 4 & 35.25 & 60.18 & 3.09 & 1.48 \\
        \hline
    \end{tabular}
    \caption{FBS detection performance breakdown (in percentage)}
    \vspace{-0.7cm}
    \label{tab:attacker-level-fpr-breakdown}
\end{table}

\section{Implementation Efforts}
The implementation effort for \system is listed in Table~\ref{tab:imp-efforts}.
\begin{table}[ht]
    \centering
    \renewcommand{\arraystretch}{1}
    \fontsize{7}{7}\selectfont
    \setlength{\tabcolsep}{2pt}
    \begin{tabular}{l|c|c|L{2cm}|R{0.8cm}}
        \hline
         \textbf{Task} & \textbf{Framework} & \textbf{Lang} & \textbf{Libraries Used} & \textbf{LoC} \\
         \hline
          Handover capability in srsUE  & srsRAN, OAI & C                 &                       & \totalLinesOfCodeAddedInsrsRANforFBS      \\
          \hline
          MSAs                          & srsRAN, OAI & C                 &                       & \totalLinesOfCodeAddedInsrsRANforMultiStep\\
          \hline
          Data Processing               &        & Python            & tshark, scikit-learn                      & \totalLinesOfCodeinDataProcessing         \\
          \hline
          Training ML Models            &        & Python            & scikit-learn, TensorFlow & \totalLinesOfCodeinTraingMLModels         \\
          \hline
          Stateful LSTM w/ Attention              &        & Python            & scikit-learn, TensorFlow                      & \totalLinesOfCodeinSequentialLSTM         \\
          \hline
          Graph Learning Pipeline       &        & Python            & networkx, pytorch-geometric                      & \totalLinesOfCodeinGraphLearning          \\
        \hline
        App Development       &  Flutter      & Dart            & mobileinsight-libs, tensorflow-lite                      & \totalLinesOfCodeinAppDevelopment          \\
        \hline
         %& & & \textbf{1042}\\
         %\hline
    \end{tabular}
    \caption{Implementation efforts}
    \label{tab:imp-efforts}
    \vspace{-0.7cm}
\end{table}

\section{Real-World Test}
\begin{figure*}[t]
     \centering
     \begin{subfigure}[b]{0.3\textwidth}
         \centering
         \includegraphics[width=\linewidth, height=0.6\linewidth]{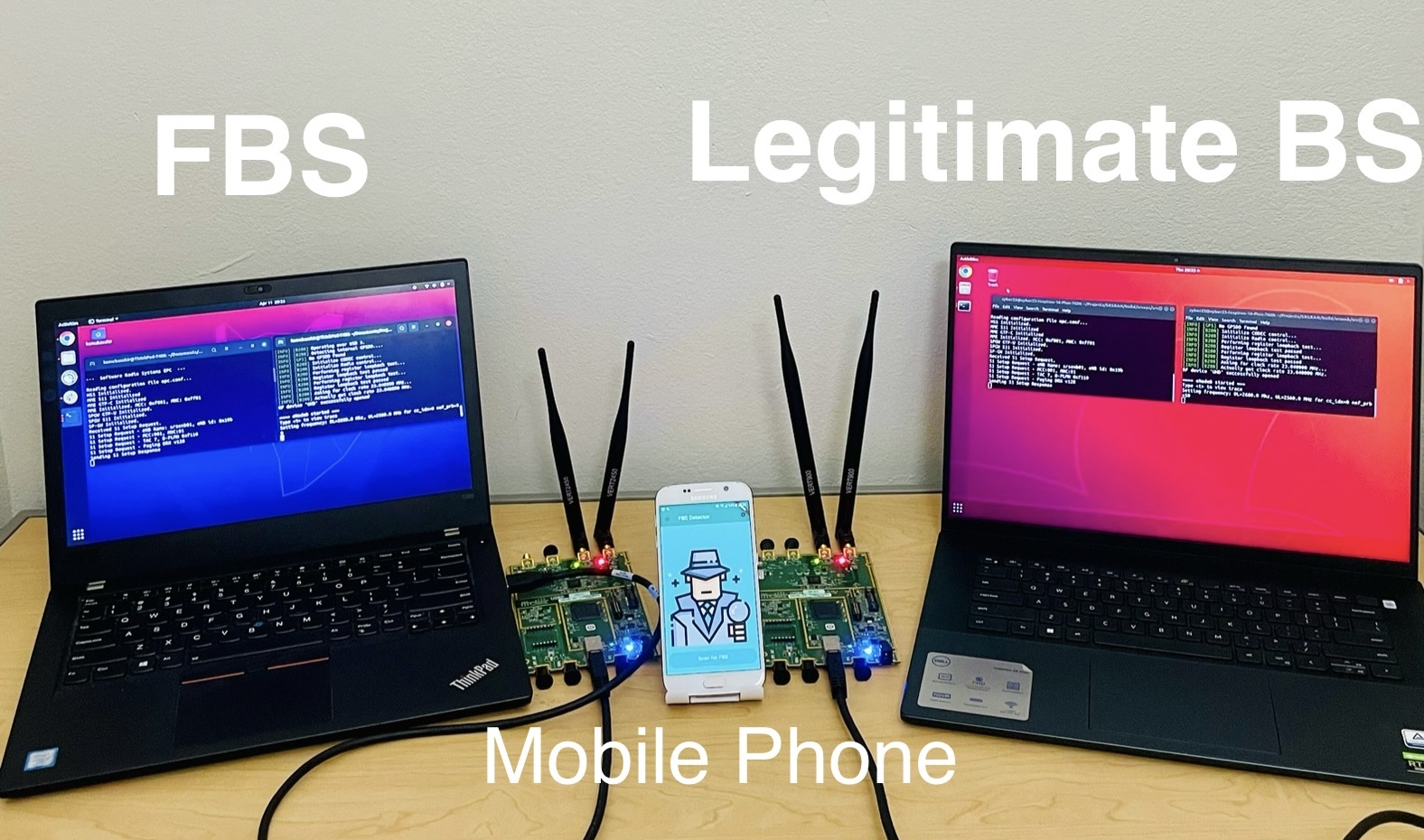}
        \caption{FBS, Legitimate BS and the Mobile device are put close to each other}
        \label{fig:real-world-config1}
     \end{subfigure}
     % \hfill
     \begin{subfigure}[b]{0.3\textwidth}
         \centering
         \includegraphics[width=\linewidth, height=0.6\linewidth]{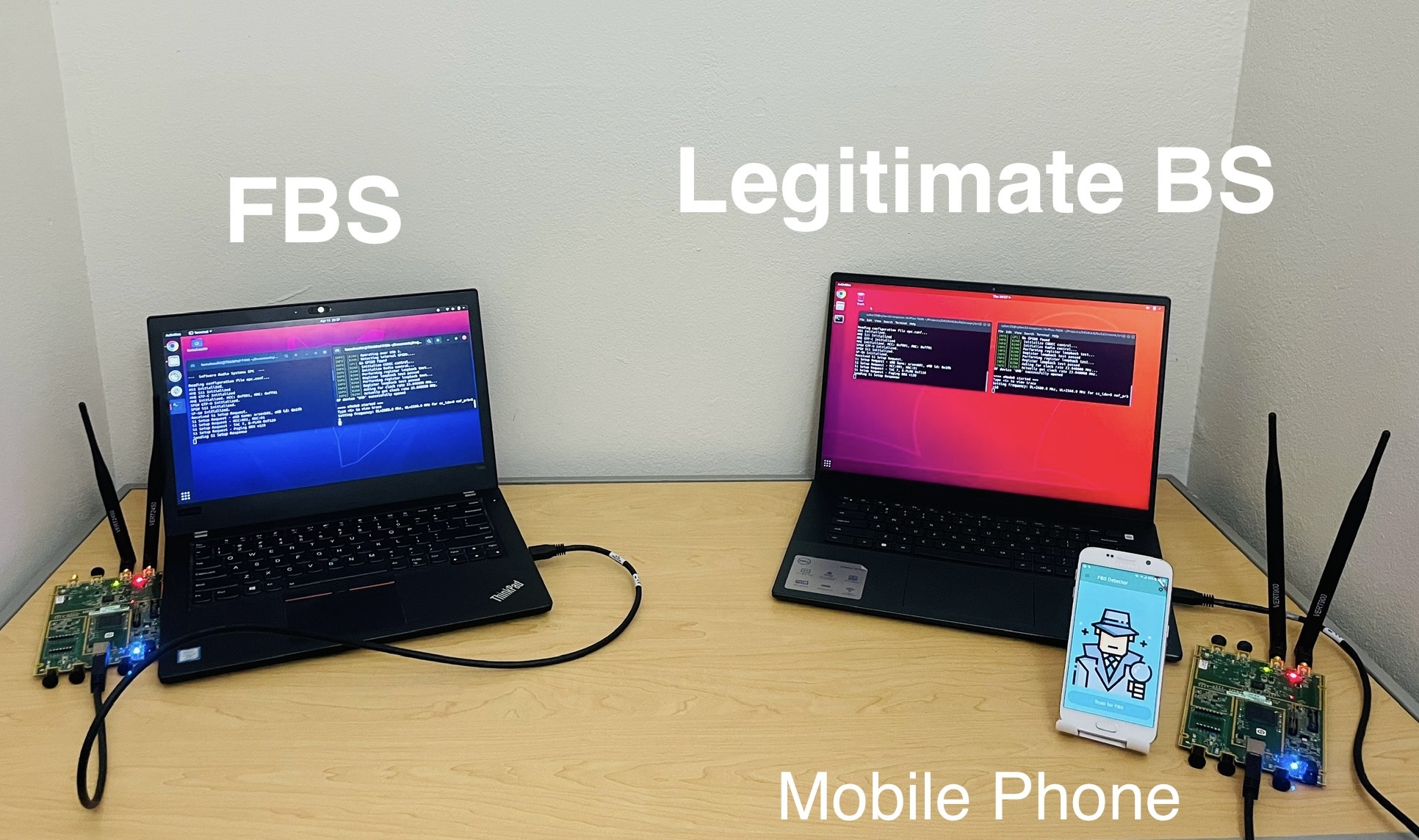}
         \caption{The mobile device is put close to the legitimate BS, far from the FBS}
         \label{fig:real-world-config2}
     \end{subfigure}
     % \hfill
     \begin{subfigure}[b]{0.3\textwidth}
         \centering
         \includegraphics[width=\linewidth, height=0.6\linewidth]{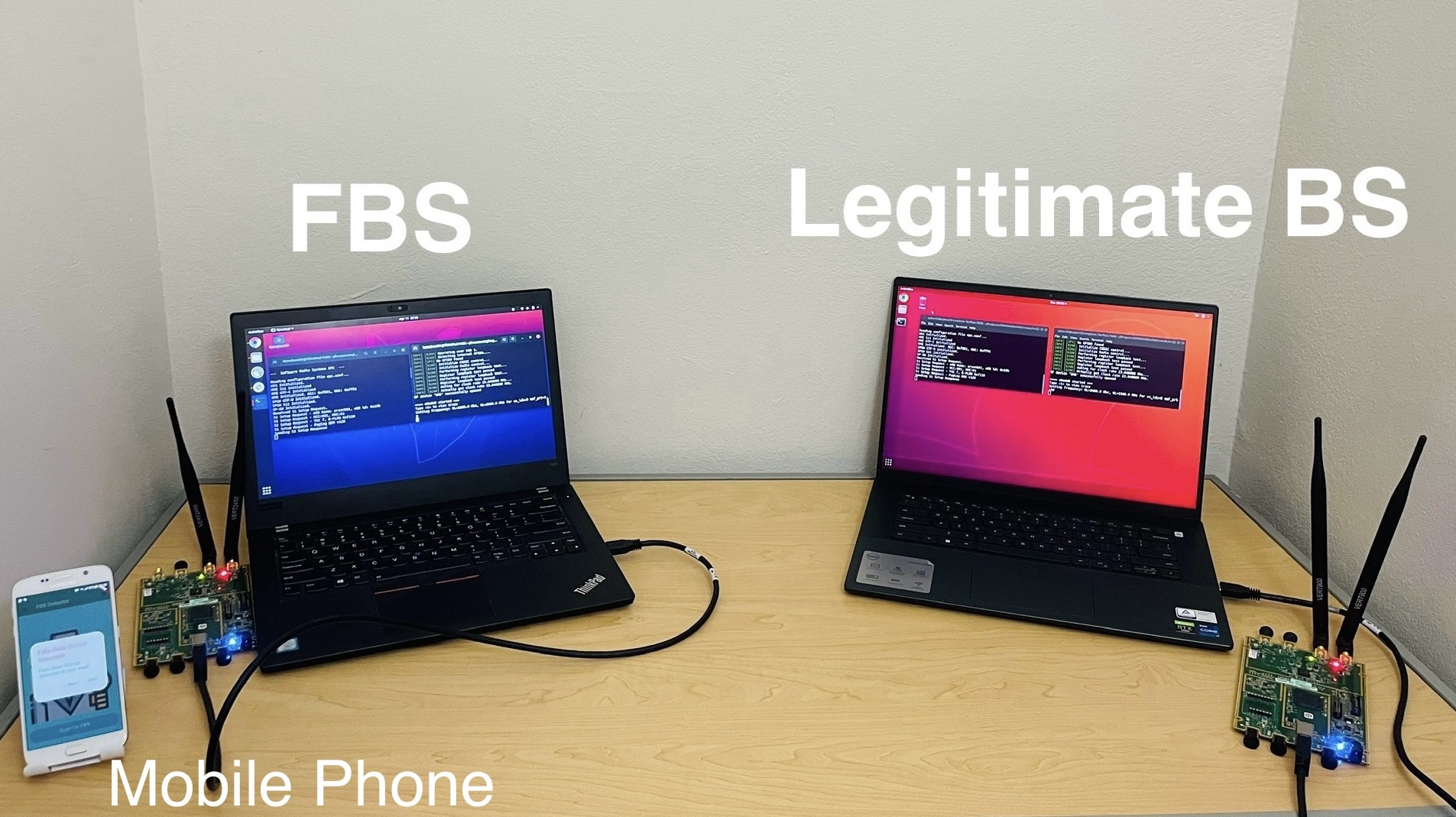}
         \caption{The mobile device is put close to the FBS}
         \label{fig:real-world-config3}
     \end{subfigure}
     \begin{subfigure}[b]{0.3\textwidth}
         \centering
         \includegraphics[width=\linewidth, height=0.6\linewidth]{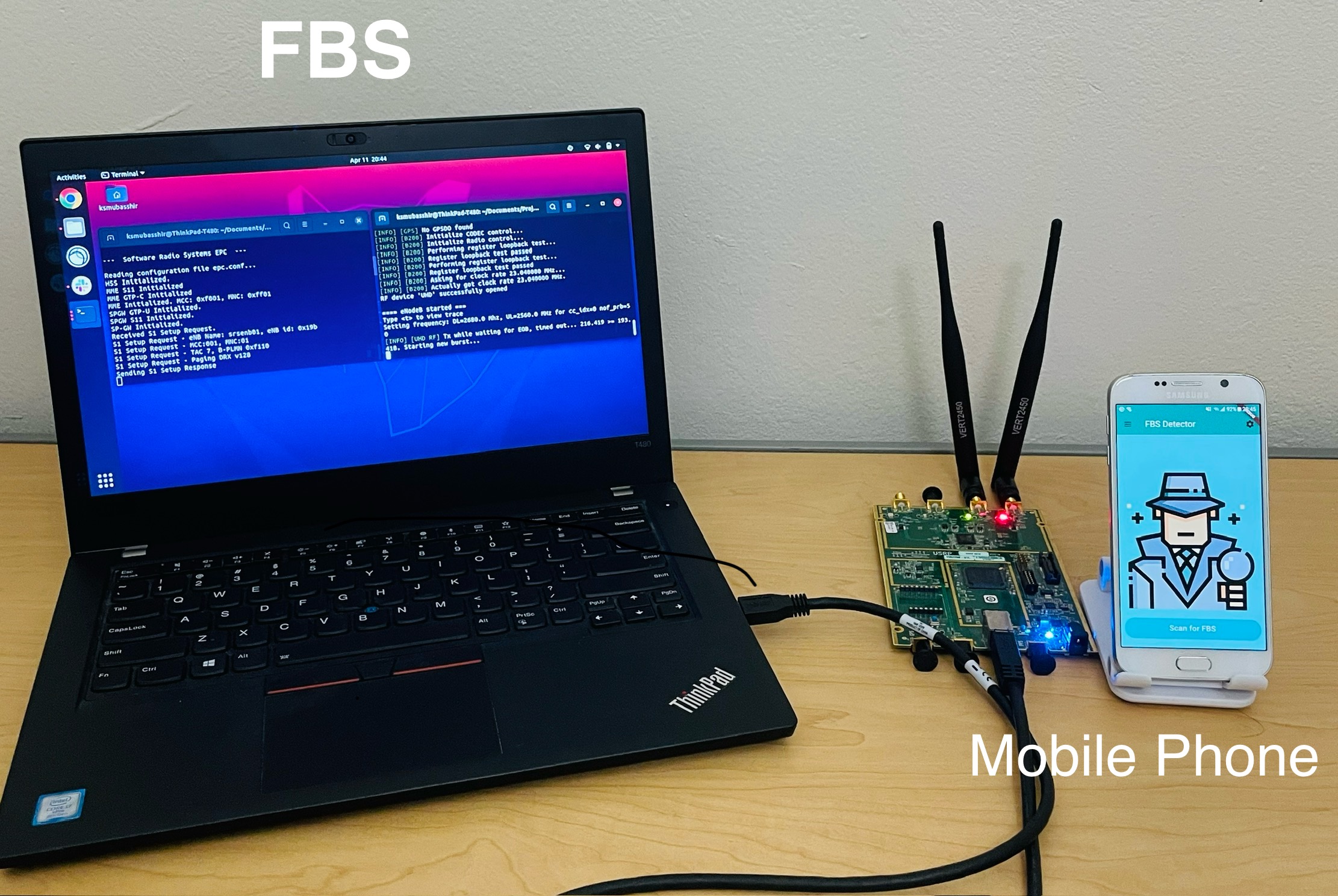}
         \caption{Mobile device (connected to a commercial network) is put close to the FBS}
         \label{fig:real-world-config4}
     \end{subfigure}
     \begin{subfigure}[b]{0.3\textwidth}
         \centering
         \includegraphics[width=\linewidth, height=0.6\linewidth]{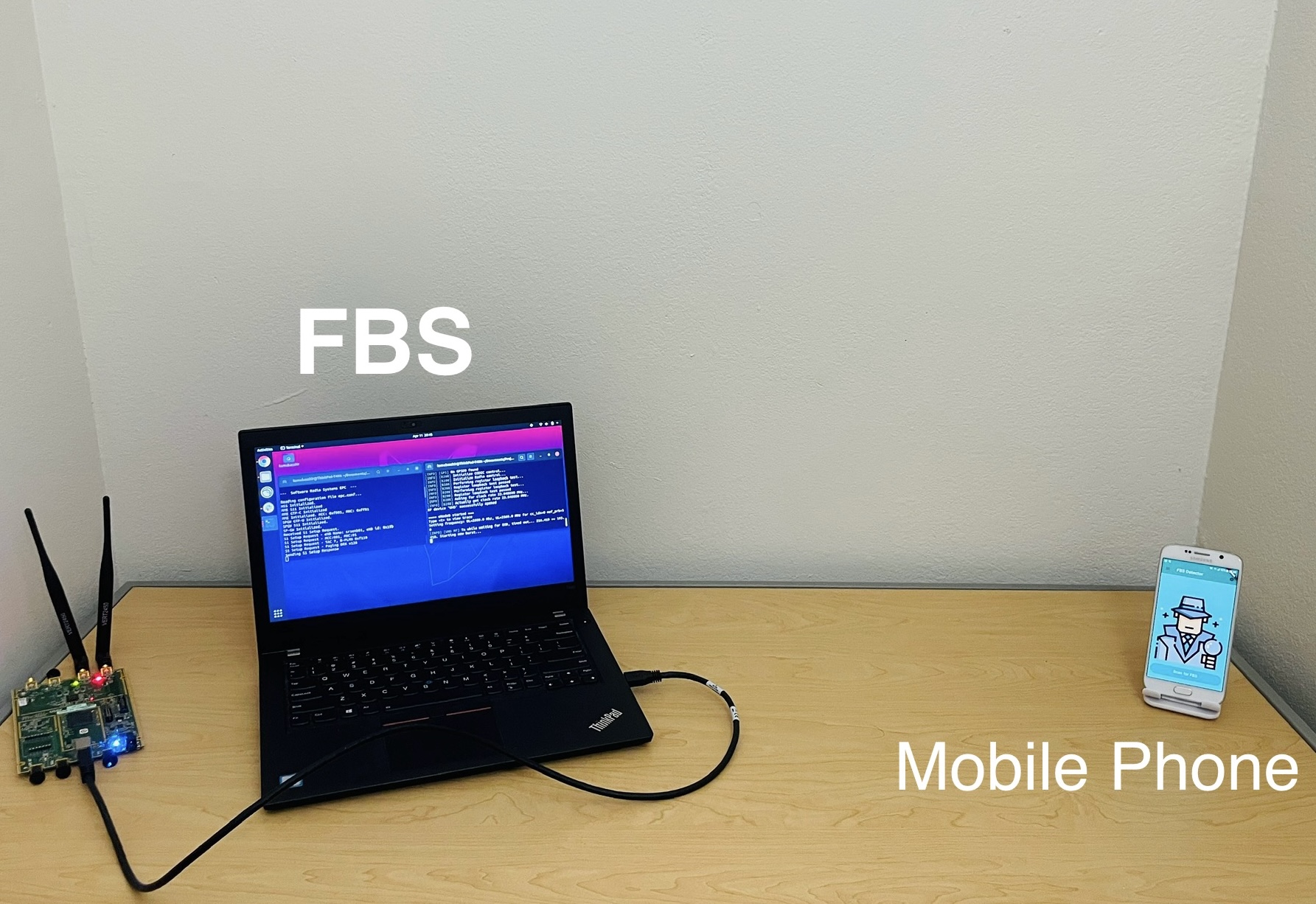}
         \caption{Mobile device (connected to a commercial network) is put far from the FBS}
         \label{fig:real-world-config5}
     \end{subfigure}
     \iffalse
     \begin{subfigure}[b]{0.3\textwidth}
         \centering
         \includegraphics[width=\linewidth, height=0.6\linewidth]{figures/real-world-test/open-world-test.png}
         \caption{Mobile device with FBSDetector mounted on a car for open world test}
         \label{fig:open-world-config1}
     \end{subfigure}
     \fi
    \caption{Real world test for \system}
    \label{fig:real-world-test}
\end{figure*}

Figure \ref{fig:real-world-test} shows our real-world test setup. Table~\ref{tab:real-world-msa-attacks-evaluation} shows the real-world attack detection performance of \system.

\section{List of Features}
Features used to train the models are listed in Table~\ref{tab:list-of-features}
\begin{table*}[]
\centering
\fontsize{7}{7}\selectfont
\begin{tabular}{l|l|l|l}
\hline
\multicolumn{2}{c}{NAS layer features} & \multicolumn{2}{c}{RRC layer features} \\
\hline
nas eps common elem id & nas eps emm EPS attach & lte rrc si WindowLength & lte rrc ue Identity \\
gsm a dtap add ci & nas eps msg auth code & lte rrc DL CCCH Message & per extension present bit \\
gsm a dtap elem id & nas eps emm eia6 & lte rrc shortMAC I & lte rrc maxHARQ Tx \\
nas eps emm eea4 & nas eps emm eia0 & lte rrc rlf InfoAvailable r10 & lte rrc phich Duration \\
gsm a extension & nas eps emm iwkn26 & lte rrc m TMSI & lte rrc mnc \\
e212 assoc imsi & nas eps emm spare half & lte rrc setup element & lte rrc rrcConnectionRequest r8 element \\
nas eps emm eps att & nas eps emm cause & lte rrc pdsch ConfigDedicated element & lte rrc message \\
nas eps emm cp ciot & nas eps emm detach type & lte rrc csg Indication & lte rrc ackNackRepetition \\
e212 tai mnc & nas eps emm detach req & lte rrc rrcConnectionRelease element & ws expert \\
nas eps emm mme grp & gsm a dtap dst adjustment & lte rrc ul SCH Config & lte rrc p a \\
nas eps emm epco & nas eps emm 15 bearers & lte rrc uplinkPowerControlDedicated element & lte rrc ulInformationTransfer element \\
nas eps emm tsc & nas eps emm er wo & lte rrc sr PUCCH ResourceIndex & lte rrc dedicatedInfoNAS \\
nas eps emm restrict dcnr & gsm a dtap timezone & lte rrc systemInfoModification & lte rrc UL DCCH Message \\
nas eps emm hc cp & nas eps emm imeisv req & lte rrc securityModeCommand r8 element & lte rrc securityModeComplete r8 element \\
e212 imsi & gsm a dtap text string & lte rrc cellAccessRelatedInfo element & lte rrc securityConfigSMC element \\
nas eps emm nas key setid & gsm a id dig 1 & lte rrc timeAlignmentTimerDedicated & lte rrc messageClassExtension element \\
e212 mnc & nas eps emm 128eea1 & lte rrc PCCH Message element & lte rrc srb ToAddModList \\
nas eps emm switch off & nas eps emm s1 u & lte rrc cqi pmi ConfigIndex & per small number bit \\
nas eps emm tai n & gsm a oddevenind & lte rrc trackingAreaCode & lte rrc q RxLevMin \\
gsm a spare bits & nas eps emm eea6 & lte rrc lateNonCriticalExtension & lte rrc pagingRecordList \\
nas eps emm eia3 & gsm a dtap autn & lte rrc plmn IdentityList & lte rrc pusch ConfigDedicated element \\
nas eps emm eea7 & nested field3 & lte rrc mcc & lte rrc rrcConnectionSetup r8 element \\
nas eps emm ims vops & gsm a gm gmm tmsi & per open type length & ws expert group \\
nas eps emm odd even & nas eps emm toi & per num sequence extensions & lte rrc reestablishmentCause \\
gsm a gm elem id & gsm a dtap rand & lte rrc p0 UE PUSCH & lte rrc rrcConnectionRequest element \\
nas eps emm eia7 & nas eps emm 128eia2 & lte rrc retxBSR Timer & lte rrc BCCH BCH Message \\
nas eps emm eia5 & nas eps emm tai tol & lte rrc sr SubframeOffset & lte rrc securityAlgorithmConfig element \\
nas eps emm toc & nas eps emm res & lte rrc BCCH DL SCH Messageelement & lte rrc integrityProtAlgorithm \\
gsm a dtap time zone & nas eps emm elem id & lte rrc rrcConnectionReconfiguration r8 element & lte rrc selectedPLMN Identity \\
nas eps emm tai tac & nas eps spare bits & lte rrc systemInfoModification eDRX r13 & lte rrc widebandCQI element \\
gsm a common elem id & ws expert & lte rrc transmissionMode & lte rrc ueCapabilityInformation element \\
nas eps emm m tmsi & gsm a dtap number of sparebits & lte rrc simultaneousAckNackAndCQI & lte rrc measResultLastServCell r9 element \\
gsm a ie mobileid type & nested field4 & lte rrc cmas Indication r9 & lte rrc cipheringAlgorithm \\
nas eps emm esr ps & nas eps emm hash mme & lte rrc mac MainConfig & lte rrc securityModeCommand element \\
gsm a imeisv & nas eps emm active flg & per extension bit & lte rrc rach ReportReq r9 \\
nas eps emm esm msg & nested field1 & lte rrc dedicatedInfoNASList & lte rrc rrcConnectionSetupComplete element \\
nas eps emm id type2 & e212 mcc & per optional field bit & lte rrc systemInfoValueTag \\
3gpp tmsi & gsm a gm gmm gprs & lte rrc drb ToAddModList & lte rrc p0 UE PUCCH \\
gsm a filler & nas eps emm eea0 & per enum index & lte rrc rlf Report r9 \\
nas eps security header type & nas eps emm 128eia1 & lte rrc explicitValue element & lte rrc radioResourceConfigDedicated element \\
nas eps emm eea5 & gsm a len & lte rrc mmec & lte rrc betaOffset CQI Index \\
e212 tai mcc & gsm a dtap coding scheme & lte rrc rlf ReportReq r9 & lte rrc cqi ReportPeriodic \\
nas eps emm eia4 & nas eps emm ebi0 & lte rrc dsr TransMax & lte rrc systemInfoUnchanged BR r15 \\
nested field2 & nas eps emm ebi13 & lte rrc cellIdentity & lte rrc ue CapabilityRAT ContainerList \\
nas eps seq no & nas eps emm ebi2 & lte rrc quantityConfig element & lte rrc sibs changing \\
nested field5 & nas eps emm ebi14 & lte rrc cqi FormatIndicatorPeriodic & lte rrc release element \\
nas eps emm cs lcs & nas eps emm ebi12 & lte rrc phich Resource & lte rrc dedicatedInfoType \\
nas eps emm restrict ec & nas eps emm ebi7 & lte rrc freqBandIndicator & lte rrc rsrpResult r9 \\
nas eps nas msg emm & nas eps emm ebi4 & lte rrc ue TransmitAntennaSelection & lte rrc cellGlobalId r10 element \\
nas eps emm guti type & nas eps emm ebi8 & lte rrc cqi ReportConfig element & lte rrc pSRS Offset \\
nas eps emm up ciot & nas eps emm ebi1 & lte rrc ueCapabilityEnquiry element & lte rrc ueCapabilityEnquiry r8 element \\
gsm a L3 protocol discriminator & nas eps emm ebi9 & lte rrc prohibitPHR Timer & lte rrc cellBarred \\
nas eps emm update type & nas eps emm ebi3 & per octet string length & lte rrc antennaInfo \\
nas eps emm type of & nas eps emm ebi10 & lte rrc UL CCCH Message & lte rrc cqi PUCCH ResourceIndex \\
nas eps emm 128eea2 & nas eps emm ebi11 & lte rrc securityModeComplete element & lte rrc sr Periodicity \\
nested field6 & nas eps emm ebi6 & lte rrc betaOffset RI Index & lte rrc ue Identity element \\
nas eps emm mme code & nas eps emm ebi15 & lte rrc ueInformationResponse r9 element & lte rrc pucch ConfigDedicated element \\
nas eps emm eea3 & nas eps emm eps update resultvalue & lte rrc connectionFailureType r10 & lte rrc deltaMCS Enabled \\
nas eps emm epc lcs & nas eps emm ebi5 & lte rrc periodicPHR Timer & lte rrc failedPCellId r10 \\
nas eps emm emc bs &  & lte rrc nomPDSCH RS EPRE & lte rrc dlInformationTransfer element \\
\hline
\end{tabular}
\caption{List of subset of features}
\label{tab:list-of-features}
\end{table*}

\section{Unseen and Reshaped Attack Detection}
The performance of \system in detecting unseen and reshaped attacks are shown in Tables~\ref{tab:cross-val-unseen-attack} and~\ref{tab:reshaped-attack}, respectively.

\begin{table*}
\centering
\renewcommand{\arraystretch}{1}
\fontsize{6}{6}\selectfont
\setlength{\tabcolsep}{1pt} 
\begin{tabular}{L{2.5cm}|C{0.6cm}|C{0.6cm}|C{0.6cm}|C{0.6cm}|C{0.6cm}|C{0.6cm}|C{0.6cm}|C{0.6cm}|C{0.6cm}|C{0.6cm}|C{0.6cm}|C{0.6cm}|C{0.6cm}|C{0.6cm}|C{0.6cm}|C{0.6cm}|C{0.6cm}|C{0.6cm}|C{0.6cm}|C{0.6cm}|C{0.6cm}|C{0.6cm}|C{0.6cm}}
\hline
\textbf{Actual $\backslash$ Predicted} & \rotatebox{90}{Benign}  & \rotatebox{90}{Authentication relay attack} & \rotatebox{90}{IMSI catching} & \rotatebox{90}{Paging Channel Hijacking Attack} & \rotatebox{90}{Location tracking via measurement reports} & \rotatebox{90}{Capability Hijacking} & \rotatebox{90}{Bidding down with \M{AttachReject}} & \rotatebox{90}{Bidding down with \M{TAUreject}} & \rotatebox{90}{Bidding down with \M{ServiceReject}} & \rotatebox{90}{Mobile Network Mapping (MNmap)} & \rotatebox{90}{Energy Depletion Attack} & \rotatebox{90}{Panic Attack} & \rotatebox{90}{Stealthy Kickoff Attack} & \rotatebox{90}{Incarceration with \M{rrcReject}} & \rotatebox{90}{Incarceration with \M{rrcReject}, \M{rrcRelease}} & \rotatebox{90}{Incarceration with \M{rrcReestablishReject}} & \rotatebox{90}{NAS Counter Desync Attack} & \rotatebox{90}{X2 signalling flood} & \rotatebox{90}{Handover hijacking} & \rotatebox{90}{RRC replay attack} & \rotatebox{90}{Lullaby attack with \M{rrcReconfiguration}} & \rotatebox{90}{Lullaby attack with \M{rrcReestablishRequest}} & \rotatebox{90}{Lullaby attack with \M{rrcResume}} \\
\hline
Authentication relay attack & 0 & 0 & 2.52 & 0 & 0 & 4.86 & 3.81 & 0.19 & 7.78 & 0 & 0 & \textbf{54.49} & 0 & 4.1 & 0 & 2.99 & 6.63 & 0.78 & 7.33 & 0 & 0 & 0 & 4.52 \\
\hline
IMSI catching &  0 & 0 & 0 & 4.38 & 5.2 & \textbf{58.77} & 0 & 0 & 2.85 & 0 & 5.49 & 0 & 4.33 & 5.5 & 5.03 & 1.23 & 4.75 & 0 & 1.87 & 0 & 0 & 0 & 0.6 \\
\hline
Paging Channel Hijacking Attack & 2.7 & 4.92 & \textbf{46.03} & 0 & 4.26 & 0 & 6.74 & 0 & 0 & 3.53 & 7.34 & 3.57 & 0 & 5.86 & 4.1 & 7.12 & 0 & 0 & 0 & 3.82 & 0 & 0 & 0 \\
\hline
Location tracking via measurement reports &  0 & 0 & 2.51 & 0 & 0 & 2.31 & 4.88 & 8.13 & 0 & \textbf{45.87} & 1.88 & 3.53 & 0 & 3.37 & 0 & 2.51 & 7.69 & 0 & 0 & 0 & 9.33 & 0 & 7.99 \\
\hline
Capability Hijacking &  0 & 5.81 & 0.12 & 2.98 & 10.76 & 0 & 2.45 & 0 & 0 & 0 & 5.04 & 0 & \textbf{44.34} & 0 & 0 & 3.31 & 7.08 & 5.24 & 8.94 & 3.93 & 0 & 0 & 0  \\
\hline
Bidding down with \M{AttachReject} &  0 & 0 & 8.37 & 0 & 0 & 6.72 & 0 & 1.85 & 0 & 2.35 & 0 & 0 & 8.69 & 0 & 0 & \textbf{33.08} & 0 & 11.05 & 5.44 & 5.76 & 8.33 & 0.74 & 7.62 \\
\hline
Bidding down with \M{TAUreject} & 0 & 3.28 & 0.49 & 3.69 & 0 & \textbf{52.98} & 0 & 0 & 3.91 & 0 & 5.04 & 3.81 & 0 & 7.89 & 7.19 & 0 & 1.59 & 0 & 2.98 & 0 & 0 & 0 & 7.14  \\
\hline
Bidding down with \M{ServiceReject} &  0 & 0 & 11.58 & 2.87 & \textbf{32.18} & 0 & 0 & 1.54 & 0 & 11.0 & 0.47 & 7.98 & 0.78 & 0 & 0 & 0 & 0 & 7.44 & 1.82 & 14.77 & 7.55 & 0 & 0  \\
\hline
Mobile Network Mapping (MNmap) & 0 & 3.81 & \textbf{32.37} & 1.18 & 9.24 & 0 & 7.91 & 0 & 14.86 & 0 & 3.62 & 0 & 3.64 & 0 & 0 & 0 & 0 & 4.82 & 0 & 4.37 & 0 & 12.3 & 1.88 \\
\hline
Energy Depletion Attack & 0 & 5.68 & 0 & 0 & 0 & 0 & 2.1 & 4.38 & 5.57 & 5.44 & 0 & 1.54 & 0 & 2.51 & 0 & 0 & 0 & 0 & 4.74 & \textbf{59.36} & 0.43 & 5.19 & 3.06 \\
\hline
Panic Attack & 9.46 & 4.94 & 10.28 & 1.79 & 4.4 & 0 & 0.49 & 3.75 & 0 & 0 & 9.97 & 0 & 2.05 & 0 & 0 & 4.85 & 0 & 0 & 9.95 & 0 & 0 & 0 & \textbf{38.08} \\
\hline
Stealthy Kickoff Attack & 0 & 0.05 & \textbf{45.69} & 4.6 & 5.93 & 3.81 & 0 & 0 & 0 & 9.76 & 0.9 & 0 & 0 & 0 & 0 & 0 & 9.45 & 0 & 5.06 & 6.17 & 0.86 & 7.72 & 0 \\
\hline
Incarceration with \M{rrcReject} & 0 & 0 & 0.91 & 0 & 1.18 & 0 & 4.69 & 8.1 & 2.89 & 0 & 3.42 & 0 & 0 & 0 & 3.6 & 0.52 & 0 & 5.43 & 0 & 0 & \textbf{59.69} & 7.81 & 1.76 \\
\hline
Incarceration with \M{rrcReject}, \M{rrcRelease} & 11.14 & 9.48 & 3.91 & 3.77 & 2.29 & 0 & 0 & 0 & 0 & 0 & 5.86 & 0 & 10.08 & 0 & 0 & 5.31 & \textbf{30.37} & 6.55 & 8.28 & 0 & 0 & 2.93 & 0 \\
\hline
Incarceration with \M{rrcReestablishReject} & 0 & 0 & 0 & 2.63 & 0 & 6.5 & 1.07 & 6.05 & 3.14 & 0 & \textbf{59.17} & 0 & 0.99 & 3.72 & 0 & 0 & 0 & 2.21 & 0 & 5.61 & 5.48 & 3.43 & 0 \\
\hline
NAS Counter Desync Attack & 0 & 6.36 & 0 & 0 & 7.47 & 8.94 & \textbf{36.38} & 0.69 & 0 & 0 & 10.8 & 0 & 9.49 & 0 & 3.69 & 0 & 0 & 2.13 & 0 & 6.21 & 3.22 & 0 & 4.61 \\
\hline
X2 signalling flood & 0 & \textbf{39.04} & 9.96 & 4.0 & 3.6 & 0 & 12.12 & 0 & 0.68 & 1.09 & 0 & 0 & 1.31 & 0 & 10.01 & 0 & 0 & 0 & 2.66 & 5.98 & 0 & 0 & 9.55 \\
\hline
Handover hijacking &  11.36 & 11.37 & 3.96 & 0.07 & 0 & 0 & 0 & 0.35 & 0 & \textbf{32.35} & 7.79 & 0 & 6.37 & 2.59 & 0 & 9.03 & 0 & 0 & 0 & 8.8 & 0 & 0 & 5.96 \\
\hline
RRC replay attack & 2.66 & 1.77 & 4.47 & 7.4 & 0 & 7.31 & 0 & 0 & 0 & 0 & 0 & 5.77 & 0 & 0 & 11.53 & 0.99 & 9.91 & 1.04 & 0 & 0 & 5.21 & 0 & \textbf{41.93} \\
\hline
Lullaby attack with \M{rrcReconfiguration} & 0 & 0 & 0 & 0 & 0 & 0 & 10.05 & 8.14 & 8.58 & 0 & 4.09 & 0 & 6.69 & \textbf{35.68} & 5.78 & 0 & 6.95 & 4.6 & 4.11 & 5.14 & 0 & 0 & 0.18 \\
\hline
Lullaby attack with \M{rrcReestablishRequest} &  0 & 0 & 7.55 & 3.97 & 0 & \textbf{52.56} & 0 & 0 & 0 & 0 & 0 & 1.59 & 3.87 & 8.35 & 3.66 & 5.54 & 5.68 & 3.19 & 1.43 & 2.62 & 0 & 0 & 0 \\
\hline
Lullaby attack with \M{rrcResume} &  5.99 & 0 & 3.77 & 0 & 0 & 8.87 & 4.76 & 4.59 & 0 & 1.15 & \textbf{48.12} & 0 & 7.36 & 4.53 & 0 & 0 & 0 & 0.71 & 0 & 0 & 8.23 & 1.92 & 0 \\
\hline
\end{tabular}
\caption{Cross validation for unseen attack detection (shown in percentage)}
\vspace{-0.6cm}
\label{tab:cross-val-unseen-attack}
\end{table*}

\begin{table*}
\centering
\renewcommand{\arraystretch}{1}
\fontsize{6}{6}\selectfont
\setlength{\tabcolsep}{1pt} 
\begin{tabular}{L{2.5cm}|c|c|C{0.6cm}|C{0.6cm}|C{0.6cm}|C{0.6cm}|C{0.6cm}|C{0.6cm}|C{0.6cm}|C{0.6cm}|C{0.6cm}|C{0.6cm}|C{0.6cm}|C{0.6cm}|C{0.6cm}|C{0.6cm}|C{0.6cm}|C{0.6cm}|C{0.6cm}|C{0.6cm}|C{0.6cm}|C{0.6cm}|C{0.76cm}|C{0.6cm}}
\hline
\textbf{Reshaped $\backslash$ Predicted} & \rotatebox{90}{Benign} & \rotatebox{90}{FBS} & \rotatebox{90}{Energy Depletion Attack} & \rotatebox{90}{NAS Counter Desync Attack} & \rotatebox{90}{X2 signalling flood} & \rotatebox{90}{Paging Channel Hijacking Attack} & \rotatebox{90}{Handover hijacking} & \rotatebox{90}{Incarceration Attack} & \rotatebox{90}{Panic Attack} & \rotatebox{90}{Stealthy Kickoff Attack} & \rotatebox{90}{Authentication relay attack} & \rotatebox{90}{IMSI catching} & \rotatebox{90}{Location tracking via measurement reports} & \rotatebox{90}{Capability Hijacking} &  \rotatebox{90}{Bidding down with \M{AttachReject}} & \rotatebox{90}{Bidding down with \M{TAUreject}} & \rotatebox{90}{Bidding down with \M{ServiceReject}} & \rotatebox{90}{Mobile Network Mapping (MNmap)} & \rotatebox{90}{Incarceration with \M{rrcReject}, \M{rrcRelease}} & \rotatebox{90}{Incarceration with \M{rrcReestablishReject}} & \rotatebox{90}{RRC replay attack} & \rotatebox{90}{Lullaby attack with \M{rrcReconfiguration}} & \rotatebox{90}{Lullaby attack with \M{rrcReestablishRequest}} & \rotatebox{90}{Lullaby attack with \M{rrcResume}}  \\
\hline
FBS & 10 & \textbf{90} & - & - & - & - & - & - & - & - & - & - & - & - & - & - & - & - & - & - & - & - & - & -  \\
\hline
Energy Depletion Attack & 7.3 & - & \textbf{44.73} & 0 & 3.0 & 0 & 0 & 2.14 & 9.55 & 0 & 6.48 & 4.24 & 0 & 0 & 5.24 & 7.62 & 2.42 & 0 & 4.14 & 0 & 0 & 3.14 & 0 & 0  \\
\hline
NAS Counter Desync Attack & 0.59 & - & 5.99 & \textbf{42.07} & 7.69 & 0 & 0 & 0 & 0 & 6.23 & 0 & 7.31 & 0 & 4.59 & 1.64 & 0 & 7.93 & 0 & 6.15 & 7.92 & 0 & 0 & 1.89 & 0  \\
\hline
X2 signalling flood & 3.04 & - & 0 & 0 & \textbf{48.18} & 7.98 & 0 & 0 & 7.47 & 0.76 & 0 & 4.49 & 0 & 0 & 0 & 8.26 & 0.87 & 1.46 & 5.87 & 8.47 & 0 & 0 & 0 & 3.14 \\
\hline
Paging Channel Hijacking Attack & 8.94 & - & 7.74 & 0.78 & 0 & \textbf{39.15} & 0 & 4.57 & 0 & 5.76 & 0 & 0 & 6.4 & 0 & 0.58 & 8.01 & 0 & 4.4 & 0 & 5.52 & 0 & 0 & 0 & 0 \\
\hline
Handover hijacking & 4.27 & - & 0 & 0.2 & 9.15 & 10.02 & \textbf{43.46} & 0 & 0 & 0 & 0 & 3.79 & 1.78 & 0 & 6.81 & 3.59 & 0 & 0 & 0 & 0 & 10.51 & 0 & 0.46 & 0 \\
\hline
Incarceration with \M{rrcReject} & 0 & - & 0 & 6.25 & 0 & 0 & 6.1 & \textbf{46.82} & 0 & 3.19 & 0 & 5.39 & 2.11 & 0 & 6.05 & 0 & 5.09 & 0 & 0 & 0 & 6.81 & 5.91 & 0 & 2.06 \\
\hline
Panic Attack & 0 & - & 8.2 & 0 & 0 & 3.73 & 0 & 0.72 & \textbf{30.35} & 7.44 & 8.5 & 0 & 5.96 & 6.96 & 6.77 & 0 & 0 & 7.85 & 0 & 0 & 10.18 & 0 & 0 & 3.34 \\
\hline
Stealthy Kickoff Attack & 0 & - & 0.19 & 1.48 & 0 & 0 & 6.25 & 8.26 & 0 & \textbf{53.01} & 4.04 & 0 & 6.35 & 0 & 3.44 & 0 & 0 & 3.92 & 1.53 & 0 & 0 & 0 & 9.55 & 1.98 \\
\hline
Authentication relay attack &  0 & - & 0 & 0 & 0 & 0 & 4.59 & 0 & 2.96 & 0 & \textbf{49.04} & 0 & 5.87 & 0 & 6.61 & 0 & 0 & 1.39 & 4.17 & 5.72 & 0 & 4.16 & 2.04 & 4.41  \\
\hline
IMSI catching & 9.49 & - & 0 & 0.65 & 0 & 4.31 & 0 & 7.61 & 5.35 & 0 & 0 & \textbf{48.84} & 0 & 0 & 0 & 5.95 & 0 & 0 & 7.6 & 0 & 0 & 1.87 & 0.71 & 2.5 \\
\hline
Location tracking via measurement reports & 4.45 & - & 8.32 & 0 & 0 & 0 & 0 & 8.76 & 0 & 0 & 8.19 & 0 & \textbf{33.69} & 0 & 0 & 0 & 9.83 & 0 & 8.15 & 0 & 7.93 & 7.57 & 1.3 & 0.04 \\
\hline
Capability Hijacking & 0 & - & 2.3 & 0 & 0 & 5.44 & 2.39 & 2.59 & 0 & 6.21 & 0 & 7.68 & 0 & \textbf{57.34} & 6.26 & 0 & 0 & 0.21 & 0.48 & 2.66 & 0 & 6.43 & 0 & 0 \\
\hline
Bidding down with \M{AttachReject} & 0 & - & 0 & 3.58 & 3.8 & 0 & 0 & 0 & 5.19 & 0 & 5.62 & 0 & 0 & 5.18 & \textbf{56.11} & 0 & 2.64 & 5.85 & 0 & 1.38 & 5.28 & 0 & 0 & 1.67 \\
\hline
Bidding down with \M{TAUreject} & 0 & - & 0 & 6.26 & 0 & 0 & 1.88 & 0 & 4.74 & 4.71 & 6.35 & 3.9 & 1.37 & 1.41 & 0 & \textbf{58.96} & 0 & 0 & 0 & 4.41 & 0 & 1.09 & 4.92 & 0 \\
\hline
Bidding down with \M{ServiceReject} & 5.45 & - & 0 & 0 & 0 & 9.07 & 7.98 & 0 & 0 & 0 & 6.88 & 0 & 8.38 & 0 & 3.98 & 0 & \textbf{38.83} & 6.67 & 0 & 0 & 1.63 & 0.89 & 5.15 & 0 \\
\hline
Mobile Network Mapping (MNmap) & 0 & - & 0.74 & 0 & 0 & 0 & 2.18 & 0 & 0 & 0 & 0 & 1.47 & 0 & 7.84 & 5.37 & 3.87 & 0 & \textbf{38.72} & 10.38 & 0 & 10.95 & 2.31 & 0 & 8.05 \\
\hline
Incarceration with \M{rrcReject}, \M{rrcRelease} & 8.34 & - & 1.45 & 0 & 0 & 0 & 8.99 & 0 & 3.63 & 12.11 & 0 & 0 & 6.2 & 11.88 & 5.06 & 0 & 0 & 1.23 & \textbf{36.67} & 0 & 1.13 & 0 & 0 & 3.32 \\
\hline
Incarceration with \M{rrcReestablishReject} & 0 & - & 0 & 0 & 4.49 & 0 & 0 & 6.84 & 9.34 & 0 & 0 & 0 & 3.34 & 0 & 5.78 & 3.15 & 0 & 0 & 0 & \textbf{39.3} & 8.67 & 1.54 & 8.14 & 6.3 \\
\hline
RRC replay attack & 0 & - & 0 & 2.24 & 4.8 & 0 & 0 & 0.15 & 0 & 6.47 & 0 & 11.33 & 7.34 & 7.02 & 7.77 & 0 & 0 & 1.89 & 0 & 10.59 & \textbf{36.14} & 0 & 0 & 0 \\
\hline
Lullaby attack with \M{rrcReconfiguration} & 0 & - & 5.65 & 7.64 & 8.75 & 0 & 6.27 & 7.53 & 0 & 0 & 5.67 & 5.45 & 0 & 0 & 0 & 0 & 8.65 & 3.47 & 3.14 & 0.3 & 0 & \textbf{37.46} & 0 & 0 \\
\hline
Lullaby attack with \M{rrcReestablishRequest} & 9.94 & - & 4.39 & 0 & 9.95 & 5.59 & 0 & 0 & 0 & 2.29 & 0 & 0 & 0 & 0 & 0.54 & 6.73 & 8.35 & 2.93 & 6.56 & 0 & 10.13 & 0 & \textbf{32.6} & 0 \\
\hline
Lullaby attack with \M{rrcResume} & 0 & - & 4.42 & 5.68 & 2.28 & 2.33 & 0 & 0 & 4.43 & 0 & 0 & 0 & 5.77 & 3.81 & 1.73 & 0 & 0 & 5.94 & 0 & 2.42 & 0 & 0 & 0 & \textbf{59.66} \\
\hline
\end{tabular}
\caption{Reshaped attack detection (shown in percentage)}
\label{tab:reshaped-attack}
\vspace{-0.6cm}
\end{table*}

% That's all, folks
\end{document}